\documentclass[pre,twocolumn,floatfix,groupedaddress]{revtex4-2}

\usepackage{microtype}
\usepackage[utf8]{inputenx}
\usepackage{physics}                            
\usepackage{bm}									
\usepackage{enumerate}							
\usepackage{amsthm}								
\usepackage{comment}							
\usepackage[caption=false]{subfig}
\usepackage{amssymb}							
\usepackage{t1enc}								
\usepackage{mathtools}							
\usepackage{tensor}								
\usepackage{wrapfig}							
\usepackage{listings}							
\usepackage{xcolor}								
\usepackage{hyperref}
\hypersetup{colorlinks=true}
\usepackage[capitalize]{cleveref}               
\usepackage{graphicx,color}
\usepackage{amsfonts,amsmath}
\usepackage{xstring}                            

\newcommand{\cor}[1]{\mathcal{#1}}									
\newcommand{\T}[1]{\text{#1}}										
\newcommand{\lgraf}{\left\lbrace}									
\newcommand{\rgraf}{\right\rbrace}									
\newcommand{\dslash}[1]{\frac{\dd[d]{#1}}{(2\pi)^d}}                
\newcommand{\ps}[2]{\left( #1 ,\, #2 \right)}                         
\def \z {^{(0)}}
\def \o {^{(1)}}
\def \t {^{(2)}}
\def \a {^{(a)}}
\def \y {^{(y)}}
\def \zz {^{(z)}}

\newcommand{\eg}{e.g.}
\newcommand{\ie}{i.e.}
\newcommand{\n}{\nonumber}
\newcommand{\ccite}[1]{\IfSubStr{#1}{,}{Refs.~}{Ref.~}\cite{#1}}

\newcommand{\smallerrel}[1]{\mathrel{\mathpalette\smallerrelaux{#1}}}
\newcommand{\smallerrelaux}[2]{\raisebox{.1ex}{\scalebox{.75}{$#1#2$}}}
\newcommand{\smallin}{\smallerrel{\in}}

\begin{document}
 
\title{Inducing oscillations of trapped particles in a near-critical Gaussian field}
\author{Davide Venturelli}
\author{Andrea Gambassi}
\affiliation{SISSA -- International School for Advanced Studies and INFN, via Bonomea 265, 34136 Trieste, Italy}

\begin{abstract}
We study the non-equilibrium dynamics of two particles confined in two spatially separated harmonic potentials and linearly coupled to the same thermally fluctuating scalar field, a cartoon for optically trapped colloids in contact with a medium close to a continuous phase transition. When an external periodic driving is applied to one of these particles, a non-equilibrium periodic state is eventually reached in which their motion synchronizes thanks to the field-mediated effective interaction, a phenomenon already observed in experiments. We fully characterize the nonlinear response of the second particle as a function of the driving frequency, and in particular far from the adiabatic regime in which the field can be assumed to relax instantaneously. We compare the perturbative, analytic solution to its adiabatic approximation, thus determining the limits of validity of the latter, and we qualitatively test our predictions against numerical simulations.
\end{abstract}

\maketitle
\section{Introduction}
Objects immersed in a fluctuating medium experience induced interactions due to the constraints they impose on its fluctuating modes. Among these interactions \cite{Casimir_1948, Casimir_book, kardar99, Ajdari_1991, Golestanian_2005, Kirkpatrick_2014, Aminov_2015} are the critical Casimir forces \cite{Krech_book,  Krech_1999, Danchev_book, GambassiCCF, Maciolek} observed in classical systems close to the critical point of a second-order phase transition: they are the thermal and classical counterpart of the well-known Casimir effect in quantum electrodynamics \cite{Casimir_1948}. Even when fluctuations are negligible, particles deforming a correlated elastic medium still experience field-mediated interactions \cite{fournier2021fieldmediated,fournier_2014}. The static properties of these forces in equilibrium are by now widely understood in terms of the free energy of the system \cite{kardar99,Krech_book,Danchev_book}, but this framework is generally unable to describe the forces arising in non-equilibrium conditions, such as those determined by a moving object. In order to circumvent the difficulties which arise when imposing boundary conditions on moving surfaces, one can alternatively introduce in the total Hamiltonian some suitable interaction potentials between the field and the included objects: actual boundary conditions might be eventually recovered in the formal limit in which the interaction strength becomes infinite \cite{SYMANZIK_1981,diehl_1986,Diehl_1997}. This approach is particularly suited for studying the effects of boundary conditions imposed on randomly fluctuating surfaces, such as those of Brownian particles interacting with a correlated medium \cite{fournier_2014,Gambassi_PRL}.

Parallel to this, studying the motion of colloidal particles in contact with thermally fluctuating environments provides a tool to probe the properties of soft-matter materials, a paradigm which is well established in the field of microrheology \cite{zia2013stress,squires2005simple}. While past studies have mostly focused on the behavior of tracer particles passively carried by a fluctuating medium, in recent years increasing attention has been paid to instances in which the particle and the medium affect each other dynamically \cite{demery2010, demery2010-2, demery2011, demery2013, demerypath, Dean_2011, fuji1, fuji2}.

Particularly interesting is the case in which the medium under consideration is a fluid near a critical point, which displays long-range spatial correlations and long relaxation times. While static field-mediated effects have long since been explored \cite{kardar99}, the dynamical behavior of such systems has rarely been addressed in the literature \cite{demery2010, demery2010-2, demery2011, demery2013, demerypath,Dean_2011, fuji1, fuji2,Gambassi_2006,gambassi2008relaxation,Kruger_2011,Kruger_2012,Rohwer_2017,Hanke_2013}. We wish to start filling this gap by analyzing a simple setup and predicting the value of dynamical observables which are easily accessible in experiments. In particular, we have in mind the case of colloidal particles trapped by optical tweezers in a near-critical fluid such as a binary liquid mixture, in which one measures the average and correlation functions of their positions obtained via, \eg, digital microscopy.

In this work we study the dynamics of two probe particles, trapped and kept at a certain distance by two confining harmonic potentials, and in contact with a fluctuating medium close to the bulk critical point of a continuous phase transition. The medium is characterized by a scalar order parameter $\phi(\vb{x})$ subject to a dissipative or conserved relaxational dynamics (the so-called models A and B \cite{halperin}) within the Gaussian approximation, while we neglect hydrodynamic effects. The two overdamped Brownian particles are then made to interact with the scalar field via a translationally invariant linear coupling. Since this coupling figures in the system Hamiltonian, the particles and the field affect each other dynamically along their stochastic evolution, in such a way that detailed balance holds at all times. Simple as it may look, this model already features nonlinear and non-Markovian effects in the resulting effective dynamics of the colloids, which make analytical predictions challenging beyond perturbation theory.

A series of works  \cite{demery2011, demery2013, demerypath, Dean_2011} focused on the dynamics of an unconfined particle stochastically diffusing in contact with a scalar Gaussian field, studying the resulting effective diffusion constant. Two recent works \cite{wellGauss,ioeferraro}, instead, considered a harmonically trapped particle immersed in a field, and explored how its dynamics is affected by the presence of the latter. In particular, they focused on the average particle position during its relaxation to equilibrium, and on the autocorrelation function of the particle as it diffuses in the trap, both of which can be determined within the weak-coupling approximation. Particularly interesting was the emergence at long times of algebraic tails superimposed to the usual exponential decay of both the average position and the autocorrelation function, the exponents of which depend only on the spatial dimensionality of the system and on the critical properties of the field and therefore are characterized by a certain degree of universality. In fact, these exponents do not depend on the details of the chosen interaction potential, as long as the coupling between the field and the particle is linear and translationally invariant.
A similar setup was analyzed in \ccite{gross}, where the steady-state and effective dynamics of a colloidal particle in contact with a critical Gaussian field were computed in the presence of spatial confinement for the field. There it was shown that the steady-state distribution of the colloid position is modified by the presence of other tracer particles interacting with the same medium.

A recent experiment \cite{ciliberto} reported the observation of a temperature-controlled synchronization of the motion of two Brownian particles immersed in a binary liquid mixture close to the critical point of its demixing transition. In particular, the two colloids were trapped by two optical tweezers and their distance was periodically modulated by spatially moving one of the two traps: the synchronization then occurred upon approaching the critical temperature of the fluid. Since the electrostatic and viscous forces acting on the system turned out to be insensitive to its critical state, they could not be responsible for the observed synchronization.
These results were then explained in terms of the instantaneous action of the static critical Casimir force arising between the two colloids at equilibrium (\ie, the one computed within the Derjaguin approximation from the equilibrium force \cite{Schlesener2003,Gambassi_2009_PRE}).

Motivated by this experimental study, we aim here at investigating the possible emergence of this behaviour in our minimal model, and how it is affected by the possible retardation in the "propagation" of the force \cite{Gambassi_PRL}. In particular, we analyze the simple setup in which the center of one of the two harmonic traps is driven periodically with a tunable frequency $\Omega$, so that the system eventually reaches a non-equilibrium periodic state. Working within a weak-coupling expansion, we first derive a master equation which fully describes the motion of the colloid in the spatially fixed trap. We then obtain, in the adiabatic limit, an effective Langevin equation for its motion by integrating out the field degrees of freedom. Upon approaching criticality, it is well known \cite{halperin,onuki} that the relaxation timescale of the field grows increasingly large, thus undermining the assumption of fast relaxation which the previous adiabatic approximation scheme hinges on. Accordingly, we first analyze the dynamics in the weak-coupling approximation and then compare it to the adiabatic solution, thus determining the limits of validity of the latter and characterizing the dynamical properties of the former.

The rest of the presentation is organized as follows. In Section \ref{par:Model} we introduce the model and the notation. In \cref{par:masterequation} we study, within a weak-coupling expansion, the induced motion of one of the trapped colloids when the other colloid is forced periodically, while in \cref{par:adiabaticapproximation} we study the same quantity but within the adiabatic approximation. In \cref{par:dynamic_analysis} we characterize the weak-coupling solution and compare it with the adiabatic approximation; a comparison with numerical simulations is presented in \cref{par:numerical}. In \cref{par:many-body} we extend our framework to the case in which more than two particles are immersed in the field. We finally summarize our results in \cref{par:conclusion}.

\begin{figure}
    \centering
    \resizebox{\linewidth}{!}{\includegraphics[]{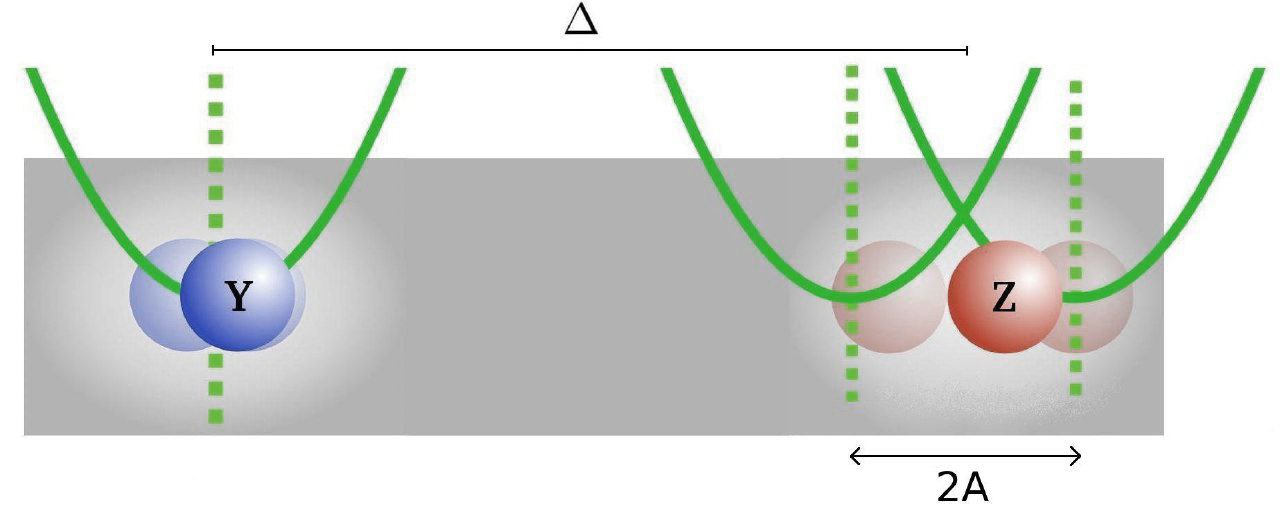}}
    \caption{Two particles of radius $R$ (blue and red spheres) are trapped in two distinct harmonic potentials spaced apart by a distance $\Delta \gg R$. The particles are immersed in a medium (grey background) represented here by a scalar Gaussian field (see Hamiltonian in \cref{eq:fullH}), and they interact with it. The centre of the trap containing the colloid $\vb{Z}$ is driven periodically in time according to \cref{eq:forcing}, with a driving amplitude $A\ll \Delta$. }
    \label{fig:setup}
\end{figure}

\section{The model} 
\label{par:Model}
The system composed by the two particles and the field is described by the Hamiltonian \cite{wellGauss,ioeferraro}
\begin{equation}
    \cor{H} = \cor{H}_\phi + \cor{U}_z +\cor{U}_y - \lambda \cor{H}_\T{int} \, ,
    \label{eq:fullH}
\end{equation}
and it is schematically represented in \cref{fig:setup}. First, the medium is modeled by a scalar Gaussian field $\phi(\vb{x},t)$ in $d$ spatial dimensions, with Hamiltonian
\begin{equation}
    \cor{H}_\phi[\phi] = \int \dd[d]{x}\left[\frac{1}{2}(\nabla\phi)^2+\frac{1}{2}r\phi^2\right] \, . \label{eq:hamiltonian_field}
\end{equation}
The parameter $r\geq 0$ measures the deviation from criticality and determines the correlation length $\xi= r^{-1/2}$ of the field fluctuations. In this simple model we neglect hydrodynamics effects and other slow variables, beyond the order parameter $\phi$, which should however be taken into account when describing the dynamics of actual fluids or binary liquid mixtures. 

The terms
\begin{equation}
    \cor{U}_y(\vb{Y}) =  \frac{k_y}{2} \vb{Y}^2 \;\;\;\;\; \T{and} \;\;\;\;\; \cor{U}_z(\vb{Z}) =  \frac{k_z}{2}\left[\vb{Z}-\vb{Z}_F(t)\right]^2
\end{equation}
in \cref{eq:fullH} represent two confining harmonic potentials with elastic constants $k_y$ and $k_z$ for the two particles. The $d$-dimensional vectors $\vb{Y}$ and $\vb{Z}$ denote the position of the centers of the particles; we will sometimes refer to them collectively as $\vb{X}_a$, with $a=y,z$. The position of the center of the second trap is externally controlled and is given by $\vb{Z}_F(t)$.

Finally, the interaction term in \cref{eq:fullH} is given by
\begin{equation}
    \cor{H}_\T{int}\left[\phi, \vb{Y},\vb{Z}\right] = \int \dd[d]{x} \phi(\vb{x})[ V^{(z)}(\vb{x}-\vb{Z})+V^{(y)}(\vb{x}-\vb{Y})]
    \label{eq:Hint}
\end{equation}
and it provides a linear and translationally invariant coupling between the particles and the field. This may physically represent, for example, the case of colloidal particles displaying preferential adsorption towards one of the two components of a binary mixture. The two interaction potentials $V\a (\vb{x})$ model the "shape" of the particles: interaction with the field occurs within the support of $V\a (\vb{x})$. For example, $V(\vb{x})=\delta(\vb{x})$ corresponds to a point-like particle, while the Gaussian potential
\begin{equation}
    V_G(\vb{x}) = (\sqrt{2\pi}R)^{-d} \exp(-|\vb{x}|^2/2 R^2)
    \label{eq:gaussianpotential}
\end{equation}
which we will mostly consider below represents a particle of radius $R$; a point-like particle is recovered in the formal limit $R \rightarrow 0$. Note that $V(\vb{x})$ is normalized so that its integral over the entire space is equal to one: this way the strength of the interaction is set only by the coupling constant $\lambda$. If the product $\lambda V\a (\vb{x})$ in \cref{eq:fullH} is chosen to be positive, then configurations are favored in which the field $\phi$ is enhanced and assumes preferentially positive values in the vicinity of and within the colloidal particles.

The field is assumed to evolve according to a relaxational dynamics \cite{halperin} involving the Hamiltonian in \cref{eq:fullH}:
\begin{align}
        &\partial_t\phi(\vb{x},t)= -D (i \grad)^\alpha \fdv{\cor{H}}{\phi(\vb{x},t)} + \zeta(\vb{x},t) \label{eq:field} \\
        &= -D(i \grad)^\alpha \left[ (r-\nabla^2)\phi-\lambda \sum_{a} V\a (\vb{x}-\vb{X}_a(t))  \right]  + \zeta \, .
        \n
\end{align}
The parameter $D$ is the field mobility, while $\alpha$ takes the value $\alpha=0$ for a non-conserved field dynamics, or $\alpha=2$ if the field is locally conserved along its evolution. Indeed, in the latter case one can rewrite $\partial_t\phi(\vb{x},t)=-\div \vb{J}(\vb{x},t) $ for a suitably chosen current $\vb{J}(\vb{x},t)$. These two choices correspond, respectively, to model A and model B dynamics in the classification of \ccite{halperin}, here considered within the Gaussian approximation. The field $\zeta(\vb{x},t)$ is a white Gaussian random noise with zero mean and variance
\begin{equation}
    \expval*{\zeta(\vb{x},t)\zeta(\vb{x}',t')}= 2DT (i \grad)^\alpha \delta^d(\vb{x}-\vb{x}')\delta(t-t') \, ,
    \label{eq:noise_corr_field}
\end{equation}
where $T$ denotes the temperature of the bath, so that the Einstein relation is satisfied. The Langevin equation for the field reads in Fourier space \footnote{We adopt here and in the following the Fourier convention $f(\vb{x}) = \int [\dd[d]{q}/(2\pi)^d] e^{i\vb{q}\cdot \vb{x}} f_{\vb{q}}$, and we normalize the delta distribution in Fourier space as $\int [\dd[d]{q}/(2\pi)^d] \delta^d(q)=1$.}
\begin{equation}
\dot{\phi}_q = -\alpha_q \phi_q +\lambda D q^\alpha \sum_a V_q\a e^{-i\vb{q}\cdot \vb{X}_a}+ \zeta_q \, ,
\label{eq:field_eom}
\end{equation}
where we introduced $\alpha_q\equiv Dq^\alpha(q^2+r)$ and where the noise satisfies
\begin{equation}
    \expval*{\zeta_q(t)\zeta_{q'}(t')} = 2DTq^\alpha  \delta^d(q+q')\delta(t-t') \, .
    \label{eq:noise_corr_field_fourier}
\end{equation}
The two particles evolve according to the overdamped Langevin equations
\begin{equation}
        \dot{\vb{Y}}(t)= -\nu_y \grad_Y \cor{H}   + \bm{\xi}^{(y)}(t) = -\gamma_y \vb{Y} + \lambda \nu_y  \vb{f}_y + \bm{\xi}^{(y)} \, , 
        \label{eq:particle_eom}
\end{equation}
where we introduced $\gamma_y \equiv \nu_y k_y$, and
\begin{align}
        \dot{\vb{Z}}(t)&= -\nu_z \grad_Z \cor{H}   + \bm{\xi}^{(z)}(t) \n \\
        &=-\gamma_z \left[\vb{Z}-\vb{Z}_F(t)\right] +  \lambda \nu_z  \vb{f}_z + \bm{\xi}^{(z)} \, .
        \label{eq:particleZ}
\end{align}
The constants $\nu_a$ denote the mobilities of the two particles, while the force $\vb{f}_a$ on each particle is given by the gradient of the interaction potential
\begin{align}
        \vb{f}_a(\vb{X}_a,\phi;t) &\equiv \nabla_{X_a}\int \dd[d]{x} \phi(\vb{x})V_a(\vb{x}-\vb{X}_a(t)) \n \\
        &= \int \dslash{q} i \vb{q}  V_{-q}\a \phi_q(t) e^{i\vb{q}\cdot \vb{X}_a(t)} \, .
    \label{eq:f}
\end{align}
Both particles are assumed to be in contact with a thermal bath at the same temperature $T$ as the field, so that $\bm{\xi}\a(t)$ are also Gaussian uncorrelated white noises satisfying the Einstein relation
\begin{equation}
    \expval*{\xi_{i}\a(t) \xi_{j}^{(b)}(t') } =  2\nu_a T \delta_{ab} \delta_{ij}\delta(t-t') \, .
    \label{eq:noise_variances}
\end{equation}
Note that, if the noise variances are chosen as in \cref{eq:noise_corr_field,eq:noise_variances}, then one expects the system to relax to a Gibbs state with the total Hamiltonian given in \cref{eq:fullH}, \ie,
\begin{equation}
    \cor{P}_\T{eq}[\phi,\vb{Y},\vb{Z}] \propto e^{-\beta \cor{H} \left[\phi,\vb{Y},\vb{Z} \right] } \, .
    \label{eq:total_boltzmann}
\end{equation}
By setting $\lambda = 0$, we obtain three non-interacting stochastic processes whose evolution is summarized in Appendix \ref{par:correlators}. They are characterized by the three relaxation timescales
\begin{align}
    &\tau_a^{-1} =\nu_a k_a \equiv \gamma_a \, , \;\;\; \T{with} \; a \in \{y,z \} \, , \label{tau:parts} \\
    &\tau_\phi^{-1}(\vb{q}) = \alpha_q = Dq^\alpha(q^2+r) \, , \label{eq:tau_phi}
\end{align}
where $\vb{q}$ is the wavevector. In particular, the relaxation time $\tau_\phi(\vb{q})$ for the long-wavelength modes of the field can become arbitrarily large for model A dynamics at criticality ($r=0$). The same happens in model B dynamics for generic values of $r$, \ie, even far from criticality ($r \neq 0$). 

In the following, we will be interested in the non-equilibrium periodic state attained at long times by the system when we apply an external periodic forcing to the center $\vb{Z}_F(t)$ of the harmonic trap of the second colloid:
\begin{equation}
    \vb{Z}_F(t) =\vb{\Delta}+ \vb{A}\sin (\Omega t) \, .
    \label{eq:forcing}
\end{equation}
Here $\bm{\Delta}$ represents the average separation between the two traps, as depicted in \cref{fig:setup}. When not specifically interested in the motion of the center $\vb{Z}(t)$ of the driven colloid, we will often adopt the \textit{deterministic} limit $k_z\rightarrow \infty$ in which the colloid follows the motion of the trap with no delay and no fluctuations, \ie, with $\vb{Z}(t)=\vb{Z}_F(t)$ (see also Appendix \ref{par:freepart}).

\section{Weak-coupling approximation}
\label{par:masterequation}
The coupled nonlinear equations \eqref{eq:field}, \eqref{eq:particle_eom} and \eqref{eq:particleZ} for the field and the two particles do not lend themselves to an analytic solution. We will then resort to a perturbative expansion of the equations of motion in powers of the coupling constant $\lambda$, and calculate the relevant observables at the lowest nontrivial order in this parameter.
One way to proceed (which has been successfully pursued in \ccite{wellGauss,ioeferraro} in the case of a single particle) is to formally expand the field and the particle coordinates as
\begin{equation}
    \phi(\vb{x},t) = \sum_{n\geqslant0} \lambda^n\phi^{(n)}(\vb{x},t) \;\;\; \T{and} \;\;\;
    \vb{X}_a(t) = \sum_{n\geqslant0} \lambda^n \vb{X}_a^{(n)}(t) \, .
    \label{eq:series_expansion}
\end{equation}
One then substitutes these expansions into the equations of motion for the field and the particles, and computes the desired observables order by order in $\lambda$; we follow this approach in Appendix \ref{par:weakcoupling} and derive the average position $\expval*{\vb{Y}(t)}$ for the sake of illustration. However, since we are mainly interested in the non-equilibrium periodic state attained by the system at long times when the colloid denoted by $\vb{Z}$ is subject to a periodic external driving, it will be convenient to work, instead, at the level of a master equation: this will make it easier to identify transient terms which play no role in the periodic state, and calculations will simplify significantly. Moreover, if one is able to derive an evolution equation for the one-point probability distribution $P_1(\vb{y},t)$, then the expectation value of any one-time observable (\eg, the variance) can be computed straightforwardly and without requiring the calculation of the corresponding perturbative series. While one generically expects the effective dynamics of the particle to be non-Markovian, and therefore not necessarily captured by a master equation for $P_1(\vb{y},t)$, we will see below that this description is however viable within the weak-coupling approximation.

\subsection{Master equation}
\label{par:masterequation_2}
Here we derive a master equation for the probability density function of the position $\vb{Y}(t)$ which is valid up to $\order{\lambda^2}$. To this aim, we start from the Langevin equation \eqref{eq:field_eom} for the field. Using the response propagator of the free field
\begin{equation}
    G_q(s_2-s_1) = e^{-\alpha_q (s_2-s_1)} \Theta(s_2-s_1) \; \label{eq:field-prop}
\end{equation}
derived in Appendix \ref{par:freefield} (where $\Theta (s)$ is the Heaviside theta function), we can solve for $\phi_q(t)$ in \cref{eq:field_eom} as 
\begin{equation}
    \phi_q (t) = \int_{t_0}^{t} \dd{s} G_q(t-s) \left[ \lambda D q^\alpha \sum_a V_q\a e^{-i\vb{q}\cdot \vb{X}_a(s)} + \zeta_q(s) \right],
    \label{eq:phi_risolto}
\end{equation}
where we set the initial condition $\phi_q(t=t_0)=0$ for simplicity, as we are interested in the long-time properties of the system. Substituting \cref{eq:phi_risolto} into \cref{eq:particle_eom}, we obtain an effective Langevin equation for the position $\vb{Y}(t)$ of the particle moving in the \textit{fixed} harmonic trap. A master equation for the associated probability distribution $P_1(\vb{y},t)$ can then be derived from its very definition
\begin{equation}
    P_1(\vb{y},t) = \expval{\delta(\vb{y}-\vb{Y}(t))} \, ,
\end{equation}
where the average is understood over all possible realizations of the stochastic noises $\zeta_q(t)$ and $\bm{\xi}_{y,z}(t)$. The equation is formally obtained by using
\begin{equation}
    \partial_t P_1(\vb{y},t) = - \grad_{\vb{y}} \cdot
    \expval{\delta(\vb{y}-\vb{Y}(t)) \dot{\vb{Y}}(t) } \, ,
    \label{eq:master_initial}
\end{equation}
and by substituting $\dot{\vb{Y}}(t)$ from the effective Langevin equation \eqref{eq:particle_eom} in which $\phi(\vb{x},t)$ has been replaced by \cref{eq:phi_risolto}. We provide the details of the calculation in Appendix \ref{par:masterequation_conti} and we report here only the final result:
\begin{align}
    &\partial_t P_1(\vb{y},t) = \cor{L}_0 P_1(\vb{y},t) + \lambda^2\cor{L}_z(t) P_1(\vb{y},t) \label{eq:master_general} \\
    &+\lambda^2 \int_{t_0}^t \dd{s}\int \dd{\vb{x}} \cor{L}(\vb{y}-\vb{x};t,s) P_2(\vb{y},t ; \vb{x}, s) + \order{\lambda^4} \, .
    \n
\end{align}
Here
\begin{equation}
    \cor{L}_0 \equiv \grad_{\vb{y}} \cdot \left(  \gamma_y \vb{y} + \nu_y T \grad_{\vb{y}} \right)
    \label{eq:L_0}
\end{equation}
is the Fokker-Planck operator for an Ornstein-Uhlenbeck particle \cite{risken}, while
\begin{equation}
    \cor{L}_z(t) \equiv \grad_{\vb{y}} \cdot \nu_y \int \dslash{q} i \vb{q} V_q\y V_{-q}\zz e^{-i\vb{q}\cdot \vb{y}} F_q\zz(t)\, , \label{eq:L_z}
\end{equation}
with
\begin{equation}
    F_q\zz(t) \equiv \int_{t_0}^t \dd{s} \chi_q(t-s) \expval*{e^{i\vb{q}\cdot \vb{Z}(s)}}_0 \, , \label{eq:F_q(z)_first}
\end{equation}
where we denoted by 
\begin{equation}
    \chi_q(s_2-s_1) = Dq^\alpha G_q(s_2-s_1) \label{eq:field-susc}
\end{equation}
the free-field susceptibility (see \cref{par:freefield}). The quantity $F_q\zz(t)$ represents an additional, nonlinear drift force due to the presence of the second colloid in position $\vb{Z}$. The average $\expval*{\dots}_0$ in \cref{eq:F_q(z)_first} is intended over the independent ($\lambda=0$) process and is computed in Appendix \ref{par:expaverages}. 
Finally, we note that \cref{eq:master_general} involves a convolution of the two-time probability distribution $P_2(\vb{y},t ; \vb{x}, s)$ with a memory kernel $\cor{L}(\vb{r};t,s)$. This is typical in non-Markovian problems, where one usually obtains a hierarchy of master equations linking the $n$-point distribution $P_n(\vb{x}_n,t_n;\vb{x}_{n-1},t_{n-1}; \dots ; \vb{x}_1,t_1)$ with $P_{n+1}$ (see for instance \ccite{Hanggi1978,Giuggioli2019}). This kernel reads (summation over the repeated indices $j$ and $k$ is implied) 
\begin{align}
    \cor{L}(\vb{r};t,s) \equiv & \, \nu_y \grad^k  \int \dslash{q} i q_k |V_q\y|^2 e^{-i\vb{q}\cdot \vb{r}} \label{eq:memory} \\
     &\times \left[ \chi_q(t-s) -i \nu_y C_q(s,t;t_0) e^{-\gamma_y (t-s)} q_j \grad^j \right] , \n
\end{align}
where $C_q(s_1,s_2;t_0)$ is the field correlator for $\lambda =0$, \ie,
\begin{equation}
    C_q(s_1,s_2) = \frac{\Omega_\phi(q)}{2\alpha_q} \left[ e^{-\alpha_q |s_2-s_1|} - e^{-\alpha_q (s_1+s_2-2t_0)} \right] \, ,
    \label{eq:freefieldcorrelator}
\end{equation}
and $\Omega_\phi(\vb{q}) \equiv 2DTq^\alpha$ (see \cref{par:freefield}). At long times, by taking the formal limit $t_0\to -\infty$, the latter renders the equilibrium form 
\begin{align}
    C_q(\tau) = \frac{T}{q^2+r} e^{-\alpha_q\abs{\tau}} \, ,
    \label{eq:C_eq}
\end{align}
with $\tau=s_2-s_1$, and the memory kernel $\cor{L}$ becomes time-translational invariant, \ie, $\cor{L}(\vb{r};t,s)=\cor{L}(\vb{r},t-s)$. Finally, in \cref{eq:memory} the notation $\grad^j$ is shorthand for $\partial / \partial r_j$.

As expected, \cref{eq:master_general} can be expressed as $\partial_t P_1(\vb{y},t)= -\grad_{\vb{y}} \cdot \vb{J}(\vb{y},t) $ for a suitably chosen current $\vb{J}(\vb{y},t)$, so that probability conservation is guaranteed. Moreover, looking at \cref{eq:L_z} one immediately observes that:
\begin{enumerate}[(i)]
    \item The contribution of the second colloid in position $\vb{Z}$ to the evolution equation of the first is only mildly non-Markovian: indeed, while $\cor{L}_z(t)$ depends on the complete past history of $\vb{Z}(t)$, it is however independent of the past history of $\vb{Y}(t)$. In the limit $k_z\rightarrow \infty$ in which the motion of $\vb{Z}(t)$ becomes deterministic, the history $\vb{Z}(t)=\vb{Z}_F(t)$ is known and the drift term in \cref{eq:L_z} becomes Markovian.
    \item The contribution of the second (and possibly of any other additional) colloid enters linearly in the master equation for $P_1(\vb{y},t)$.
\end{enumerate}
These observations may appear surprising, but in fact they apply only to the effective dynamics up to $\order{\lambda^2}$. Indeed, as discussed in \cref{par:masterequation_conti}, $P_2(\vb{y},t ; \vb{x}, s)$ at the next perturbative order in $\lambda$ satisfies a master equation completely analogous to \cref{eq:master_general} involving both $\vb{Z}(t)$ and $P_3(\vb{y},t ; \vb{x}, s; \vb{x}', s')$.

\subsection{Non-equilibrium periodic state}
\label{par:PS}
We are interested in the non-equilibrium periodic state reached at long times by the system when a periodic forcing is applied to the colloid with position $\vb{Z}(t)$, as in \cref{eq:forcing}. The task is significantly simplified when one realizes that the term containing the memory kernel $\cor{L}(t,s)$ in the master equation \eqref{eq:master_general} can be discarded in the periodic state: we prove this fact in Appendix \ref{par:machefortuna}. We are thus left with the (Markovian) master equation
\begin{equation}
    \partial_t P_1(\vb{y},t) = \cor{L}_0 P_1(\vb{y},t) + \lambda^2\cor{L}_z(t) P_1(\vb{y},t)  + \order{\lambda^4} \, ,
    \label{eq:master_PS}
\end{equation}
with $\cor{L}_z(t)$ defined in \cref{eq:L_z} and 
\begin{equation}
    F_q\zz(t) \equiv \int_0^\infty \dd{u} \chi_q(u) \expval*{e^{i\vb{q}\cdot \vb{Z}(t-u)}}_0 \, .
    \label{eq:F_q(t)}
\end{equation}
The latter coincides with \cref{eq:F_q(z)_first} after taking the limit for $t_0 \to -\infty$.
A perturbative solution of \cref{eq:master_PS} can now be found by expanding in powers of the coupling constant
\begin{equation}
    P_1(\vb{y},t) = P_1^\T{(0)}(\vb{y},t) + \lambda^2 P_1^\T{(2)}(\vb{y},t) + \order{\lambda^4} \, .
\end{equation}
This is done in Appendix \ref{par:solutionME}, where we derive an expression for $P_1^\T{(2)}(\vb{y},t)$ which can be used to compute expectation values of quantities such as the average colloid displacement from the trap center, \ie,
\begin{align}
    \expval{\vb{Y}(t)} = & -\nu_y \lambda^2 \int \dslash{q} i \vb{q} v(\vb{q}) e^{-Tq^2/(2k_y)} \n\\ & \times \int^t_{-\infty} \dd{t'} F_q\zz(t')  e^{-\gamma_y (t-t')} +\order{\lambda^4} \, ,
    \label{eq:dyn_avg_prior}
\end{align}
where we introduced for brevity 
\begin{equation}
    v(\vb{q})\equiv V\y_q V_{-q}\zz \, .
    \label{eq:v(q)}
\end{equation}
When a periodic external forcing is applied to the particle in $\vb{Z}(t)$, we expect the induced response of the particle in $\vb{Y}(t)$ to be in general nonlinear (as it is clear from \cref{eq:F_q(t)}) and therefore anharmonic, but still periodic. This suggests to look for an expression of $\expval{\vb{Y}(t)}$ in the form of a Fourier series: this is done in Appendix \ref{par:solutionME}, where we compute, up to $\order{\lambda^2}$, the cumulant generating function of the particle position
\begin{align}
    &\log \expval{e^{-i\vb{p}\cdot \vb{Y}(t)}} = -\frac{Tp^2}{2k_y}     \label{eq:cgf} \\
    & -\nu_y \lambda^2 \sum_{n\smallin\mathbb{Z}} \left[  \int \dslash{q}  e^{-\frac{Tq^2}{2k_y}} v(\vb{q}) a_n(\vb{q}) A_n(\vb{p}\cdot \vb{q}) \right] e^{i n\Omega t} \, , \n
\end{align}
where $a_n(\vb{q})$ is the $n$-th Fourier coefficient of the function $F_q\zz(t)$ defined in \cref{eq:F_q(t)}, while $A_n(\vb{p}\cdot \vb{q})$ reads 
\begin{align}
    &A_n(\vb{p}\cdot \vb{q}) \label{eq:A_n} \\
    &\equiv (\vb{p}\cdot \vb{q}) \int_0^\infty \dd{\tau} \exp[-i n\Omega \tau-\gamma_y \tau  -\frac{T}{k_y}(\vb{p}\cdot \vb{q}) e^{-\gamma_y \tau}] \, . \n
\end{align}
When a pure sinusoidal forcing is applied to the system as in \cref{eq:forcing}, the expectation value which appears in \cref{eq:F_q(z)_first} takes the simple form (see \cref{par:freepart})
\begin{align}
    \expval*{ e^{i\vb{q}\cdot \vb{Z}(t)}}_0 = \exp{-\frac{Tq^2}{2k_z}+i \vb{q}\cdot \left[ \bm{\Delta}+ \vb{A}\sin (\Omega t-\theta_z)\right] } \, .
    \label{eq:expZ}
\end{align}
For convenience we have introduced the phase shift 
\begin{equation}
    \theta_a=\arctan(\Omega/\gamma_a) \, ,
    \label{eq:phase_shift}
\end{equation}
here with $a\equiv z$, which is a measure of the delay accumulated by the colloid at point $\vb{Z}$ while following the motion of the center $\vb{Z}_F(t)$ of its harmonic trap of finite strength $k_z$. We can then use the cumulant generating function in \cref{eq:cgf} to compute the expectation value of the position and the variance of the particle $\vb{Y}$, which read
\begin{widetext}
\begin{align}
    \expval{\vb{Y}(t)} &=  \lambda^2 \sum_{n\smallin\mathbb{Z}} \frac{-i\nu_y D}{\gamma_y + i n \Omega} \left[\int \dslash{q}  \frac{\vb{q} q^\alpha  v(\vb{q})J_n(\vb{q}\cdot \vb{A})}{\alpha_q+ i n \Omega}e^{-Tq^2/(2k_p)+i\vb{q}\cdot \bm{\Delta}} \right]e^{i n(\Omega t-\theta_z)} + \order{\lambda^4} \, , \label{eq:dyn_avg} \\ 
    \expval{Y_j^2(t)}_c &=\frac{T}{k_y}\lgraf 1 -\lambda^2 \sum_{n\smallin\mathbb{Z}} \frac{\nu_y D}{2\gamma_y + i n \Omega}\left[\int \dslash{q}  \frac{q_j^2 q^\alpha  v(\vb{q})J_n(\vb{q}\cdot \vb{A})}{\alpha_q+ i n \Omega}e^{-Tq^2/(2k_p)+i\vb{q}\cdot \bm{\Delta}} \right]e^{i n(\Omega t-\theta_z)} \rgraf + \order{\lambda^4} \, ,
    \label{eq:dyn_var}
\end{align}
\end{widetext}
where $J_n$ is the modified Bessel function of the first kind. In the expressions above we introduced $k_p$ such that $1/k_p = 1/k_z+1/k_y$; in the \textit{deterministic} limit $k_z \rightarrow \infty$, one has $k_p\rightarrow k_y$ and $\theta_z \rightarrow 0$ (see \cref{eq:phase_shift}). One can also check that, since the integrand functions in \cref{eq:dyn_avg,eq:dyn_var} have a definite parity in $\vb{q}$, then the resulting expressions are real-valued.

\subsection{Effective field interpretation}
\label{par:effective_field_interpretation}
The form of the master equation \eqref{eq:master_general}, obtained in the limit of small coupling $\lambda$, lends itself to a simple physical interpretation. The original problem consisted of two colloidal particles whose reciprocal interactions are mediated by the field $\phi$, and the strength of such interactions is controlled by the coupling $\lambda$. Applying a periodic driving of $\order{\lambda^0}$ on the colloid $\vb{Z}$ induces a displacement of $\order{\lambda^2}$ on the colloid $\vb{Y}$, as shown by \cref{eq:dyn_avg,eq:dyn_var}. By the same token, any feedback reaction of $\vb{Z}$ due to $\vb{Y}$ will be at least of $\order{\lambda^4}$ and, as such, it will not contribute to the expressions discussed here, which are valid up to and including $\order{\lambda^2}$. We also noticed above that the motion of the colloid $\vb{Z}$ does not affect the memory kernel in the master equation \eqref{eq:master_general}, whose presence is thus only to be ascribed to the self-interaction of the colloid $\vb{Y}$, again mediated by the field $\phi$. Once this contribution has faded out and the long-time periodic state is reached (see the discussion in Appendix \ref{par:machefortuna}), the colloid $\vb{Y}$ is essentially moving within the mean effective field $\expval{\phi^\T{eff}}$ obtained by treating the colloid $\vb{Z}$ as a source term, \ie,
\begin{equation}
    \expval{\phi_q^\T{eff} (t)} = \lambda \int_{-\infty}^{t} \dd{s} \chi_q(t-s) V_q\zz \expval{e^{-i\vb{q}\cdot \vb{Z}(s)}}  \, ,
    \label{eq:phi_eff_avg}
\end{equation}
where again $\chi_q(u)$ is the linear susceptibility of the field reported in \cref{eq:field-susc}. Indeed, we show in Appendix \ref{par:effectivefield} how \cref{eq:dyn_avg,eq:dyn_var} for the average displacement and variance of the colloid $\vb{Y}$ can be retrieved by studying the dynamics of $\vb{Y}$ as if it were immersed into the mean effective field in \cref{eq:phi_eff_avg}, but in the absence of the second colloid $\vb{Z}$.

We can build an analogy with Casimir force calculations \cite{kardar99}, in which the Casimir energy in the presence of two surfaces can be computed by taking into account the multiple scatterings of the freely propagating field between the two surfaces -- \ie, by first considering its free propagator, which propagates fluctuations from one surface to the other, and then summing over all possible numbers of round-trip reflections \cite{Bimonte_2022}. Our perturbative calculation up to $\order{\lambda^2}$ corresponds to restricting this sum to the first scattering.

By extension, one can convince oneself that, within this weak-coupling expansion where multiple scatterings are neglected, the effect of the presence of any other particle within the same medium would simply add up to that of the particle $\vb{Z}$ in generating the effective field in \cref{eq:phi_eff_avg}. This is in contrast with other types of fluctuation-induced interactions such as Casimir forces \cite{kardar99}, which have a non-additive nature. Although we have drawn here this conclusion on the basis of a weak-coupling expansion, we will in fact verify in \cref{par:many-body} that this pairwise additivity persists beyond the perturbative regime.

\subsection{A physical bound on the value of $\lambda$}
\label{par:small_lambda}
The coupling constant $\lambda$ around which we constructed a perturbative expansion is not dimensionless: dimensional analysis of $\cor{H}_\phi$ in \cref{eq:hamiltonian_field} gives $[\phi]= d/2-1$ and accordingly $[\lambda]= 1-d/2$ for the dimensions $[\phi]$ and $[\lambda]$ of the field and the coupling, respectively, in units of inverse length.
It is thus useful to clarify what we mean by weak coupling. Hereafter, let us choose for definiteness a Gaussian interaction potential $V\a(\vb{x})$ as in \cref{eq:gaussianpotential} for both particles; assume that they have the same radius $R$, so that $v(\vb{q})=\exp(-q^2R^2)$ (see \cref{eq:v(q)}). In fact, the specific choice of the interaction potential is in general largely irrelevant \cite{ioeferraro,wellGauss} and what really matters is its characteristic lengthscale $R$, which sets a UV cutoff on the field fluctuations (see also Appendix \ref{par:effectivepotential}).

In order to obtain an upper bound on the value of the coupling constant $\lambda$ for which the perturbative expansion leads to reliable predictions, we may inspect the variance derived in \cref{eq:dyn_var} which, by definition, cannot become negative. A simple calculation (see Appendix \ref{par:upperbound}) shows that this requirement is always fulfilled if one chooses
\begin{equation}
    \lambda^2 \leq 2 d k_y \left( 2\sqrt{\pi} \widetilde{R} \right)^d \, ,
    \label{eq:upperbound}
\end{equation}
where we introduced the effective colloid radius
\begin{equation}
    \widetilde{R}^2 \equiv \frac{T}{2k_p}+R^2 \, .
    \label{eq:rtilde}
\end{equation}
Note that, in fact, this effective radius appears in \cref{eq:dyn_avg} rather than $R$ or $T$ separately. This implies that the only effect of temperature on the average particle position $\expval{\vb{Y}(t)}$ is that of renormalizing the radius $R$ of the particle by the average mean square displacement of the particle in the trap alone, which follows from equipartition theorem as $\expval*{Y_j^2}_0\sim T/k_p$.

\begin{figure}
    \centering
    \resizebox{\linewidth}{!}{\includegraphics[]{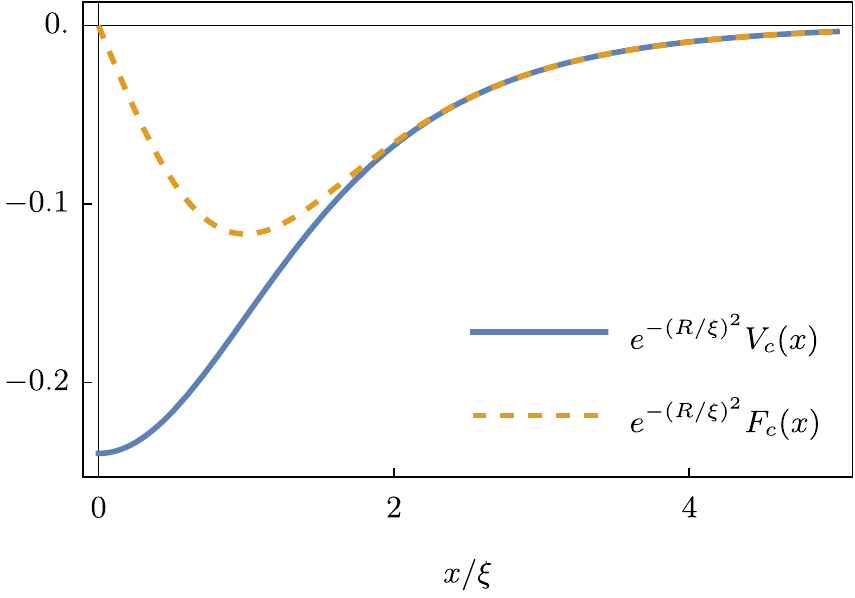}}
    \caption{Field-induced effective potential $V_c(x)$ and force $F_c(x)$ within the adiabatic approximation (in spatial dimension $d=1$), as a function of the "center-to-center" distance $x$ between the particles. They are plotted in units of the field correlation length $\xi=r^{-1/2}$ and rescaled by the $R$-dependent part of their asymptotic amplitude computed in \cref{eq:potential_asymptotic}. Here $R$ corresponds to the linear size of the colloids which characterizes the interaction potentials $V^{(a)}(\vb{x})$, chosen to be Gaussian as in \cref{eq:gaussianpotential}. The force shows a maximum at a distance $x_\T{max}$ implicitly defined by the condition in \cref{eq:max_condition}, while it approaches zero for both small and large values of $x/\xi$. The parameters used in the plot are $R=0.5$ and $r=1$.}
    \label{fig:potential}
\end{figure}

\section{Adiabatic approximation}
\label{par:adiabaticapproximation}
Any adiabatic elimination scheme \cite{risken,ioeferraro} of the field degrees of freedom $\phi_q(t)$ from the coupled equations of motion \eqref{eq:field}, \eqref{eq:particle_eom}, and \eqref{eq:particleZ} relies on the assumption that the motion of the two colloids is much slower than the relaxation timescales of the field. Note that, due to critical slowing down, this is expected to happen only sufficiently far from criticality (we will make this statement more precise later). When this is the case, the field effectively equilibrates around the instantaneous positions of the two colloids, hence distributing according to
\begin{equation}
    \cor{P}_\T{st}\left[\phi| \vb{Y},\vb{Z}\right] = \frac{1}{\cor{Z}_\T{st}(\vb{Y},\vb{Z})} e^{-\beta \left( \cor{H}_\phi - \lambda \cor{H}_\T{int} \right) } \, ,
    \label{eq:stationarydist}
\end{equation}
where $\cor{H}_\phi$ and $\cor{H}_\T{int}$ were given in \cref{eq:hamiltonian_field,eq:Hint}, respectively, and where we introduced the partition function
\begin{equation}
    \cor{Z}_\T{st}(\vb{Y},\vb{Z}) \equiv \int \cor{D}\phi \, e^{-\beta \left( \cor{H}_\phi - \lambda \cor{H}_\T{int} \right) } \, .
    \label{eq:Z_st}
\end{equation}
An effective Hamiltonian $\cor{H}_\T{eff}(\vb{Y},\vb{Z})$ describing the distribution of the particles alone can thus be obtained by marginalizing the equilibrium Boltzmann distribution in \cref{eq:total_boltzmann} over the field degrees of freedom, \ie,
\begin{align}
        \cor{P}_\T{eq}(\vb{Y},\vb{Z}) &\propto e^{-\beta \cor{H}_\T{eff}(\vb{Y},\vb{Z}) } \equiv \int \cor{D}\phi \, e^{-\beta \cor{H} \left[\phi,\vb{Y},\vb{Z} \right] } \n \\
        &= e^{-\beta \left( \cor{U}_y + \cor{U}_z \right)  } \int \cor{D}\phi \, e^{-\beta \left( \cor{H}_\phi - \lambda \cor{H}_\T{int} \right)  } \, ,
        \label{eq:canonical}
\end{align}
where the last integral is nothing but $\cor{Z}_\T{st}(\vb{Y},\vb{Z})$ in \cref{eq:Z_st}. From this partition function one can naturally derive the effective interaction potential $V_c(\vb{x})$ as
\begin{equation}
    \cor{Z}_\T{st}(\vb{Y},\vb{Z}) \propto e^{-\beta \lambda^2 V_c(\vb{Z}-\vb{Y}) } \, ,
    \label{eq:Vc_partitionfunction}
\end{equation}
and therefore from \cref{eq:canonical} it follows that
\begin{equation}
    \cor{H}_\T{eff}(\vb{Y},\vb{Z}) = \cor{U}_y(\vb{Y}) + \cor{U}_z(\vb{Z}) + \lambda^2 V_c(\vb{Z}-\vb{Y}) \, .
    \label{eq:effectiveH}
\end{equation}
The coupling to the field in the exponential of \cref{eq:Z_st} is linear, so the Gaussian integral can be performed easily (see Appendix \ref{par:effectivepotential}), resulting in
\begin{equation}
    V_c(\vb{x}) = -\int \dslash{q} \frac{v(\vb{q})}{q^2+r}  e^{i \vb{q}\cdot \vb{x}} \, .
    \label{eq:inducedpotential}
\end{equation}
In this expression we have already subtracted the self-energy contributions, \ie, the energy needed to bring each of the two particles (separately) from an infinite distance into the field: as a result, $V_c(\vb{x}\to \infty)= 0$.
An analysis of the latter is presented in Appendix \ref{par:effectivepotential} for the case of particles with rotationally invariant interaction with the field. The effective potential $V_c(\vb{x})$ is plotted in \cref{fig:potential}, together with the corresponding induced force $\vb{F}_c(\vb{x}) = -\lambda^2 \grad_{\vb{x}} V_c(\vb{x})$, in one spatial dimension and for the choice of identical Gaussian interaction potentials $V^{(a)}(\vb{x})$ between the field and the colloids. A similar qualitative behavior is observed in higher spatial dimensions and for different interaction potentials characterized by the same cutoff scale $R$. The induced force $\vb{F}_c(\vb{x})$ features a maximum at a distance $x_\T{max}$ implicitly defined by the condition in \cref{eq:max_condition}, while it decays to zero both for small and large values of $x=|\vb{x}|$. Both $V_c(\vb{x})$ and $\vb{F}_c(\vb{x})$ decay as $\exp(-x/\xi)$ when $x$ is large compared to the correlation length $\xi = r^{-1/2}$ (see \cref{eq:potential_asymptotic}). One expects in general $V_c(\vb{x})$ and $\vb{F}_c(\vb{x})$ to exhibit an algebraic decay for $r=0$ (see \cref{par:effectivepotential}), but we will not explore this issue further since we will assume that the medium has a finite (although possibly very small) correlation length $\xi$.

The colloid dynamics at the lowest order in the adiabatic approximation is then obtained by averaging the equations of motion \eqref{eq:particle_eom} and \eqref{eq:particleZ} for $\vb{Y}(t)$ and $\vb{Z}(t)$ over the stationary distribution $\cor{P}_\T{st}\left[\phi; \vb{Y},\vb{Z}\right]$ of the field $\phi$ for fixed $\vb{Y}$ and $\vb{Z}$, given in \cref{eq:stationarydist}. The resulting effective \textit{adiabatic} Langevin equation for the colloid $\vb{Y}$ subject to the fixed trap, derived in Appendix \ref{par:adiabatic_dynamics}, is
\begin{align}
    \dot{\vb{Y}}(t) &= -\nu_y k_y \vb{Y} - \nu_y \lambda^2 \nabla_y V_c(\vb{Z}-\vb{Y})  + \bm{\xi}\y \n \\
    & = -\nu_y \nabla_y \left[ \cor{U}_y(\vb{Y}) + \lambda^2  V_c(\vb{Z}-\vb{Y}) \right] + \bm{\xi}\y \, ,
    \label{eq:adiabaticlangevin}
\end{align}
which (as expected) we recognize as an overdamped Langevin dynamics computed as if the two particles interact via the effective, field-independent Hamiltonian computed in \cref{eq:effectiveH}.
We will denote as $\vb{Y}_\T{ad}(t)$ the solution of the Langevin equation \eqref{eq:adiabaticlangevin}, which reads, for small $\lambda$ (see the details in Appendix \ref{par:adiabatic_dynamics}),
\begin{align}
    \expval{\vb{Y}_\T{ad}(t)} =& - \lambda^2\nu_y \int \frac{\dd[d] {q}}{(2\pi)^d}  \frac{i \vb{q} v(\vb{q} )}{q^2+r} e^{-Tq^2/(2k_y)} \label{eq:y_adiabatic} \\
    &\times \int_{0}^\infty \dd{u} e^{-\gamma_y u}\expval{e^{i \vb{q}\cdot \vb{Z}(t-u)}}_0 +\order{\lambda^4} \, . \n
\end{align}
This expression should be compared to the actual solution of the dynamics in \cref{eq:dyn_avg}.
In Appendix \ref{par:adiabaticlimit} we show how we may recover this result starting from the dynamical expression in \cref{eq:dyn_avg_prior} and taking the formal limit $D\rightarrow \infty$ of extremely fast field relaxation, which however is only meaningful if we assume $q^\alpha(q^2+r) \neq 0$ (see \cref{eq:tau_phi}). Clearly this last condition is not fulfilled in the presence of slow modes: recalling the discussion about timescales in Section \ref{par:Model}, these modes appear in model A at criticality, but also off-criticality in model B.

\begin{figure}
\centering
  \centering
  \includegraphics[width=\columnwidth]{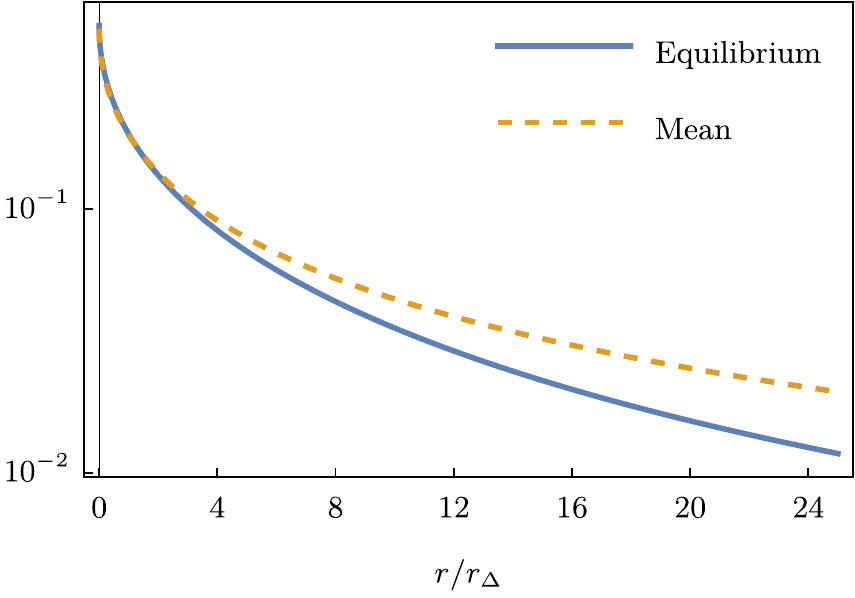}
\caption{Equilibrium position of the particle $\vb{Y}$ in the fixed-traps limit (solid line), and temporal mean value of the average position $\expval{Y(t)}$ of the particle in the fixed trap (dashed line, indicated by $b_0$ and $c_0$ in, c.f., \cref{par:mean_value}). The two curves refer to one spatial dimension, and show the behavior as a function of $r/r_\Delta=\left(\Delta/\xi \right)^2$ (see \cref{eq:r_delta}).
The position of the particle $\vb{Y}$ when it is only subject to the equilibrium attraction to the particle $\vb{Z}$ is described by \cref{eq:equilibriumposition}.
The temporal mean value $\expval{Y(t)}$ is the same in the adiabatic ($b_0$) and in the dynamical response ($c_0$), as predicted by \cref{eq:meanvalue}, and it is $\Omega$-independent. The parameters used in the plot are $\gamma_y=1$, $D=10$, $\widetilde{R}=0.7$, $\Delta=3$, and $A=1$.}
\label{fig:time_avg}
\end{figure}

\subsection{Fixed traps}
In the absence of a time-dependent external forcing, both the dynamical expression in \cref{eq:dyn_avg_prior} and the adiabatic expression in \cref{eq:y_adiabatic} describe the simple equilibrium attraction between the two particles, mediated by the field. This can be seen explicitly by fixing the position of the particle in $\vb{Z}(t)$ to a constant value $\vb{Z}\equiv \bm{\Delta}$: in both equations, the time integral can be simply computed and we get
\begin{equation}
     \expval{\vb{Y}(t)},\expval{\vb{Y}_\T{ad}(t)} \xrightarrow[\vb{Z}\equiv \bm{\Delta}]{} \frac{\lambda^2}{k_y} \int \frac{\dd[d]{q}}{(2\pi)^d}  \frac{\vb{q}\, e^{- \widetilde{R}^2 q^2}}{q^2+r} \sin(\vb{q} \cdot \bm{\Delta}) \, ,
     \label{eq:equilibriumposition}
\end{equation}
for both model A and B. This expression can be alternatively obtained (up to $\order{\lambda^2}$) by requiring that the total force $\vb{F}_\T{tot}$ acting on the colloid at position $\vb{Y}$ vanishes, \ie,
\begin{equation}
    \vb{F}_\T{tot} = -k_y \vb{Y} - \lambda^2 \grad_{\vb{y}} V_c(\vb{Z}-\vb{Y}) \equiv 0 \, ,
    \label{eq:equilibrium_condition}
\end{equation}
which corresponds to the condition of mechanical equilibrium reached when the force derived from the field-induced potential $V_c$ given in \cref{eq:inducedpotential} counterbalances the restoring attraction of the harmonic trap of strength $k_y$. In \cref{fig:time_avg} we plot the resulting equilibrium position of the particle $\vb{Y}$ as a function of $r/r_\Delta=\left(\Delta/\xi \right)^2$, having defined
\begin{equation}
    r_\Delta\equiv \Delta^{-2} \, .
    \label{eq:r_delta}
\end{equation}
The plot shows that the attraction is maximum at criticality and it decays monotonically as we increase the parameter $r$.

\subsection{Periodic driving}
Let us specialize \cref{eq:y_adiabatic} to the case in which a sinusoidal forcing is applied to one of the colloids ($\vb{Z}$) as in \cref{eq:forcing}. As for the dynamical case, we expect the response of the other colloid ($\vb{Y}$, in the static trap) to be periodic, but not harmonic. We can then expand $\expval{\vb{Y}_\T{ad}(t)}$ in Fourier series as
\begin{align}
    \expval{\vb{Y}_\T{ad}(t)} &= \sum_{n=-\infty}^\infty \vb{b}_n e^{i n \Omega t} \n \\
    &= \vb{b}_0 + 2 \sum_{n=1}^\infty |\vb{b}_n| \cos( n \Omega t  + \theta_n) \, ,
    \label{eq:Fourier_series}
\end{align}
where $|\vb{b}_n|$ and $\theta_n$ indicate the complex modulus and the phase, respectively, of the Fourier coefficients
\begin{align}
    \vb{b}_n \equiv \frac{\Omega}{2\pi} \int_0^{\frac{2\pi}{\Omega}} \dd{t} e^{-i n \Omega t}\expval{\vb{Y}_\T{ad}(t)} \, ,
    \label{eq:bn_coefficient}
\end{align}
with the property $\vb{b}_{-n}=\vb{b}^*_n$.
These coefficients can be easily computed by means of \cref{eq:expZ,eq:fourier1}, yielding
\begin{equation}
    \vb{b}_n = \frac{-i\lambda^2 e^{-i n\theta_z} }{k_y(1 + i n \Omega/\gamma_y)} \int \frac{\dd[d] {q}}{(2\pi)^d}  \frac{\vb{q} J_n(\vb{q}\cdot \vb{A}) }{q^2+r}  e^{-q^2 \widetilde{R}^2 +i\vb{q}\cdot \Delta} \, ,
    \label{eq:coeff_adiabatic}
\end{equation}
where $\widetilde{R}$ is the effective colloid radius defined in \cref{eq:rtilde}.
They are to be compared with the analogous coefficients $\vb{c}_n$ of the expansion of the \textit{dynamical} response $\expval{\vb{Y}(t)}$ which we can read from \cref{eq:dyn_avg}, \ie, 
\begin{equation}
    \vb{c}_n = \frac{-i\lambda^2 D  e^{-i n\theta_z}}{k_y(1 + i n \Omega/\gamma_y)} \int \frac{\dd[d] {q}}{(2\pi)^d}  \frac{\vb{q} q^\alpha  J_n(\vb{q}\cdot \vb{A}) }{\alpha_q+i n \Omega}  e^{-q^2 \widetilde{R}^2 +i\vb{q}\cdot \Delta}.
    \label{eq:coeff_dynamic}
\end{equation}
We discuss this comparison in Section \ref{par:dynamic_analysis}, while we focus below on the adiabatic response. In the following, we will often indicate by $b_n$, $c_n$ their vector norm $b_n \equiv \norm{\vb{b}_n}$, $b_n \equiv \norm{\vb{c}_n}$; however, one can check that their only nonzero component is the one along the direction of $\vb{A}$ and $\bm{\Delta}$. 

\subsection{Analysis of the adiabatic response}
\label{par:adiabatic_analysis}
We are interested here in studying the behavior of the adiabatic response in \cref{eq:y_adiabatic} as we vary the external driving frequency $\Omega$. To this end, it is useful to rewrite the corresponding Fourier coefficients $\vb{b}_n$ in \cref{eq:coeff_adiabatic} as
\begin{equation}
    \vb{b}_n(\Omega) = \frac{\vb{b}_n(\Omega=0)}{1+in\Omega/\gamma_y} \, ,
    \label{eq:adiabatic_reduced_coeff}
\end{equation}
where $\vb{b}_n(\Omega=0)=-i \lambda^2  e^{-i n\theta_z} \cor{I}_n/k_y$, having defined
\begin{align}
    \cor{I}_n \equiv \int \frac{\dd[d] {q}}{(2\pi)^d} \vb{q}  \frac{e^{-q^2 \widetilde{R}^2 } }{q^2+r} J_n(\vb{q}\cdot \vb{A}) e^{i\vb{q}\cdot \Delta} \, . 
    \label{eq:In}
\end{align}

\subsubsection{Mean value}
\label{par:mean_value}
The temporal mean value $\vb{b}_0\equiv \vb{b}_0(\Omega) = \vb{b}_0(\Omega=0)$ around which the oscillations occur is the same in the adiabatic and dynamical response, \ie, $\vb{c}_0=\vb{b}_0$: from \cref{eq:coeff_dynamic,eq:adiabatic_reduced_coeff}, it amounts to
\begin{equation}
    \vb{b}_0=\vb{c}_0 = \frac{\lambda^2}{k_y} \int \frac{\dd[d] {q}}{(2\pi)^d} \vb{q}  \frac{e^{-q^2 \widetilde{R}^2 } }{q^2+r} J_0(\vb{q}\cdot \vb{A}) \sin(\vb{q}\cdot \Delta) \, .
    \label{eq:meanvalue}
\end{equation}
This quantity is plotted in \cref{fig:time_avg} as a function of the correlation length $\xi$ of the field: the average is maximum at criticality, $r=0$, and it decays monotonically as $\sim r^{-1}$ as one moves away from the critical point.

We note that the temporal mean value $\vb{b}_0$ of the (anharmonic) oscillations is $\Omega$-independent, but it does \textit{not} coincide with the position of mechanical equilibrium in \cref{eq:equilibriumposition} as long as the driving amplitude $\vb{A}$ does not vanish. This is expected, since the field-induced attraction is nonlinear (see, c.f., \cref{eq:force_scaling} in Appendix \ref{par:effectivepotential} and \cref{fig:potential}). Indeed, let us analyze a single oscillation in one spatial dimension, and consider the second derivative of the induced force $h\equiv \eval{\partial_x^2 F_c(x)}_{x=x_\T{eq}} \neq 0$ computed in correspondence of the equilibrium interparticle distance $x=x_\T{eq}$ (see \cref{eq:equilibrium_condition}).
When the two particles approach each other, if $h<0$ ($h>0$), they experience a stronger (weaker) attraction which is not completely counterbalanced by a proportionally weaker (stronger) attraction felt while they are further away from each other. The net result is that they spend more (less) time close to one another than they would if the attraction were the same during the two phases of the oscillation (as it happens in a linear force gradient, for which $h=0$).

In Appendix \ref{par:linear_response} we derive again, using linear response theory, the value of the temporal average of the oscillations for small driving amplitudes $\vb{A}$: its expression is given in \cref{eq:staticpart} but it does not coincide with the value of $\vb{b}_0$ in \cref{eq:meanvalue} if not for $\vb{A}=0$. Indeed, linear response theory cannot capture the effect of the dynamical perturbation on the mean value of the oscillations, which is quadratic in $\vb{A}$ (being $J_0(x)\simeq 1-x^2/4$ for small $x$ in \cref{eq:meanvalue}).

\begin{figure}
    \centering
    \resizebox{\linewidth}{!}{\includegraphics[]{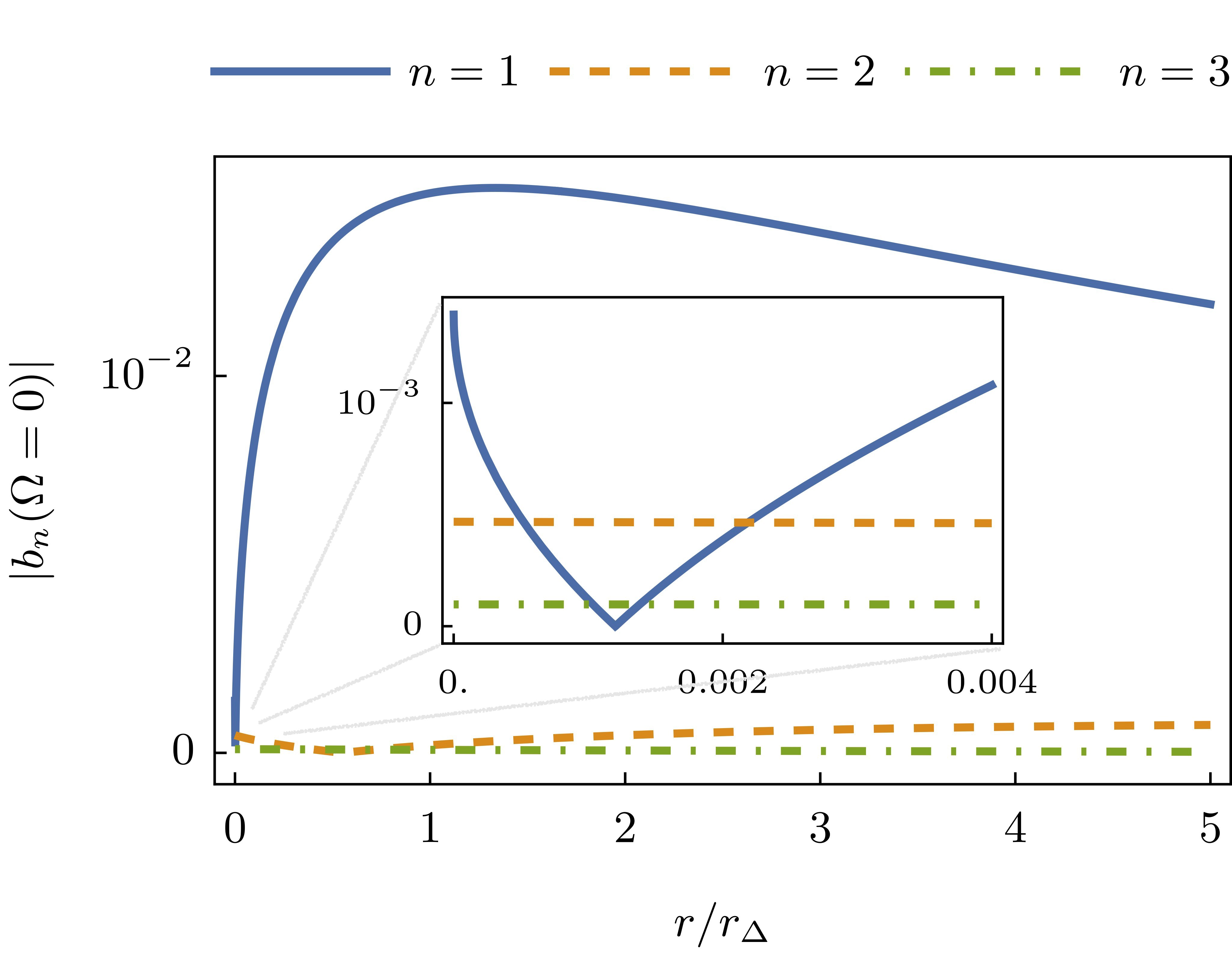}}
    \caption{Amplitude of the first three Fourier harmonics (indexed by $n$) of the adiabatic response in \cref{eq:coeff_adiabatic}, in spatial dimension $d=1$ and for $\Omega=0$. This provides an indication on the ratio of their magnitudes also for $\Omega\neq 0$, see \cref{eq:adiabatic_abs_coeff}. They are plotted as a function of $r/r_\Delta=\left(\Delta/\xi \right)^2$ (see \cref{eq:r_delta}). The adiabatic response is in general dominated by the first harmonic, but the latter is suppressed in correspondence of a specific value $r_1$ of $r$ (see the main text). The parameters used in the plot are $\gamma_y=1$, $\widetilde{R}=0.7$, $\Delta=3$, and $A=0.5$.}
    \label{fig:adiabatic_harmonics}
\end{figure}
\begin{figure}
    \centering
    \resizebox{\linewidth}{!}{\includegraphics[]{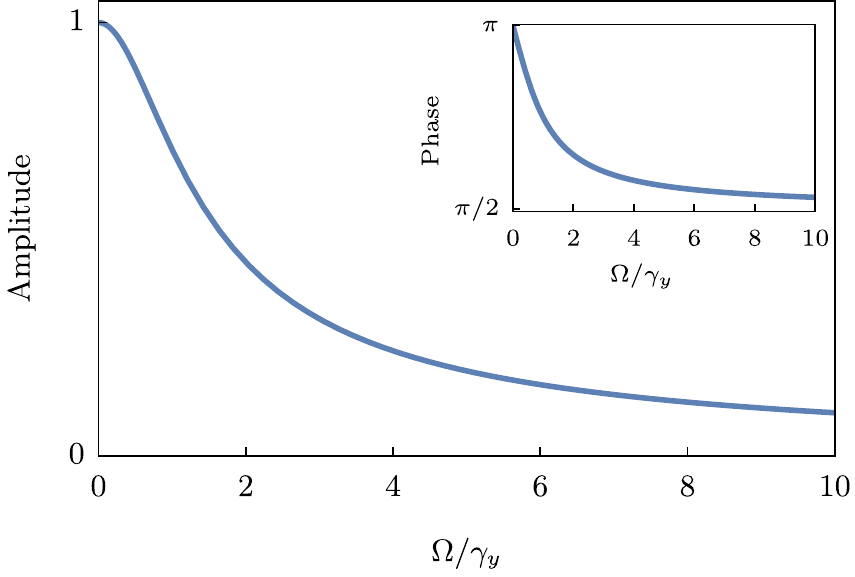}}
    \caption{Amplitude $|\vb{b}_1|$ and relative phase $\delta\theta$ of the first Fourier harmonic in the adiabatic response, see \cref{eq:coeff_adiabatic}. The amplitude is normalized by $|\vb{b}_1(\Omega=0)|$, see \cref{eq:adiabatic_abs_coeff}, and $\delta\theta$ is the phase difference with respect to the mean position of the driven colloid $\expval*{\vb{Z}(t)}_0$, see \cref{eq:phase_difference}. The curves in this plot are then independent of all the other parameters.}
    \label{fig:adiabatic_omega}
\end{figure}

\subsubsection{Amplitude}
\label{par:adiabatic_amplitude}
The amplitude of the $n$-th harmonic of $\expval{\vb{Y}_\T{ad}}$ is found by inspecting \cref{eq:adiabatic_reduced_coeff}, and it reads
\begin{equation}
    |\vb{b}_n(\Omega)| =  \frac{|\vb{b}_n(\Omega=0)|}{\sqrt{1 + (n \Omega/\gamma_y)^2}} \, . 
    \label{eq:adiabatic_abs_coeff} 
\end{equation}
It is interesting first to compare the relative magnitude of $|\vb{b}_n(\Omega=0)|$ for various $n$: they are plotted in \cref{fig:adiabatic_harmonics} as a function of the ratio $r/r_\Delta=\left(\Delta/\xi \right)^2$ (see \cref{eq:r_delta}). For $r\simeq r_\Delta$ the amplitude of the first harmonic attains a maximum: this corresponds to the correlation length $\xi$ of the field being of the same order as the average separation $\Delta$ between the two traps, \ie, $\xi \sim \Delta$. 

In general, it appears from \cref{fig:adiabatic_harmonics} that the adiabatic response is essentially and generically determined by its dominant first harmonic. Although higher harmonics become more relevant when the amplitude $A$ of the driving is much larger than the effective colloid radius $\widetilde{R}$, they still remain small compared to the first harmonic as long as $A$ and $\widetilde{R} \ll \Delta$. As an exception, however, \cref{fig:adiabatic_harmonics} shows that the first harmonic is significantly reduced at a small value of $r$ which we denote by $r_1$. Expanding for small forcing amplitudes $\vb{A}$ the equation $|(\vb{b}_1)_i|\equiv 0$ which defines $r_1$, one finds
\begin{equation}
    A^j \int \dslash{q} \frac{e^{-q^2 \widetilde{R}^2}}{q^2+r_1} q_i q_j e^{i \vb{q}\cdot \bm{\Delta}} \equiv 0 \, .
    \label{eq:def_r1}
\end{equation}
This equation turns out to be the same as the condition in \cref{eq:minimum_condition_d}, which defines the distance $\vb{x}_\T{max}$ at which the field-induced interparticle force $\vb{F}_c(\vb{x})$ is maximum (see \cref{fig:potential}), as it is clear by identifying $\vb{x}\equiv \bm{\Delta}$ and $v(\vb{q})\equiv \exp (-q^2 \widetilde{R}^2)$. The physical interpretation is the following: for $r=r_1$ and small $\vb{A}$, the average interparticle distance $\bm{\Delta}$ actually coincides with the distance $\vb{x}=\vb{x}_\T{max}$ at which the field-induced force $\vb{F}_c(\vb{x})$ is maximum. Expanding $\vb{F}_c(\vb{x})$ at the leading order around $\vb{x}=\vb{x}_\T{max}$ gives a force gradient which is at least quadratic in $|\vb{x}-\vb{x}_\T{max}|$, so that the response loses its linear component (\ie, the first harmonic in its Fourier expansion - for example, feeding $\sin(\Omega t)$ into a quadratic force gradient would render $\sin^2(\Omega t)$, whose frequency is doubled). Notice that the identification between \cref{eq:def_r1,eq:minimum_condition_d} is not accidentally due to the choice of a Gaussian interaction potential $V_q=\exp(-q^2 R^2/2)$: the generalization to another interaction potential $V'_q$ is straightforwardly obtained by replacing $\exp(-q^2 \widetilde{R}^2)\to |V'_q|^2 \exp(-q^2 T/2\kappa_p)$ in \cref{eq:def_r1} (see \cref{eq:v(q),eq:rtilde}). In both cases, we see that the only effect of the temperature $T$ is to renormalize the parameter $R$ (which characterizes $V_q$) by the mean-square displacement of the colloid in the trap; in the case in which $V_q$ is Gaussian, $R$ gets simply replaced by $\widetilde{R}$ defined in \cref{eq:rtilde}.

In Appendix \ref{par:freq_doubling} we determine the value $r_1$ of $r$ at which this frequency doubling occurs for the case $d=1$ (see \cref{eq:frequencydoubling}). However, from the above discussion it emerges that a similar qualitative behavior holds also for $d>1$, as we check within linear response theory in Appendix \ref{par:linear_response}. Indeed, the occurrence of frequency doubling relies only on the existence of a local maximum in the induced force (see \cref{fig:potential}), a feature which goes possibly beyond our particular choice of a Gaussian interaction potential $V(\vb{x})$ (see, for instance, the analysis of the theta-potential in \cref{par:effectivepotential} and that of the critical Casimir force in \ccite{dietrich98}). We anticipate here that frequency doubling is actually a feature of the adiabatic response which is observed in the full dynamical response only when the adiabatic approximation is applicable -- this will be shown below in Section \ref{par:onset}.

Finally, for any given value of $r$, \cref{eq:adiabatic_abs_coeff} shows that the amplitude $|\vb{b}_n|$ is maximum at low driving frequencies $\Omega$, while it decays as $\sim \Omega^{-1}$ upon increasing $\Omega$ beyond values which are larger than $\tau_y^{-1} \equiv \gamma_y$: this is shown in \cref{fig:adiabatic_omega}, where the amplitude $|\vb{b}_1|$ of the first harmonic is plotted as a function of $\Omega/\gamma_y$. We recall that $\tau_y$ is the timescale which characterizes the relaxation of the particle $\vb{Y}$ in its harmonic trap.

\subsubsection{Phase}
\label{par:adiabatic_phase}
When $r>r_1$ the adiabatic response is dominated by its first harmonic, which is completely characterized by its amplitude $|\vb{b}_1|$ studied above and by its phase $\theta_1$ (see \cref{eq:Fourier_series}), which we analyze here. This phase can be extracted from the complex Fourier coefficient $\vb{b}_1$ in \cref{eq:adiabatic_reduced_coeff} as
\begin{equation}
    \theta_1 = - \left( \theta_y +\theta_z +\pi/2 \right) +\pi \times \T{sign}(\cor{I}_1 )  \, ,
    \label{eq:phase_adiabatic}
\end{equation}
where $\theta_a$ is given in \cref{eq:phase_shift} and $\T{sign}(\cor{I}_1 )=\pm 1$, depending on the sign of $\cor{I}_1$ given in \cref{eq:In}.
In $d=1$ and for $r>r_1$, the integral $\cor{I}_1$ is negative: this can be checked via a numerical evaluation of \cref{eq:In} within a range of parameters compatible with our physical setting in \cref{fig:setup}, \ie, $\Delta\gg A\, ,\, \widetilde{R}$. We recall that the average motion of the driven colloid is given, at lowest order in $\lambda$, by (see Appendix \ref{par:freepart}) 
\begin{equation}
    \expval{\vb{Z}(t)}_0 = \vb{\Delta} + \vb{A}\cos (\Omega t - \theta_z - \pi/2) \, ,
    \label{eq:forced_phase}
\end{equation}
where the average is computed over the independent ($\lambda=0$) process.
By comparing \cref{eq:phase_adiabatic,eq:forced_phase} with \cref{eq:Fourier_series}, we can extract the actual phase difference $\delta \theta$ between $\expval*{\vb{Y}(t)}$ and $\expval*{\vb{Z}(t)}_0$, \ie,
\begin{equation}
    \delta \theta \equiv \theta_1 -(-\theta_z -\pi/2) = -\theta_y -\pi \, .
    \label{eq:phase_difference}
\end{equation}
In the slow-forcing limit $\Omega \ll \gamma_y$ it is $\theta_y\rightarrow 0$, and from \cref{eq:phase_difference} we deduce that the particle $\vb{Y}(t)$ moves in counterphase with respect to $\vb{Z}(t)$. This is physically expected, as the particle $\vb{Y}$ feels a stronger attraction when the particle $\vb{Z}$ is closer to it than when it is further apart. In the fast-forcing limit $\Omega \gg \gamma_y$, where $\theta_y\rightarrow \pi/2$, we get instead $\delta \theta =-3\pi/2 $: the particle $\vb{Y}(t)$ develops a $\pi/2$ phase shift with respect to the driven colloid $\vb{Z}(t)$. The situation is depicted in \cref{fig:adiabatic_omega} (inset), where we plot the phase difference $\delta \theta$ and we show that it varies smoothly by $\pi/2$ over a scale determined by $\gamma_y$.

We mention that a richer phenomenology is expected in spatial dimension $d>1$, where the direction of the driving $\vb{A}$ could in principle be chosen to be orthogonal to that of the average separation $\bm{\Delta}$ between the two traps. In this setup, one can check that the sign of the integral $\cor{I}_1$ in \cref{eq:In} is positive, so that \cref{eq:phase_difference} reads $\delta \theta = -\theta_y$. In the slow-forcing limit in which $\theta_y \to 0$, the two particles would then move in phase ($\delta \theta = 0$), as physically expected by arguing again that their attraction is stronger when they are spatially close to one another, than when they are further apart.

\section{Analysis of the dynamical response}
\label{par:dynamic_analysis}
In this section we analyze the dynamical response $\expval*{\vb{Y}(t)}$ of the particle in the fixed well, within the weak-coupling approximation given in \cref{eq:dyn_avg}. All the figures we present and discuss below refer for simplicity to the case $d=1$, but the main qualitative features of the response persist in higher spatial dimensions. 

We start by focusing on the Fourier coefficients of the dynamical response given in \cref{eq:coeff_dynamic} and by comparing them to those of the adiabatic response given in \cref{eq:coeff_adiabatic}. First and not surprisingly, they coincide for a vanishing driving frequency, \ie, $\vb{c}_n(\Omega=0)=\vb{b}_n(\Omega=0)$: their difference is only manifest in the dynamics. Secondly, a common factor $(1+in\Omega/\gamma_y)^{-1}$ multiplies both sets of coefficients, and this is the only place where the relaxation timescale $\tau_y^{-1} = \gamma_y$ of the fixed trap appears. We have seen in Section \ref{par:adiabatic_analysis} how it is this factor alone which determines the properties of the adiabatic response as a function of $\Omega$, see \cref{eq:adiabatic_reduced_coeff}; its qualitative features (amplitude, phase) are analogous to those of a low-pass filter in circuit electronics. Even though the dependence on $\Omega$ is more complicated in \cref{eq:coeff_dynamic}, this ``filter'' remains and it characterizes the dynamical response for frequencies $\Omega \geq \gamma_y$.

We noticed in \cref{par:adiabatic_amplitude} that, in general, the first Fourier harmonic dominates the adiabatic response (see \cref{fig:adiabatic_harmonics}). One can check that this is also the case for the dynamical response, both at low $\Omega$ (which is not surprising, since for $\Omega=0$ the two sets of Fourier coefficients $\vb{b}_n$ and $\vb{c}_n$ coincide) and for higher driving frequencies because, for large $\Omega$, one has $|\vb{c}_n|\sim (n\Omega)^{-2}$ from \cref{eq:coeff_dynamic}. In the following, we will then focus mostly on the analysis of the first harmonic, bearing in mind that the zeroth harmonic, \ie, the average value around which the colloid $\vb{Y}$ oscillates, is the same as that of the adiabatic approximation (see \cref{eq:meanvalue}), whose features have been described in \cref{par:mean_value}.

\subsection{Adiabatic limit}
\label{par:onset}
\begin{figure}[t]
  \centering
  \includegraphics[width=\columnwidth]{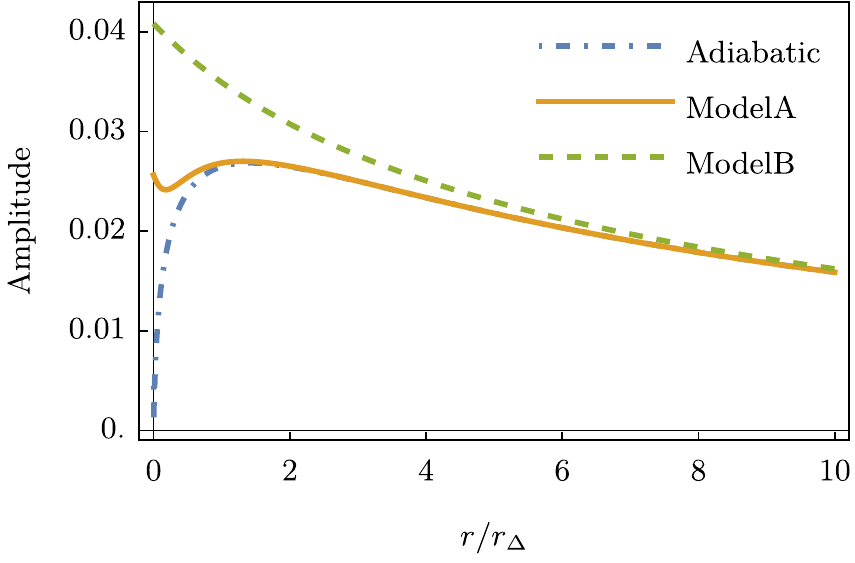}
    \caption{Amplitude $|b_1|$ and $|c_1|$ of the first (and most relevant) Fourier components in the adiabatic and dynamical response for model A and B, plotted as a function of $r/r_\Delta=\left(\Delta/\xi \right)^2$ (see \cref{eq:r_delta}). The amplitudes of the oscillations in the two cases are particularly different upon decreasing $r$ for $r<r_\Delta$, where the adiabatic response reaches its maximum before decreasing towards zero (see also \cref{fig:adiabatic_harmonics} and the discussion in \cref{par:onset}). Here the amplitude is plotted for a driving frequency $\Omega<\Omega_\T{peak}$ (see discussion in \cref{par:dynamic_amplitude}). The parameters used in the plot are $\gamma_y=1$, $D=10$, $\widetilde{R}=0.7$, $\Delta=3$, $A=1$, and $\Omega = 0.35$.}
\label{fig:amp}
\end{figure}

Let us first compare the dynamical response to the adiabatic one. Looking at \cref{fig:amp}, which shows the amplitude of the first harmonic as a function of $r=\xi^{-2}$, it appears that for any fixed value of the driving frequency $\Omega$ there exists a threshold value $r_A$ or $r_B$ (depending on the model considered) such that for $r\geq r_{A,B}(\Omega)$ the system dynamics becomes effectively adiabatic. When this happens, the amplitude of the dynamical response in model A/B is very well approximated by that of the adiabatic response, and the corresponding curves in \cref{fig:amp} coincide.

\begin{figure}[t]
    \centering
    \includegraphics[]{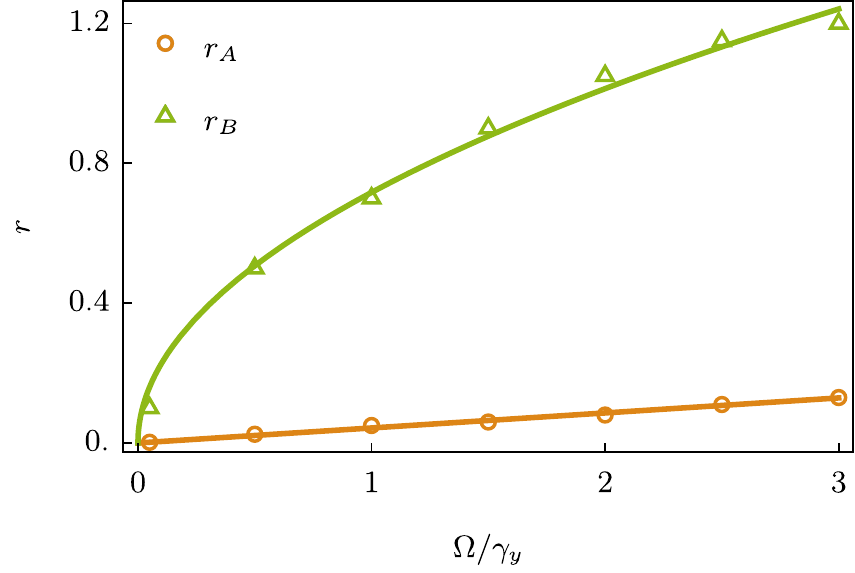}
    \caption{Values $r_{A,B}$ of the parameter $r$ such that, for a given value of the driving frequency $\Omega$, the amplitude of the dynamical response in model A, B matches that of the adiabatic approximation (see the main text for further explanations). By scaling arguments, we expect $r_A \sim \Omega$ and $r_B \sim \sqrt{\Omega}$ (see \cref{eq:r_AB}). The parameters used in the graph are $\gamma_y=1$, $D=100$, $\widetilde{R}=0.7$, $\Delta=3$, and $A=1$.}
    \label{fig:r_w}
\end{figure}

This can be understood in terms of the competition between the relaxation timescale $\tau_\phi$ of the field, which is given in \cref{eq:tau_phi}, and the one set by the external periodic driving, \ie, $\tau_\Omega \sim \Omega^{-1}$. Typical field fluctuations are those with wavevector $q\sim \xi^{-1}$, where $\xi\sim r^{-1/2}$ is the field correlation length. We expect the adiabatic approximation to be accurate when the timescale $\tau_\phi^\T{typ}$ of these typical fluctuations is much shorter than $\tau_\Omega$, \ie, $\tau_\phi^\T{typ}\equiv \tau_\phi(q\sim \xi^{-1}) \ll \tau_\Omega$: a simple calculation indicates that the threshold values $r_{A,B}$ are given by
\begin{equation}
    \begin{cases}
     r_A \sim  \Omega/D \, ,\\
     r_B \sim \sqrt{\Omega/D} \, .
    \end{cases} 
    \label{eq:r_AB}
\end{equation}
This is verified in \cref{fig:r_w}, where we plot $r_A$ and $r_B$ as a function of the driving frequency $\Omega$. The symbols correspond to numerical estimates of $r_{A,B}$ obtained by inspecting plots analogous to that of \cref{fig:amp}, while the solid lines correspond to \cref{eq:r_AB}.

Note that the timescale $\tau_y \sim \gamma_y^{-1}$, which characterizes the relaxation of the colloid $\vb{Y}$ in its harmonic trap, does not affect this interplay between $\tau_\Omega$ and $\tau_\phi$. As anticipated above, it merely contributes a common scaling factor $[1+(\Omega / \gamma_y)^2]^{-1/2}$ to the amplitude of the first harmonic and results into a phase shift $\theta_y$ given by \cref{eq:phase_shift}. This is in fact consistent with the effective field interpretation we gave in Section \ref{par:effective_field_interpretation}: the colloid $\vb{Y}$ moves under the effect of the excitations generated on the field $\phi$ by the motion of the colloid $\vb{Z}$. Any feedback of the colloid $\vb{Y}$ on the field is neglected, because we are considering only the lowest nontrivial order in a perturbative expansion in the coupling $\lambda$.
Accordingly, adiabaticity depends on how faithfully the field $\phi$ (which relaxes on a finite timescale) is able to transmit the excitation generated by the motion of the colloid $\vb{Z}$: the smaller the driving frequency $\Omega$, the more accurate this transmission becomes. What happens to the colloid $\vb{Y}$ after the ``message'' is received will only eventually depend on its characteristic timescale $\tau_y$.

Outside the adiabatic regime, the adiabatic and dynamical responses are qualitatively different especially for $r<r_\Delta = \Delta^{-2}$, the latter being the value of $r$ around which the adiabatic response reaches its maximum (see \cref{fig:amp} and the discussion in \cref{par:adiabatic_amplitude}). This also marks the point at which the correlation length of the field becomes of the same order of magnitude as the average separation between the two traps, \ie, $\xi \sim \Delta$. In \cref{par:adiabatic_amplitude} we described the phenomenon of frequency doubling in the adiabatic response: the amplitude of its first harmonic decreases upon decreasing $r$ below $r_\Delta$, and vanishes at $r=r_1$ (see \cref{fig:adiabatic_omega}). We can conclude that, in general, frequency doubling is not observed in the dynamical response, unless the adiabatic approximation is accurate (\ie, at small driving frequency $\Omega$ and large field mobility $D$, according to the discussion above).


\subsection{Frequency dependence of the dynamical response}
\label{par:dynamical_response}
The behavior of the actual dynamical response in \cref{eq:dyn_avg} as a function of the driving frequency $\Omega$ is richer than that of the adiabatic response. The limiting cases of slow and fast driving are analytically accessible, while for intermediate values of the driving frequency $\Omega$ we can evaluate numerically the integrals which appear in \cref{eq:dyn_avg}. We can then use the insight we gained in \cref{par:adiabatic_analysis} in order to rationalize the qualitative behavior observed in the plots.

In order to simplify the discussion by enforcing a separation of timescales, we consider in this Section a large value of the inverse timescale $\gamma_y=\tau_y^{-1}$. Indeed, as anticipated above, the amplitude of $\expval{\vb{Y}(t)}$ is significantly reduced at frequencies $\Omega \gg \gamma_y$ and this would make the features of the dynamical response hardly appreciable. Let us also set the parameter $r\ll r_\Delta$ (see \cref{eq:r_delta}), a choice which we will motivate further below.

\begin{figure}
    \centering
    \resizebox{\linewidth}{!}{\includegraphics[]{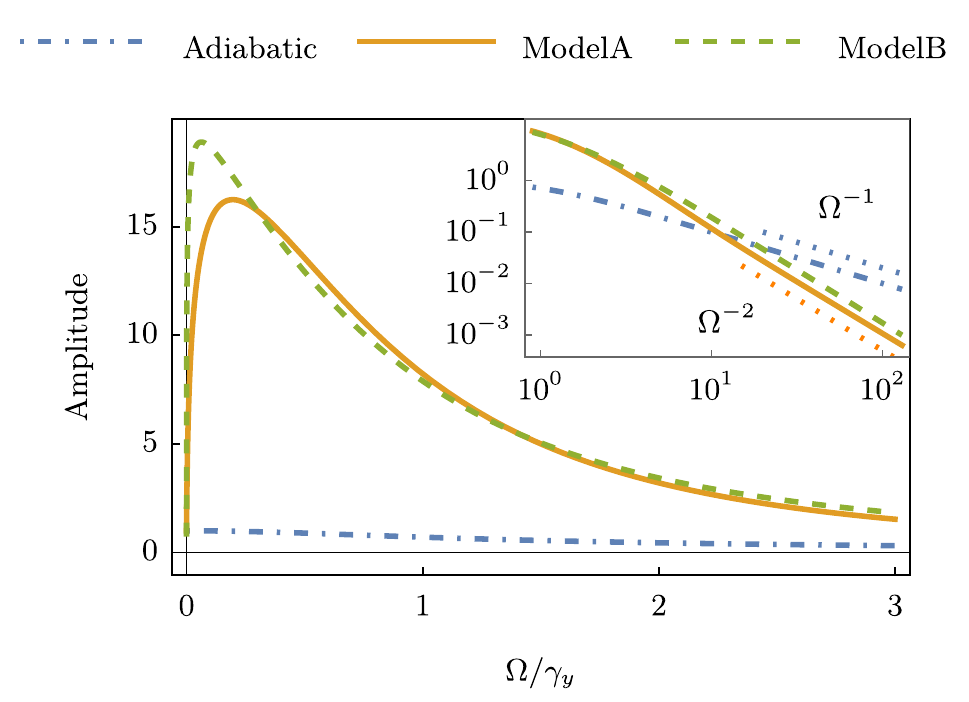}}
    \caption{Amplitudes $|b_1|$ and $|c_1|$ of the first Fourier harmonic in the adiabatic and dynamical responses, respectively, shown as functions of the driving frequency $\Omega$ in $d=1$, for both models A and B. For large $\Omega$, the amplitude decays as $\Omega^{-1}$ for the adiabatic response and as $\Omega^{-2}$ in the dynamical case (see the inset in log-log scale, where we indicated the asymptotic behaviors with dotted lines). For small values of $\Omega$, the dynamical response is typically larger than the one predicted by the adiabatic approximation, and it is peaked around $\Omega_\T{peak}$ given in \cref{eq:omega_peak}. Close to $\Omega\sim 0$, both responses must collapse on their static amplitude given in \cref{eq:staticamp}; all the curves in this plot are normalized by this value. The parameters used in the graphs are $\nu_y=1$, $k_y=1$, $D=1$, $\widetilde{R}=0.7$, $\Delta=3$, $A=1$, and $r = 10^{-4}$.}
    \label{fig:amp_omega}
\end{figure}

\subsubsection{Amplitude}
\label{par:dynamic_amplitude}
The main qualitative features of the dynamical response are displayed in \cref{fig:amp_omega}, where we plot the amplitude $|\vb{c}_1|$ of the first Fourier harmonic (see \cref{eq:coeff_dynamic}) as a function of $\Omega$ for models A and B, and we compare it to the amplitude of the adiabatic response. For vanishing $\Omega$ both responses must collapse on a common quasi-static curve, which follows from \cref{eq:Fourier_series,eq:coeff_adiabatic,eq:coeff_dynamic,eq:adiabatic_reduced_coeff,eq:In} as
\begin{align}
    2|\vb{b}_1(\Omega=0)| =2|\vb{c}_1(\Omega=0)| = 2\lambda^2 \cor{I}_1/\kappa_y \, .
    \label{eq:staticamp}
\end{align}
For small but nonzero $\Omega$, on the other hand, the dynamical response is typically larger than the one predicted within the adiabatic approximation. The former appears to be peaked around a frequency $\Omega_\T{peak}$ which can be identified as the inverse relaxation timescale of the field $\phi$ over a distance comparable with the average separation $\Delta$ between the two traps. This can be obtained from \cref{eq:tau_phi} by setting $q\simeq 1/\Delta$: for $r\ll r_\Delta=\Delta^{-2}$, we find
\begin{equation}
    \Omega_\T{peak} \sim \tau_\phi^{-1}(q\simeq 1/\Delta) \simeq D/\Delta^z \, ,
    \label{eq:omega_peak}
\end{equation}
where $z=2+\alpha$ is the dynamical critical exponent of the field $\phi$ (we recall that $\alpha=0$ and $2$ for model A and B respectively \cite{Tauber}). Accordingly, $\Omega_\T{peak}$ is different for model A and model B dynamics.

Finally, for large $\Omega$, the amplitude of the dynamical response decays as $\Omega^{-2}$, at odds with the adiabatic response which decays as $\Omega^{-1}$, so that the former becomes eventually smaller than the latter. This is shown in the inset of \cref{fig:amp_omega}, where the amplitude is plotted as a function of $\Omega$ in log-log scale, together with the asymptotic decays mentioned above.

Let us now motivate the choice $r\ll r_\Delta$. The argument we gave in \cref{par:onset} when discussing the adiabatic limit can be reversed: for every fixed value of the parameter $r$, there will be a driving frequency $\Omega_{A,B}(r)$ such that when $\Omega \leq \Omega_{A,B}(r)$ the dynamics of the system is well approximated by the adiabatic one. Their value can be found by inverting \cref{eq:r_AB}, \ie,
\begin{equation}
\begin{cases}
     \Omega_A \sim  Dr \, ,\\
     \Omega_B \sim Dr^2 \, .
\end{cases} 
\end{equation}
Since the characteristic frequency scale of the dynamical response is given by $\Omega_\T{peak}$ (see \cref{fig:amp_omega}), in order to appreciate the difference with respect to the adiabatic response we must require $\Omega_{A,B}(r) \ll \Omega_\T{peak}$. By choosing $r \ll r_\Delta$ this requirement is automatically satisfied, as it can be checked by using the definition of $\Omega_\T{peak}$ in \cref{eq:omega_peak}. If, on the contrary, one chooses $r \gtrsim r_\Delta$, then intermediate cases occur in which the peak shifts towards larger values of $\Omega$, while still remaining far from the adiabatic limit.

Similarly, in plotting the amplitude of the dynamical response as a function of $r$ in \cref{fig:amp} we chose $\Omega\ll \Omega_\T{peak}$. In fact, had we chosen instead $\Omega\gg \Omega_\T{peak}$, the dynamical amplitude would have been smaller than the adiabatic amplitude, and it would have approached the latter from \textit{below} in correspondence of $r_{A,B}(\Omega)$.

\subsubsection{Phase}
\label{par:dyn_phase_main}
\begin{figure}
    \centering
    \resizebox{\linewidth}{!}{\includegraphics[]{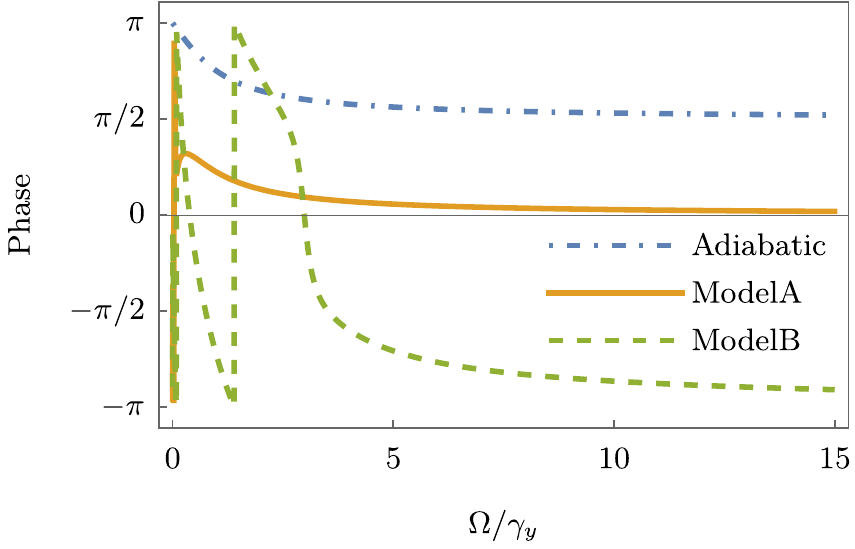}}
    \caption{Phase of the adiabatic and dynamic responses, shown as a function of the driving frequency $\Omega$ in $d=1$. In both cases the relative phases $\delta \theta$ and $\delta \varphi$, respectively, are measured with respect to the motion of the driven colloid $\expval*{\vb{Z}(t)}_0$ (see \cref{eq:phase_difference,eq:phase_difference_dynamic}). For large values of $\Omega$, the response in model A is in phase with the motion of the driven colloid (\ie, $\delta \varphi \to 0$), while in model B it is in counterphase (\ie, $\delta \varphi \to -\pi$). They are both in contrast with the adiabatic approximation, which predicts a $\pi/2$ phase shift $\delta \theta$. For sufficiently small $\Omega$, the three responses must coincide and we recover the physically familiar picture in which the motion is in counterphase with respect to $\expval*{\vb{Z}(t)}_0$ with $\delta\varphi = \delta \theta =\pi$. For intermediate values of $\Omega$, the phase in the dynamical response varies rapidly and non-monotonically, if $\widetilde{R}\ll \Delta$, before reaching its asymptotic value. The parameters used in the plot are $\gamma_y=1$, $D=10^{-3}$, $\widetilde{R}=0.4$, $\Delta=3$, $A=0.1$, and $r = 10^{-3}$.}
    \label{fig:phase}
\end{figure}

In analogy with what we did for the adiabatic response discussed in \cref{par:adiabatic_phase}, from the Fourier coefficient $\vb{c}_1$ in \cref{eq:coeff_dynamic} one can determine the phase of the dynamical response which we indicate by $\varphi_1$, so as to distinguish it from the phase $\theta_1$ of the adiabatic response. In particular, one finds
\begin{equation}
    \varphi_1 = -\left( \theta_y+\theta_z +  \pi/2 \right) +\arg(I_1) \, ,
    \label{eq:varphi1}
\end{equation}
where $\arg(I_1)$ indicates the argument of the complex integral
\begin{equation}
    I_1 \equiv \int \frac{\dd[d] {q}}{(2\pi)^d}  \frac{q_{||} q^\alpha  J_1(\vb{q}\cdot \vb{A}) }{\alpha_q+i \Omega}  e^{-q^2 \widetilde{R}^2 +i\vb{q}\cdot \Delta} \, .
    \label{eq:varphi1_integral}
\end{equation}
In the expression above $q_{||}$ indicates the component of $\vb{q}$ along $\vb{A}$ and $\bm{\Delta}$.
For $\Omega \to 0$, we notice that $I_1 \simeq \cor{I}_1/D$ (see \cref{eq:In}) and we recover the adiabatic limit with $\varphi_1 \simeq \theta_1$. For $\Omega \to \infty$, instead, one finds
\begin{equation}
    I_1 \simeq \frac{1}{i \Omega}\int \frac{\dd[d] {q}}{(2\pi)^d} q_{||} q^\alpha  J_1(\vb{q}\cdot \vb{A})  e^{-q^2 \widetilde{R}^2 +i\vb{q}\cdot \Delta}\, .
    \label{eq:I1_largeOmega}
\end{equation}
In analogy with \cref{par:adiabatic_phase}, we focus on the phase difference $\delta \varphi$ with respect to the motion of the driven colloid $\expval*{\vb{Z}(t)}_0$, \ie,
\begin{equation}
    \delta \varphi \equiv \varphi_1 -(-\theta_z -\pi/2) = -\theta_y + \arg(I_1) \, .
    \label{eq:phase_difference_dynamic}
\end{equation}
Recalling that $\theta_y\to \pi/2$ for large $\Omega$, it follows from \cref{eq:I1_largeOmega} that $\delta \varphi \simeq \pi/2 \pm \pi/2$, where the sign of the last term can be determined by performing the integration over $\vb{q}$ in \cref{eq:I1_largeOmega} and it is in general different for model A or B (see \cref{app:dynamical_phase} -- in $d=1$, the plus sign corresponds to model A, and the minus sign to model B). The motion of $\vb{Y}$ for large $\Omega$ is thus either in phase or in counterphase with the motion of the driven colloid, depending on the model: in both cases, this is in sharp contrast with the adiabatic approximation, which predicts a $\pi/2$ phase shift (see \cref{fig:adiabatic_omega} in the same limit). 
However, the approximation we used to derive \cref{eq:I1_largeOmega} can only be accurate if $\Omega$ is larger than all the physical frequencies involved in the problem. If we assume that the system is sufficiently close to criticality so that $\xi\gg \widetilde{R}$, then the effective colloid radius $\widetilde{R}$ plays the role of a cutoff and the fastest timescale is represented by $\tau_\phi(q\sim1/\widetilde{R})$. Accordingly, we expect the dynamical phase to reach its asymptotic value for
\begin{equation}
    \Omega \gg \Omega_\T{cutoff} \equiv \tau_\phi^{-1}(q\sim1/\widetilde{R}) \sim D/\widetilde{R}^z \, .
    \label{eq:phase_asymptotic_condition}
\end{equation}
Recall that the amplitude $|\vb{c}_1|$ of the dynamical response starts decreasing for $\Omega \gg \Omega_\T{peak}$ (see \cref{par:dynamic_amplitude} and \cref{eq:omega_peak}), and within our setup of \cref{fig:setup} with $\widetilde{R}\ll \Delta$ it is $\Omega_\T{cutoff} \gg \Omega_\T{peak}$. As a result, the asymptotic value of $\varphi_1$ will not be reached in practice if not for vanishing values of the amplitude $|\vb{c}_1|$, and one observes instead a phase which is rapidly changing as a function of $\Omega$, different in general from the adiabatic phase $\theta_1$ (if not by coincidence). This can be seen in \cref{fig:phase}, where the relative phase $\delta\varphi$ of the dynamical response is plotted as a function of the driving frequency $\Omega$ and is compared to the relative phase $\delta \theta$ of the adiabatic response. Moreover, since $\widetilde{R}$ (which enters in the integral $I_1$ defined in \cref{eq:varphi1_integral}) depends on the temperature $T$ via \cref{eq:rtilde}, an interesting outcome of the analysis presented above is that the phase $\varphi_1$ itself is $T$-dependent in our model. This was not the case for the phase $\theta_1$ within the adiabatic approximation, see \cref{eq:phase_adiabatic}.

\begin{figure}
    \centering
    \resizebox{\linewidth}{!}{\includegraphics[]{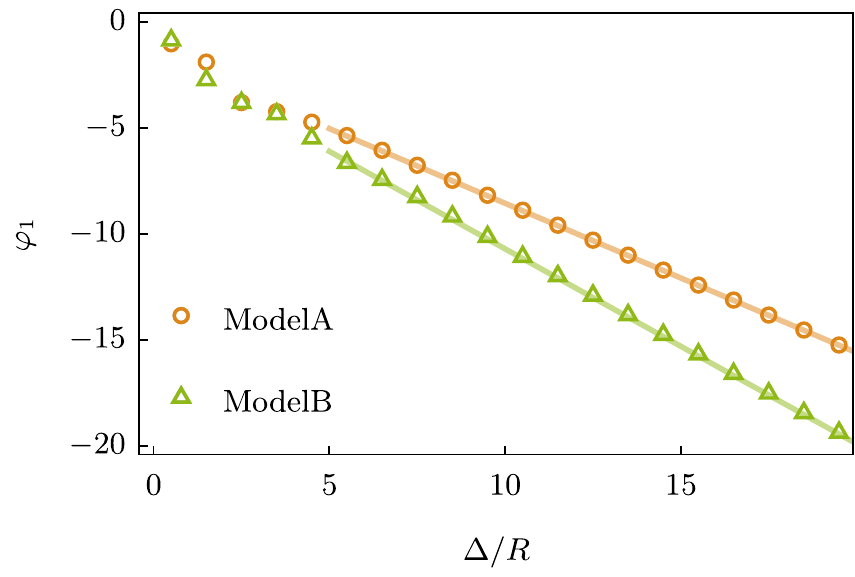}}
    \caption{Phase $\varphi_1$ of the dynamical response, shown as a function of the distance $\Delta$ between the two traps, for small values of the driving frequency $\Omega$ (see the main text). The behavior of $\varphi_1$ as a function of $\Delta$ is asymptotically linear, with a slope $\kappa$ which is independent of the spatial dimensionality $d$; for the case of model A, it is predicted by \cref{eq:effective_wavenumber}. The parameters used in the graph are $\gamma_y=1$, $D=0.1$, $\widetilde{R}=1$, $A=0.1$, and $r = 10^{-3}$.}
    \label{fig:phidelta}
\end{figure}

Finally, in \cref{fig:phidelta} we plot the phase $\varphi_1$ as a function of the average separation $\Delta$ between the traps and for small values of the driving frequency $\Omega$: the dependence of $\varphi_1$ on $\Delta$ turns out to be linear for sufficiently large $\Delta$. The corresponding slope $\kappa$ is independent of the spatial dimensionality $d$, and it can be extracted explicitly in the case of model A by using the method of steepest descent: this is done in Appendix \ref{app:slope}, where we show that
\begin{equation}
    \kappa \equiv - \pdv{\varphi_1}{\Delta} = \left[ r^2 +(\Omega/D)^2 \right]^{1/4} \sin \left(\frac12 \arctan(\frac{\Omega}{Dr})\right) \, . 
    \label{eq:effective_wavenumber}
\end{equation}
This fact suggests an interesting interpretation within the effective field picture presented in \cref{par:effective_field_interpretation}. Indeed, the response of the colloid $\vb{Y}$ to a small sinusoidal perturbation generated by the colloid $\vb{Z}$ at a distance $\Delta$ apart effectively reads
\begin{equation}
    \expval*{\vb{Y}(t)} \simeq \vb{R}(\Omega) \cos(\Omega t-\kappa \Delta +\varphi_\kappa ) \, ,
    \label{eq:carrier_signal}
\end{equation}
where the phase shift $\varphi_\kappa$ and $\vb{R}(\Omega)\simeq |\vb{c}_1|$ (see \cref{eq:coeff_dynamic}) depend in general on the various parameters of the problem. Equation \eqref{eq:carrier_signal} describes a wave propagating out of the source $\vb{Z}(t)$, and in this analogy the parameter $\kappa$ plays the role of an effective wavenumber. This simplified picture does not apply when $\Omega$ becomes large compared to the other characteristic frequencies of the system, because then we have seen that $\varphi_1$ must saturate to a constant limiting value (which is, in particular, independent of $\Delta$). Moreover, albeit small, the contribution of higher harmonics will still modify the first harmonic contribution described by \cref{eq:carrier_signal}.


\begin{figure*}
\centering
\subfloat[]{
  \centering
  \includegraphics[width=\columnwidth]{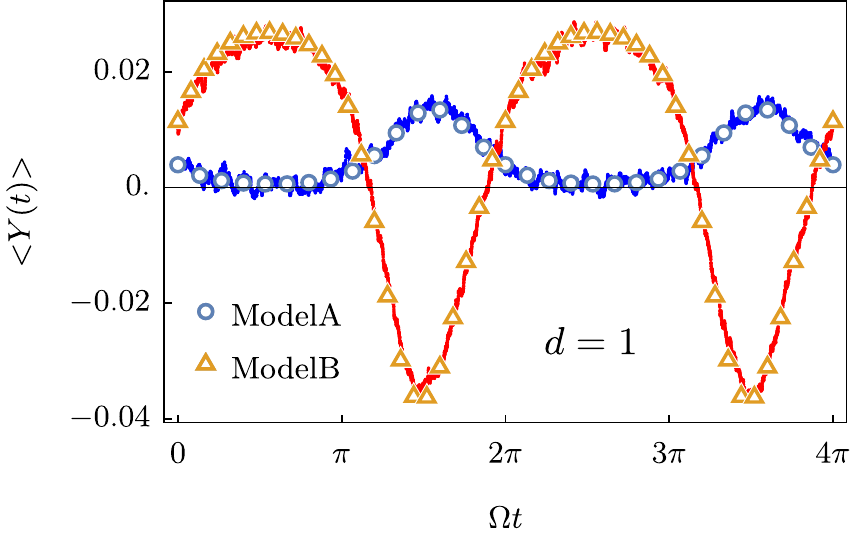}
  \label{fig:sim_1d}
  }
\subfloat[]{
  \centering
  \includegraphics[width=\columnwidth]{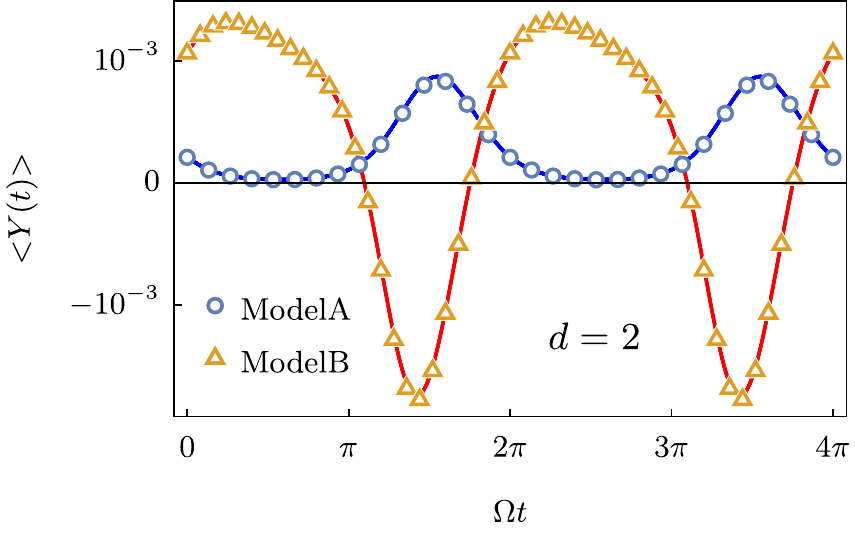}
  \label{fig:sim_2d}
  }
\caption{Average position $\expval*{Y(t)}$ of the colloid in the fixed trap, (a) in $d=1$, and (b) in $d=2$. The results of the numerical simulations (colored lines) are reported together with the analytical prediction in \cref{eq:dyn_avg} (symbols), showing excellent agreement. The parameters used in both graphs are $\nu_y=1$, $D=1$, $k_y=0.1$, $r=0.1$, $\lambda=0.5$, $\Delta=20$, $A=5$, $\Omega = 2\pi \times 10^{-3}$, lattice side $L=128$, and integration time step $\delta t=0.01$. In panel (a) we set $R=1.5$, $T=0.01$, and we averaged over $N=10^5$ realizations of the thermal noise appearing in \cref{eq:field,eq:particle_eom}. In panel (b) the noise is absent (corresponding to $T=0$), and we take a single realization of the dynamics with the effective particle radius $\widetilde{R}=1.5165$ (see the main text), corresponding to the values of $R$ and $T$ considered in panel (a) and as obtained from \cref{eq:rtilde}.}
\label{fig:simulation}
\end{figure*}

\section{Numerical simulation}
\label{par:numerical}
In this Section we investigate the validity of our analytical predictions, derived within the weak-coupling expansion, by direct integration of the coupled Langevin equations of motion of the field in \cref{eq:field}, and of the two particles in \cref{eq:particle_eom,eq:particleZ}. To this end, we discretize the field $\phi$ over a lattice of side $L$ in $d=1$ or $d=2$ spatial dimensions, as described in Appendix \ref{par:appnumerical}, and we assume periodic boundary conditions. We consider, for simplicity, the limit $k_z \rightarrow \infty$ for the driven colloid $\vb{Z}(t)$, which thus evolves deterministically according to \cref{eq:forcing}, while the second colloid $\vb{Y}(t)$ undergoes Brownian diffusion under the effects of its fixed trap.

We first simulate the system in $d=1$ in the presence of noise. Figure \ref{fig:sim_1d} compares the average over many realizations of the simulated trajectories of the particles with the analytical predictions in \cref{eq:dyn_avg}, showing a good agreement for both model A and model B. For this simulation we chose a set of parameters which poses model A close to the adiabatic regime, while model B is actually far from it. As a result, the curve corresponding to model A is (almost) in counterphase with respect to the external driving $\sim A\sin(\Omega t)$, while the curve corresponding to model B has a generic phase. We chose a large value of the driving amplitude $A$ so as to emphasize also the contribution of higher Fourier harmonics, although the first harmonic still dominates the response, as expected.

A further conclusion we can draw from this agreement between theoretical predictions and numerical simulations is the following. As we emphasized in \cref{par:small_lambda}, the prediction for $\expval*{\vb{Y}(t)}$ in \cref{eq:dyn_avg} does not distinguish the separate effects of having a larger particle radius $R$ from those of a higher temperature $T$, being them tangled into the effective radius $\widetilde{R}$ defined in \cref{eq:rtilde}. This observation actually simplifies the task of performing numerical simulations in higher spatial dimension $d$, where they become longer and more resource-demanding: we simply set $T=0$ and simulate the noiseless (\ie, deterministic) equations of motion, correcting $\widetilde{R}$ accordingly. Figure \ref{fig:sim_2d} exemplifies this in $d=2$, for the same set of parameters as those used in \cref{fig:sim_1d}. The curves we observe are qualitatively similar to those in $d=1$, and again they are in good agreement with the analytical prediction. In this second plot it appears even more evidently that the oscillations of the probe particle are not harmonic, as a result of the nonlinear interaction.

\section{Extension to many particles}
\label{par:many-body}
In \cref{par:masterequation} we noted that the contribution of any additional particle enters linearly in the master equation \eqref{eq:master_general} which describes the one-point probability $P_1(\vb{y},t)$ of the position $\vb{Y}(t)$ of the particle. In \cref{par:effective_field_interpretation} we further commented that the \textit{effective field} in which the particle $\vb{Y}$ evolves can be obtained by simply summing the contributions of all the other particles, which are acting as source terms for the field $\phi$. It would thus appear that multi-body effects are absent in our model, and that the induced interactions are indeed pairwise-additive, at odds with other types of fluctuation-induced interactions such as Casimir forces. Similar conclusions have been recently reached in Ref.~\cite{fournier2021fieldmediated}, where it was shown that field-mediated forces between point-like particles linearly coupled to a Gaussian field in \textit{equilibrium} are indeed pairwise-additive, independently of the strength of the linear coupling.
However, this is in principle not the case for non-equilibrium settings, such as the one considered in this work. Since our analysis was based on a perturbative description valid for a small coupling $\lambda$, it is then natural at this point to ask whether pairwise-additivity holds beyond the perturbative regime. In order to answer this question, we now assume that $N$ particles $\{ \vb{X}_1,\vb{X}_2,\dots,\vb{X}_N\}$ are in contact with the field $\phi$ as in \cref{par:Model}, so that
\begin{equation}
    \cor{H} = \cor{H}_\phi + \sum_{a=1}^N \cor{U}_a(\vb{X}_a) - \lambda \cor{H}_\T{int} \, ,
\end{equation}
where $\cor{U}_a$ are generic confining potentials, and 
\begin{equation}
    \cor{H}_\T{int} = \sum_{a=1}^N \int \dd[d]{\vb{x}} \phi(\vb{x})  V^{(a)}(\vb{x}-\vb{X}_a)
\end{equation}
generalizes \cref{eq:Hint} to many particles.
The field $\phi$ still evolves according to \cref{eq:field}, while the particles follow
\begin{equation}
        \dot{\vb{X}}_a(t)=  \vb{F}_a(\vb{X}_a,t)   + \lambda \nu_a \vb{f}_a(\vb{X}_a,\phi) + \bm{\xi}^{(a)}(t) \, ,
        \label{eq:langevin_many_parts}
\end{equation}
where we denoted by $\nu_a$ the mobility coefficients, $\bm{\xi}^{(a)}(t)$ are independent white Gaussian noises with the same variance as in \cref{eq:noise_variances}, and $\vb{f}_a$ is defined as in \cref{eq:f}. To make contact with \cref{eq:particle_eom} we can choose
$\vb{F}_a(\vb{X}_a,t) \equiv - \nu_a \grad_{X_a} \cor{U}_a(\vb{X}_a(t))$, so as to describe the equilibrium fluctuations of the particles in their confining potentials $\cor{U}_a(\vb{X}_a)$ and in contact with the field. However, $\vb{F}_a(\vb{X}_a,t)$ can also be explicitly time-dependent (\eg, as in \cref{eq:particleZ}), so that the problem is in general out of equilibrium (and similar to the one discussed above).

In order to study the dynamics induced by the set of Langevin equations \cref{eq:field,eq:langevin_many_parts}, it is convenient to consider the corresponding Martin-Siggia-Rose \cite{MSR,DeDominicis,Janssen1976,demerypath} dynamical functional $\cor{S}[\phi,\tilde{\phi},\{ \vb{X}_a, \widetilde{\vb{X}}_a\} ]$, as detailed in \cref{app:many_body}. Here we indicated by $\tilde{\phi}(\vb{x},t)$ and $\widetilde{\vb{X}}_a(t)$ the variables dynamically conjugate to $\phi(\vb{x},t)$ and $\vb{X}_a(t)$, respectively. Integrating out the fields $\phi$ and $\tilde{\phi}$ from the dynamical functional $\cor{S}$ formally yields an effective functional $\cor{S}_\T{eff}[\{ \vb{X}_a, \widetilde{\vb{X}}_a\} ]$: any expectation value over the realization of the noises of quantities such as $\cor{O}\left[\{ \vb{X}_a\}\right]$, involving the particles but not the field, can then be expressed as
\begin{align}
    &\expval*{\cor{O}\left[\{ \vb{X}_a\}\right]} \\
    &= \int \left(\prod_{a=1}^N \cor{D}\vb{X}_a \,  \cor{D}\widetilde{\vb{X}}_a \right) \cor{O}\left[\{ \vb{X}_a\}\right] e^{- \cor{S}_\T{eff}[\{ \vb{X}_a, \widetilde{\vb{X}}_a\} ]}  \, ,\n
\end{align}
where $\cor{D}\vb{X}_a$ indicates a path integral over the realizations of $\vb{X}_a$ (and similarly for $\cor{D}\widetilde{\vb{X}}_a$).

The integration over the fields $\phi$ and $\tilde{\phi}$ in the dynamical functional $\cor{S}$ given in \cref{eq:total_action} is possible for any value of $\lambda$, because the field Hamiltonian $\cor{H}_\phi$ in \cref{eq:hamiltonian_field} is Gaussian and the field-particles coupling is linear. This results in the effective functional
\begin{equation}
    \cor{S}_\T{eff}[\{ \vb{X}_a, \widetilde{\vb{X}}_a\} ] = \cor{S}_0[\{ \vb{X}_a,\widetilde{\vb{X}}_a\} ]-\lambda^2\cor{S}_\lambda[\{ \vb{X}_a, \widetilde{\vb{X}}_a\} ] \, ,
    \label{eq:effective_dynamical_action}
\end{equation}
where the free part $\cor{S}_0$ can be expressed as a sum of single-particle contributions (see \cref{eq:free_action_parts}),
\begin{equation}
    \cor{S}_0[\{ \vb{X}_a, \widetilde{\vb{X}}_a\} ] = \sum_{a=1}^N \cor{S}_a[ \vb{X}_a, \widetilde{\vb{X}}_a ] \, ,
    \label{eq:free_action}
\end{equation}
while the interacting part $\cor{S}_\lambda$ contains a sum over two-particle contributions (see \cref{eq:int_action_parts}),
\begin{equation}
    \cor{S}_\lambda[\{ \vb{X}_a, \widetilde{\vb{X}}_a\} ] = \sum_{a,b=1}^N \cor{S}_{ab}[ \vb{X}_a, \widetilde{\vb{X}}_a ,\vb{X}_b, \widetilde{\vb{X}}_b ] \, ,
    \label{eq:int_action}
\end{equation}
where the explicit form of $\cor{S}_{ab}$ is provided in \cref{eq:int_action_parts}. The dynamical action in \cref{eq:effective_dynamical_action} is markedly pairwise additive, as it is only written in terms of one- and two-body terms. Moreover, it is exact for any value of the coupling $\lambda$. We can thus conclude that higher-order corrections which we have not included in our perturbative calculation will have the effect of renormalizing the (pairwise) interaction potential, but they will not introduce any additional multi-body interaction. In this respect, the conclusions of Ref.~\cite{fournier2021fieldmediated} readily extend also out of equilibrium.

\section{Summary and conclusions}
\label{par:conclusion}
In this work we considered two Brownian particles interacting with the same fluctuating field, which are therefore subject to field-mediated forces: these might be used to induce synchronization when one of the two particles is externally driven. In equilibrium, these forces can be obtained by integrating out the field degrees of freedom from the system composed by the particles and the field: in this \textit{adiabatic} approximation, the effective Langevin dynamics of the particles remains Markovian. The same holds if the medium is not instantaneously in equilibrium, but still characterized by a relaxation timescale which is short compared to that characterizing the motion of the particles. However, if the relaxation time of the medium becomes longer, then the adiabatic approximation fails and different techniques are needed to study the (non-equilibrium) dynamics of the tracer particles.

We exemplified these facts by studying a simple model in which a scalar Gaussian field is linearly coupled to two overdamped Brownian particles kept spatially separated by two confining harmonic traps (\cref{fig:setup}). One of the two traps is driven periodically with a tunable frequency $\Omega$, which allows us to probe the dynamical response of the other particle over a range of frequencies which spans across the various timescales of the system. As the field approaches its critical point $r=0$, its relaxation timescale diverges and one observes a gradual departure from the condition of adiabatic response presented above.

Within a weak-coupling expansion, we derived the master equation \eqref{eq:master_PS} which describes the dynamics of the non-driven particle in the non-equilibrium periodic state attained by the system at long times. This can be used to determine the cumulant generating function of the particle position reported in \cref{eq:cgf}, from which one can deduce, inter alia, the average and variance of the actual \textit{dynamical} response of the particle given in \cref{eq:dyn_avg,eq:dyn_var}, respectively.

The latter has to be compared to the adiabatic response in \cref{eq:y_adiabatic}, which we derived in \cref{par:adiabaticapproximation} under the assumption of fast field relaxation. Its behavior as a function of the driving frequency $\Omega$ is analogous to that of a low-pass filter in circuit electronics (\cref{fig:adiabatic_omega}), and therefore we focus on its dependence on the field correlation length $\xi=r^{-1/2}$ (\cref{fig:adiabatic_harmonics}): the amplitude of the oscillations induced on the particle in the fixed trap presents a peak when $\xi \sim \Delta$, being $\Delta$ the average separation between the two traps, while it decays to zero for both larger and smaller values of $\xi$. Observing the response of such a particle then becomes a way to probe the effective potential $V_c(\vb{x})$ induced between the two particles by the presence of the field, see \cref{eq:inducedpotential,fig:potential}. Being $V_c(\vb{x})$ non-linear, interesting phenomena such as frequency doubling can occur under periodic driving (see \cref{par:adiabatic_amplitude}).

Conversely, the behavior of the actual dynamical response as a function of $\Omega$ is significantly richer and it is determined by the interplay between the various timescales characterizing the system. In particular, these are the relaxation time of the colloid in its trap (see \cref{tau:parts}), the timescale set by the external driving $\Omega$, and the relaxation times of the field (see \cref{eq:tau_phi}) across the typical length scales of the system: the field correlation length $\xi$, the average separation $\Delta$ between the two traps, the radius $R$ and the mean square displacement of the colloid in the trap (see \cref{eq:rtilde}). In \cref{par:dynamic_analysis} we study in detail the amplitude (\cref{fig:amp,fig:amp_omega}) and the phase (\cref{fig:phase,fig:phidelta}) of this dynamical response. In particular, the amplitude of the oscillations displays a peak when the driving frequency $\Omega$ matches the relaxation timescale of the field over a length scale of the order of $\Delta$ (see \cref{fig:amp_omega}). Moreover, for sufficiently slow driving, the phase $\varphi_1$ is shown to display a linear dependence on $\Delta$ (see \cref{fig:phidelta} and \cref{eq:effective_wavenumber}). Both these features are not captured by the adiabatic response, whose amplitude decays monotonically upon increasing $\Omega$, and whose phase $\theta_1$ is $\Delta$-independent. Finally, a clear effect of retardation is visible in the behavior of the phase $\varphi_1$ in the limit of fast driving $\Omega$, where the dynamical response predicts a $\pi/2$ phase shift with respect to the adiabatic approximation (see \cref{fig:phase}).

In passing, we interpret these results in terms of the \textit{effective field} (see \cref{par:effective_field_interpretation}): within the weak-coupling approximation, one can study the dynamics of a tracer particle as if it were immersed in the effective field generated by the motion of all the other particles coupled to the same field, which can be treated as source terms. In fact, it turns out that the excitations generated by each of these moving particles contribute additively to the average effective field given in \cref{eq:phi_eff_avg}. This feature persists beyond the perturbative regime, as we verified in \cref{par:many-body} by computing the dynamical functional which describes the many-particle dynamics for any value of the coupling constant $\lambda$, and checking that it does not give rise to genuine many-body effects.

We finally checked the accuracy of the perturbative approach by comparing its analytical predictions with the results of the numerical integration of the coupled equations of motion, finding in general a good agreement (see \cref{fig:simulation}).

We conclude by noting that not only the kind of systems investigated here are well within the reach of current experiments \cite{casimirColloids}, but a similar setup has in fact already been studied in \ccite{ciliberto}, where the motion of silica particles immersed in a near-critical binary liquid mixture was observed by video-microscopy, and synchronization of their motion under external driving was reported upon approaching the critical point.

The simplified model considered here does not account for hydrodynamic effects, which are expected to be relevant in actual fluid media, and moreover one should go beyond the Gaussian approximation in order to describe the dynamics of a binary liquid mixture in the vicinity of a critical point. Future works will then address these issues and possibly include also the effects of activity \cite{fournier2021fieldmediated,fournier_2018,active} or anisotropies which the particles may additionally display. Addressing the case of a quadratic instead of a linear field-particle coupling is also relevant \cite{demery2013}, since it is closer to the effect of imposing Dirichlet boundary conditions on the field fluctuations, which is another typical setting for critical Casimir forces \cite{kardar99,GambassiCCF}.

\begin{acknowledgments}
We thank U. Basu for useful discussions, and B. Walter, who is co-author of the code used for numerical simulations. DV would like to thank L. Correale, J.-B. Fournier and S. Loos for fruitful conversations. AG acknowledges support from MIUR PRIN project “Coarse-grained description for non-equilibrium systems and transport phenomena (CO-NEST)” n. 201798CZL.
\end{acknowledgments}

\bibliography{references}

\begin{thebibliography}{65}%
\makeatletter
\providecommand \@ifxundefined [1]{%
 \@ifx{#1\undefined}
}%
\providecommand \@ifnum [1]{%
 \ifnum #1\expandafter \@firstoftwo
 \else \expandafter \@secondoftwo
 \fi
}%
\providecommand \@ifx [1]{%
 \ifx #1\expandafter \@firstoftwo
 \else \expandafter \@secondoftwo
 \fi
}%
\providecommand \natexlab [1]{#1}%
\providecommand \enquote  [1]{``#1''}%
\providecommand \bibnamefont  [1]{#1}%
\providecommand \bibfnamefont [1]{#1}%
\providecommand \citenamefont [1]{#1}%
\providecommand \href@noop [0]{\@secondoftwo}%
\providecommand \href [0]{\begingroup \@sanitize@url \@href}%
\providecommand \@href[1]{\@@startlink{#1}\@@href}%
\providecommand \@@href[1]{\endgroup#1\@@endlink}%
\providecommand \@sanitize@url [0]{\catcode `\\12\catcode `\$12\catcode
  `\&12\catcode `\#12\catcode `\^12\catcode `\_12\catcode `\%12\relax}%
\providecommand \@@startlink[1]{}%
\providecommand \@@endlink[0]{}%
\providecommand \url  [0]{\begingroup\@sanitize@url \@url }%
\providecommand \@url [1]{\endgroup\@href {#1}{\urlprefix }}%
\providecommand \urlprefix  [0]{URL }%
\providecommand \Eprint [0]{\href }%
\providecommand \doibase [0]{https://doi.org/}%
\providecommand \selectlanguage [0]{\@gobble}%
\providecommand \bibinfo  [0]{\@secondoftwo}%
\providecommand \bibfield  [0]{\@secondoftwo}%
\providecommand \translation [1]{[#1]}%
\providecommand \BibitemOpen [0]{}%
\providecommand \bibitemStop [0]{}%
\providecommand \bibitemNoStop [0]{.\EOS\space}%
\providecommand \EOS [0]{\spacefactor3000\relax}%
\providecommand \BibitemShut  [1]{\csname bibitem#1\endcsname}%
\let\auto@bib@innerbib\@empty
\bibitem [{\citenamefont {Casimir}(1948)}]{Casimir_1948}%
  \BibitemOpen
  \bibfield  {author} {\bibinfo {author} {\bibfnamefont {H.~B.~G.}\
  \bibnamefont {Casimir}},\ }\bibfield  {title} {\bibinfo {title} {{On the
  attraction between two perfectly conducting plates}},\ }\href@noop {}
  {\bibfield  {journal} {\bibinfo  {journal} {Kon. Ned. Akad. Wetensch. Proc.}\
  }\textbf {\bibinfo {volume} {51}},\ \bibinfo {pages} {793} (\bibinfo {year}
  {1948})}\BibitemShut {NoStop}%
\bibitem [{\citenamefont {Dalvit}\ \emph {et~al.}(2011)\citenamefont {Dalvit},
  \citenamefont {Milonni}, \citenamefont {Roberts},\ and\ \citenamefont
  {Rosa}}]{Casimir_book}%
  \BibitemOpen
  \bibfield  {author} {\bibinfo {author} {\bibfnamefont {D.}~\bibnamefont
  {Dalvit}}, \bibinfo {author} {\bibfnamefont {P.}~\bibnamefont {Milonni}},
  \bibinfo {author} {\bibfnamefont {D.}~\bibnamefont {Roberts}},\ and\ \bibinfo
  {author} {\bibfnamefont {F.}~\bibnamefont {Rosa}},\ }\href
  {https://doi.org/10.1007/978-3-642-20288-9} {\emph {\bibinfo {title} {Casimir
  Physics}}}\ (\bibinfo  {publisher} {Springer Berlin},\ \bibinfo {address}
  {Heidelberg},\ \bibinfo {year} {2011})\BibitemShut {NoStop}%
\bibitem [{\citenamefont {Kardar}\ and\ \citenamefont
  {Golestanian}(1999)}]{kardar99}%
  \BibitemOpen
  \bibfield  {author} {\bibinfo {author} {\bibfnamefont {M.}~\bibnamefont
  {Kardar}}\ and\ \bibinfo {author} {\bibfnamefont {R.}~\bibnamefont
  {Golestanian}},\ }\bibfield  {title} {\bibinfo {title} {The ``friction'' of
  vacuum, and other fluctuation-induced forces},\ }\href
  {https://doi.org/10.1103/RevModPhys.71.1233} {\bibfield  {journal} {\bibinfo
  {journal} {Rev. Mod. Phys.}\ }\textbf {\bibinfo {volume} {71}},\ \bibinfo
  {pages} {1233} (\bibinfo {year} {1999})}\BibitemShut {NoStop}%
\bibitem [{\citenamefont {Ajdari}\ \emph {et~al.}(1991)\citenamefont {Ajdari},
  \citenamefont {Peliti},\ and\ \citenamefont {Prost}}]{Ajdari_1991}%
  \BibitemOpen
  \bibfield  {author} {\bibinfo {author} {\bibfnamefont {A.}~\bibnamefont
  {Ajdari}}, \bibinfo {author} {\bibfnamefont {L.}~\bibnamefont {Peliti}},\
  and\ \bibinfo {author} {\bibfnamefont {J.}~\bibnamefont {Prost}},\ }\bibfield
   {title} {\bibinfo {title} {Fluctuation-induced long-range forces in liquid
  crystals},\ }\href {https://doi.org/10.1103/PhysRevLett.66.1481} {\bibfield
  {journal} {\bibinfo  {journal} {Phys. Rev. Lett.}\ }\textbf {\bibinfo
  {volume} {66}},\ \bibinfo {pages} {1481} (\bibinfo {year}
  {1991})}\BibitemShut {NoStop}%
\bibitem [{\citenamefont {Golestanian}(2005)}]{Golestanian_2005}%
  \BibitemOpen
  \bibfield  {author} {\bibinfo {author} {\bibfnamefont {R.}~\bibnamefont
  {Golestanian}},\ }\bibfield  {title} {\bibinfo {title} {Fluctuation-induced
  forces in and out of equilibrium},\ }\href
  {https://doi.org/10.1007/BF02704165} {\bibfield  {journal} {\bibinfo
  {journal} {Pramana}\ }\textbf {\bibinfo {volume} {64}},\ \bibinfo {pages}
  {1029} (\bibinfo {year} {2005})}\BibitemShut {NoStop}%
\bibitem [{\citenamefont {Kirkpatrick}\ \emph {et~al.}(2014)\citenamefont
  {Kirkpatrick}, \citenamefont {de~Z\'arate},\ and\ \citenamefont
  {Sengers}}]{Kirkpatrick_2014}%
  \BibitemOpen
  \bibfield  {author} {\bibinfo {author} {\bibfnamefont {T.~R.}\ \bibnamefont
  {Kirkpatrick}}, \bibinfo {author} {\bibfnamefont {J.~M.~O.}\ \bibnamefont
  {de~Z\'arate}},\ and\ \bibinfo {author} {\bibfnamefont {J.~V.}\ \bibnamefont
  {Sengers}},\ }\bibfield  {title} {\bibinfo {title} {Fluctuation-induced
  pressures in fluids in thermal nonequilibrium steady states},\ }\href
  {https://doi.org/10.1103/PhysRevE.89.022145} {\bibfield  {journal} {\bibinfo
  {journal} {Phys. Rev. E}\ }\textbf {\bibinfo {volume} {89}},\ \bibinfo
  {pages} {022145} (\bibinfo {year} {2014})}\BibitemShut {NoStop}%
\bibitem [{\citenamefont {Aminov}\ \emph {et~al.}(2015)\citenamefont {Aminov},
  \citenamefont {Kafri},\ and\ \citenamefont {Kardar}}]{Aminov_2015}%
  \BibitemOpen
  \bibfield  {author} {\bibinfo {author} {\bibfnamefont {A.}~\bibnamefont
  {Aminov}}, \bibinfo {author} {\bibfnamefont {Y.}~\bibnamefont {Kafri}},\ and\
  \bibinfo {author} {\bibfnamefont {M.}~\bibnamefont {Kardar}},\ }\bibfield
  {title} {\bibinfo {title} {Fluctuation-induced forces in nonequilibrium
  diffusive dynamics},\ }\href {https://doi.org/10.1103/PhysRevLett.114.230602}
  {\bibfield  {journal} {\bibinfo  {journal} {Phys. Rev. Lett.}\ }\textbf
  {\bibinfo {volume} {114}},\ \bibinfo {pages} {230602} (\bibinfo {year}
  {2015})}\BibitemShut {NoStop}%
\bibitem [{\citenamefont {Krech}(1994)}]{Krech_book}%
  \BibitemOpen
  \bibfield  {author} {\bibinfo {author} {\bibfnamefont {M.}~\bibnamefont
  {Krech}},\ }\href {https://doi.org/10.1142/2434} {\emph {\bibinfo {title}
  {The Casimir Effect in Critical Systems}}}\ (\bibinfo  {publisher} {World
  Scientific},\ \bibinfo {year} {1994})\BibitemShut {NoStop}%
\bibitem [{\citenamefont {Krech}(1999)}]{Krech_1999}%
  \BibitemOpen
  \bibfield  {author} {\bibinfo {author} {\bibfnamefont {M.}~\bibnamefont
  {Krech}},\ }\bibfield  {title} {\bibinfo {title} {Fluctuation-induced forces
  in critical fluids},\ }\href {https://doi.org/10.1088/0953-8984/11/37/201}
  {\bibfield  {journal} {\bibinfo  {journal} {J. Phys.-Condens. Mat.}\ }\textbf
  {\bibinfo {volume} {11}},\ \bibinfo {pages} {R391} (\bibinfo {year}
  {1999})}\BibitemShut {NoStop}%
\bibitem [{\citenamefont {Brankov}\ \emph {et~al.}(2000)\citenamefont
  {Brankov}, \citenamefont {Danchev},\ and\ \citenamefont
  {Tonchev}}]{Danchev_book}%
  \BibitemOpen
  \bibfield  {author} {\bibinfo {author} {\bibfnamefont {J.~G.}\ \bibnamefont
  {Brankov}}, \bibinfo {author} {\bibfnamefont {D.~M.}\ \bibnamefont
  {Danchev}},\ and\ \bibinfo {author} {\bibfnamefont {N.~S.}\ \bibnamefont
  {Tonchev}},\ }\href {https://doi.org/10.1142/4146} {\emph {\bibinfo {title}
  {Theory of Critical Phenomena in Finite-Size Systems}}}\ (\bibinfo
  {publisher} {World Scientific},\ \bibinfo {year} {2000})\BibitemShut
  {NoStop}%
\bibitem [{\citenamefont {Gambassi}(2009)}]{GambassiCCF}%
  \BibitemOpen
  \bibfield  {author} {\bibinfo {author} {\bibfnamefont {A.}~\bibnamefont
  {Gambassi}},\ }\bibfield  {title} {\bibinfo {title} {{The Casimir effect:
  From quantum to critical fluctuations}},\ }\href
  {https://doi.org/10.1088/1742-6596/161/1/012037} {\bibfield  {journal}
  {\bibinfo  {journal} {J. Phys. Conf. Ser.}\ }\textbf {\bibinfo {volume}
  {161}},\ \bibinfo {pages} {012037} (\bibinfo {year} {2009})}\BibitemShut
  {NoStop}%
\bibitem [{\citenamefont {Macio\l{}ek}\ and\ \citenamefont
  {Dietrich}(2018)}]{Maciolek}%
  \BibitemOpen
  \bibfield  {author} {\bibinfo {author} {\bibfnamefont {A.}~\bibnamefont
  {Macio\l{}ek}}\ and\ \bibinfo {author} {\bibfnamefont {S.}~\bibnamefont
  {Dietrich}},\ }\bibfield  {title} {\bibinfo {title} {{Collective behavior of
  colloids due to critical Casimir interactions}},\ }\href
  {https://doi.org/10.1103/RevModPhys.90.045001} {\bibfield  {journal}
  {\bibinfo  {journal} {Rev. Mod. Phys.}\ }\textbf {\bibinfo {volume} {90}},\
  \bibinfo {pages} {045001} (\bibinfo {year} {2018})}\BibitemShut {NoStop}%
\bibitem [{\citenamefont {Fournier}(2021)}]{fournier2021fieldmediated}%
  \BibitemOpen
  \bibfield  {author} {\bibinfo {author} {\bibfnamefont {J.-B.}\ \bibnamefont
  {Fournier}},\ }\href@noop {} {\bibinfo {title} {Field-mediated interactions
  of passive and conformation-active particles: multibody and retardation
  effects}} (\bibinfo {year} {2021}),\ \Eprint
  {https://arxiv.org/abs/2112.14184} {arXiv:2112.14184 [cond-mat.soft]}
  \BibitemShut {NoStop}%
\bibitem [{\citenamefont {Fournier}(2014)}]{fournier_2014}%
  \BibitemOpen
  \bibfield  {author} {\bibinfo {author} {\bibfnamefont {J.-B.}\ \bibnamefont
  {Fournier}},\ }\bibfield  {title} {\bibinfo {title} {Dynamics of the force
  exchanged between membrane inclusions},\ }\href
  {https://doi.org/10.1103/PhysRevLett.112.128101} {\bibfield  {journal}
  {\bibinfo  {journal} {Phys. Rev. Lett.}\ }\textbf {\bibinfo {volume} {112}},\
  \bibinfo {pages} {128101} (\bibinfo {year} {2014})}\BibitemShut {NoStop}%
\bibitem [{\citenamefont {Symanzik}(1981)}]{SYMANZIK_1981}%
  \BibitemOpen
  \bibfield  {author} {\bibinfo {author} {\bibfnamefont {K.}~\bibnamefont
  {Symanzik}},\ }\bibfield  {title} {\bibinfo {title} {{Schrödinger
  representation and Casimir effect in renormalizable quantum field theory}},\
  }\href {https://doi.org/https://doi.org/10.1016/0550-3213(81)90482-X}
  {\bibfield  {journal} {\bibinfo  {journal} {Nucl. Phys. B}\ }\textbf
  {\bibinfo {volume} {190}},\ \bibinfo {pages} {1} (\bibinfo {year}
  {1981})}\BibitemShut {NoStop}%
\bibitem [{\citenamefont {Diehl}(1986)}]{diehl_1986}%
  \BibitemOpen
  \bibfield  {author} {\bibinfo {author} {\bibfnamefont {H.-W.}\ \bibnamefont
  {Diehl}},\ }\href@noop {} {\emph {\bibinfo {title} {Phase transitions and
  critical phenomena}}},\ Vol.~\bibinfo {volume} {10}\ (\bibinfo  {publisher}
  {Academic Press, London},\ \bibinfo {year} {1986})\ p.~\bibinfo {pages}
  {75}\BibitemShut {NoStop}%
\bibitem [{\citenamefont {Diehl}(1997)}]{Diehl_1997}%
  \BibitemOpen
  \bibfield  {author} {\bibinfo {author} {\bibfnamefont {H.~W.}\ \bibnamefont
  {Diehl}},\ }\bibfield  {title} {\bibinfo {title} {The theory of boundary
  critical phenomena},\ }\href {https://doi.org/10.1142/S0217979297001751}
  {\bibfield  {journal} {\bibinfo  {journal} {Int. J. Mod. Phys. B}\ }\textbf
  {\bibinfo {volume} {11}},\ \bibinfo {pages} {3503} (\bibinfo {year}
  {1997})}\BibitemShut {NoStop}%
\bibitem [{\citenamefont {Furukawa}\ \emph {et~al.}(2013)\citenamefont
  {Furukawa}, \citenamefont {Gambassi}, \citenamefont {Dietrich},\ and\
  \citenamefont {Tanaka}}]{Gambassi_PRL}%
  \BibitemOpen
  \bibfield  {author} {\bibinfo {author} {\bibfnamefont {A.}~\bibnamefont
  {Furukawa}}, \bibinfo {author} {\bibfnamefont {A.}~\bibnamefont {Gambassi}},
  \bibinfo {author} {\bibfnamefont {S.}~\bibnamefont {Dietrich}},\ and\
  \bibinfo {author} {\bibfnamefont {H.}~\bibnamefont {Tanaka}},\ }\bibfield
  {title} {\bibinfo {title} {{Nonequilibrium critical Casimir effect in binary
  fluids}},\ }\href {https://doi.org/10.1103/PhysRevLett.111.055701} {\bibfield
   {journal} {\bibinfo  {journal} {Phys. Rev. Lett.}\ }\textbf {\bibinfo
  {volume} {111}},\ \bibinfo {pages} {055701} (\bibinfo {year}
  {2013})}\BibitemShut {NoStop}%
\bibitem [{\citenamefont {Zia}\ and\ \citenamefont
  {Brady}(2013)}]{zia2013stress}%
  \BibitemOpen
  \bibfield  {author} {\bibinfo {author} {\bibfnamefont {R.~N.}\ \bibnamefont
  {Zia}}\ and\ \bibinfo {author} {\bibfnamefont {J.~F.}\ \bibnamefont
  {Brady}},\ }\bibfield  {title} {\bibinfo {title} {Stress development,
  relaxation, and memory in colloidal dispersions: Transient nonlinear
  microrheology},\ }\href {https://doi.org/10.1122/1.4775349} {\bibfield
  {journal} {\bibinfo  {journal} {J. Rheol.}\ }\textbf {\bibinfo {volume}
  {57}},\ \bibinfo {pages} {457} (\bibinfo {year} {2013})}\BibitemShut
  {NoStop}%
\bibitem [{\citenamefont {Squires}\ and\ \citenamefont
  {Brady}(2005)}]{squires2005simple}%
  \BibitemOpen
  \bibfield  {author} {\bibinfo {author} {\bibfnamefont {T.~M.}\ \bibnamefont
  {Squires}}\ and\ \bibinfo {author} {\bibfnamefont {J.~F.}\ \bibnamefont
  {Brady}},\ }\bibfield  {title} {\bibinfo {title} {A simple paradigm for
  active and nonlinear microrheology},\ }\href
  {https://doi.org/10.1063/1.1960607} {\bibfield  {journal} {\bibinfo
  {journal} {Phys. Fluids}\ }\textbf {\bibinfo {volume} {17}},\ \bibinfo
  {pages} {073101} (\bibinfo {year} {2005})}\BibitemShut {NoStop}%
\bibitem [{\citenamefont {D\'emery}\ and\ \citenamefont
  {Dean}(2010)}]{demery2010}%
  \BibitemOpen
  \bibfield  {author} {\bibinfo {author} {\bibfnamefont {V.}~\bibnamefont
  {D\'emery}}\ and\ \bibinfo {author} {\bibfnamefont {D.~S.}\ \bibnamefont
  {Dean}},\ }\bibfield  {title} {\bibinfo {title} {Drag forces in classical
  fields},\ }\href {https://doi.org/10.1103/PhysRevLett.104.080601} {\bibfield
  {journal} {\bibinfo  {journal} {Phys. Rev. Lett.}\ }\textbf {\bibinfo
  {volume} {104}},\ \bibinfo {pages} {080601} (\bibinfo {year}
  {2010})}\BibitemShut {NoStop}%
\bibitem [{\citenamefont {D{\'e}mery}\ and\ \citenamefont
  {Dean}(2010)}]{demery2010-2}%
  \BibitemOpen
  \bibfield  {author} {\bibinfo {author} {\bibfnamefont {V.}~\bibnamefont
  {D{\'e}mery}}\ and\ \bibinfo {author} {\bibfnamefont {D.~S.}\ \bibnamefont
  {Dean}},\ }\bibfield  {title} {\bibinfo {title} {Drag forces on inclusions in
  classical fields with dissipative dynamics},\ }\href
  {https://doi.org/10.1140/epje/i2010-10640-1} {\bibfield  {journal} {\bibinfo
  {journal} {Eur. Phys. J. E}\ }\textbf {\bibinfo {volume} {32}},\ \bibinfo
  {pages} {377} (\bibinfo {year} {2010})}\BibitemShut {NoStop}%
\bibitem [{\citenamefont {D\'emery}\ and\ \citenamefont
  {Dean}(2011{\natexlab{a}})}]{demery2011}%
  \BibitemOpen
  \bibfield  {author} {\bibinfo {author} {\bibfnamefont {V.}~\bibnamefont
  {D\'emery}}\ and\ \bibinfo {author} {\bibfnamefont {D.~S.}\ \bibnamefont
  {Dean}},\ }\bibfield  {title} {\bibinfo {title} {{Thermal Casimir drag in
  fluctuating classical fields}},\ }\href
  {https://doi.org/10.1103/PhysRevE.84.010103} {\bibfield  {journal} {\bibinfo
  {journal} {Phys. Rev. E}\ }\textbf {\bibinfo {volume} {84}},\ \bibinfo
  {pages} {010103} (\bibinfo {year} {2011}{\natexlab{a}})}\BibitemShut
  {NoStop}%
\bibitem [{\citenamefont {D\'emery}(2013)}]{demery2013}%
  \BibitemOpen
  \bibfield  {author} {\bibinfo {author} {\bibfnamefont {V.}~\bibnamefont
  {D\'emery}},\ }\bibfield  {title} {\bibinfo {title} {Diffusion of a particle
  quadratically coupled to a thermally fluctuating field},\ }\href
  {https://doi.org/10.1103/PhysRevE.87.052105} {\bibfield  {journal} {\bibinfo
  {journal} {Phys. Rev. E}\ }\textbf {\bibinfo {volume} {87}},\ \bibinfo
  {pages} {052105} (\bibinfo {year} {2013})}\BibitemShut {NoStop}%
\bibitem [{\citenamefont {D\'emery}\ and\ \citenamefont
  {Dean}(2011{\natexlab{b}})}]{demerypath}%
  \BibitemOpen
  \bibfield  {author} {\bibinfo {author} {\bibfnamefont {V.}~\bibnamefont
  {D\'emery}}\ and\ \bibinfo {author} {\bibfnamefont {D.~S.}\ \bibnamefont
  {Dean}},\ }\bibfield  {title} {\bibinfo {title} {Perturbative path-integral
  study of active- and passive-tracer diffusion in fluctuating fields},\ }\href
  {https://doi.org/10.1103/PhysRevE.84.011148} {\bibfield  {journal} {\bibinfo
  {journal} {Phys. Rev. E}\ }\textbf {\bibinfo {volume} {84}},\ \bibinfo
  {pages} {011148} (\bibinfo {year} {2011}{\natexlab{b}})}\BibitemShut
  {NoStop}%
\bibitem [{\citenamefont {Dean}\ and\ \citenamefont
  {D{\'{e}}mery}(2011)}]{Dean_2011}%
  \BibitemOpen
  \bibfield  {author} {\bibinfo {author} {\bibfnamefont {D.~S.}\ \bibnamefont
  {Dean}}\ and\ \bibinfo {author} {\bibfnamefont {V.}~\bibnamefont
  {D{\'{e}}mery}},\ }\bibfield  {title} {\bibinfo {title} {Diffusion of active
  tracers in fluctuating fields},\ }\href
  {https://doi.org/10.1088/0953-8984/23/23/234114} {\bibfield  {journal}
  {\bibinfo  {journal} {J. Phys.-Condens. Mat.}\ }\textbf {\bibinfo {volume}
  {23}},\ \bibinfo {pages} {234114} (\bibinfo {year} {2011})}\BibitemShut
  {NoStop}%
\bibitem [{\citenamefont {Fujitani}(2016)}]{fuji1}%
  \BibitemOpen
  \bibfield  {author} {\bibinfo {author} {\bibfnamefont {Y.}~\bibnamefont
  {Fujitani}},\ }\bibfield  {title} {\bibinfo {title} {{Fluctuation amplitude
  of a trapped rigid sphere immersed in a near-critical binary fluid mixture
  within the regime of the Gaussian model}},\ }\href
  {https://doi.org/10.7566/JPSJ.85.044401} {\bibfield  {journal} {\bibinfo
  {journal} {J. Phys. Soc. Jpn.}\ }\textbf {\bibinfo {volume} {85}},\ \bibinfo
  {pages} {044401} (\bibinfo {year} {2016})}\BibitemShut {NoStop}%
\bibitem [{\citenamefont {Fujitani}(2017)}]{fuji2}%
  \BibitemOpen
  \bibfield  {author} {\bibinfo {author} {\bibfnamefont {Y.}~\bibnamefont
  {Fujitani}},\ }\bibfield  {title} {\bibinfo {title} {{Osmotic suppression of
  positional fluctuation of a trapped particle in a near-critical binary fluid
  mixture in the regime of the Gaussian model}},\ }\href
  {https://doi.org/10.7566/JPSJ.86.114602} {\bibfield  {journal} {\bibinfo
  {journal} {J. Phys. Soc. Jpn.}\ }\textbf {\bibinfo {volume} {86}},\ \bibinfo
  {pages} {114602} (\bibinfo {year} {2017})}\BibitemShut {NoStop}%
\bibitem [{\citenamefont {Gambassi}\ and\ \citenamefont
  {Dietrich}(2006)}]{Gambassi_2006}%
  \BibitemOpen
  \bibfield  {author} {\bibinfo {author} {\bibfnamefont {A.}~\bibnamefont
  {Gambassi}}\ and\ \bibinfo {author} {\bibfnamefont {S.}~\bibnamefont
  {Dietrich}},\ }\bibfield  {title} {\bibinfo {title} {Critical dynamics in
  thin films},\ }\href {https://doi.org/10.1007/s10955-006-9089-8} {\bibfield
  {journal} {\bibinfo  {journal} {J. Stat. Phys.}\ }\textbf {\bibinfo {volume}
  {123}},\ \bibinfo {pages} {929} (\bibinfo {year} {2006})}\BibitemShut
  {NoStop}%
\bibitem [{\citenamefont {Gambassi}(2008)}]{gambassi2008relaxation}%
  \BibitemOpen
  \bibfield  {author} {\bibinfo {author} {\bibfnamefont {A.}~\bibnamefont
  {Gambassi}},\ }\bibfield  {title} {\bibinfo {title} {Relaxation phenomena at
  criticality},\ }\href
  {https://link.springer.com/article/10.1140%2Fepjb%2Fe2008-00043-y} {\bibfield
   {journal} {\bibinfo  {journal} {Eur. Phys. J. B}\ }\textbf {\bibinfo
  {volume} {64}},\ \bibinfo {pages} {379} (\bibinfo {year} {2008})}\BibitemShut
  {NoStop}%
\bibitem [{\citenamefont {Krüger}\ \emph {et~al.}(2011)\citenamefont
  {Krüger}, \citenamefont {Emig}, \citenamefont {Bimonte},\ and\ \citenamefont
  {Kardar}}]{Kruger_2011}%
  \BibitemOpen
  \bibfield  {author} {\bibinfo {author} {\bibfnamefont {M.}~\bibnamefont
  {Krüger}}, \bibinfo {author} {\bibfnamefont {T.}~\bibnamefont {Emig}},
  \bibinfo {author} {\bibfnamefont {G.}~\bibnamefont {Bimonte}},\ and\ \bibinfo
  {author} {\bibfnamefont {M.}~\bibnamefont {Kardar}},\ }\bibfield  {title}
  {\bibinfo {title} {{Non-equilibrium Casimir forces: Spheres and
  sphere-plate}},\ }\href {https://doi.org/10.1209/0295-5075/95/21002}
  {\bibfield  {journal} {\bibinfo  {journal} {Europhys. Lett.}\ }\textbf
  {\bibinfo {volume} {95}},\ \bibinfo {pages} {21002} (\bibinfo {year}
  {2011})}\BibitemShut {NoStop}%
\bibitem [{\citenamefont {Kr\"uger}\ \emph {et~al.}(2012)\citenamefont
  {Kr\"uger}, \citenamefont {Bimonte}, \citenamefont {Emig},\ and\
  \citenamefont {Kardar}}]{Kruger_2012}%
  \BibitemOpen
  \bibfield  {author} {\bibinfo {author} {\bibfnamefont {M.}~\bibnamefont
  {Kr\"uger}}, \bibinfo {author} {\bibfnamefont {G.}~\bibnamefont {Bimonte}},
  \bibinfo {author} {\bibfnamefont {T.}~\bibnamefont {Emig}},\ and\ \bibinfo
  {author} {\bibfnamefont {M.}~\bibnamefont {Kardar}},\ }\bibfield  {title}
  {\bibinfo {title} {{Trace formulas for nonequilibrium Casimir interactions,
  heat radiation, and heat transfer for arbitrary objects}},\ }\href
  {https://doi.org/10.1103/PhysRevB.86.115423} {\bibfield  {journal} {\bibinfo
  {journal} {Phys. Rev. B}\ }\textbf {\bibinfo {volume} {86}},\ \bibinfo
  {pages} {115423} (\bibinfo {year} {2012})}\BibitemShut {NoStop}%
\bibitem [{\citenamefont {Rohwer}\ \emph {et~al.}(2017)\citenamefont {Rohwer},
  \citenamefont {Kardar},\ and\ \citenamefont {Kr\"uger}}]{Rohwer_2017}%
  \BibitemOpen
  \bibfield  {author} {\bibinfo {author} {\bibfnamefont {C.~M.}\ \bibnamefont
  {Rohwer}}, \bibinfo {author} {\bibfnamefont {M.}~\bibnamefont {Kardar}},\
  and\ \bibinfo {author} {\bibfnamefont {M.}~\bibnamefont {Kr\"uger}},\
  }\bibfield  {title} {\bibinfo {title} {{Transient Casimir forces from
  quenches in thermal and active matter}},\ }\href
  {https://doi.org/10.1103/PhysRevLett.118.015702} {\bibfield  {journal}
  {\bibinfo  {journal} {Phys. Rev. Lett.}\ }\textbf {\bibinfo {volume} {118}},\
  \bibinfo {pages} {015702} (\bibinfo {year} {2017})}\BibitemShut {NoStop}%
\bibitem [{\citenamefont {Hanke}(2013)}]{Hanke_2013}%
  \BibitemOpen
  \bibfield  {author} {\bibinfo {author} {\bibfnamefont {A.}~\bibnamefont
  {Hanke}},\ }\bibfield  {title} {\bibinfo {title} {{Non-equilibrium Casimir
  force between vibrating plates}},\ }\href
  {https://doi.org/10.1371/journal.pone.0053228} {\bibfield  {journal}
  {\bibinfo  {journal} {Plos One}\ }\textbf {\bibinfo {volume} {8}},\ \bibinfo
  {pages} {1} (\bibinfo {year} {2013})}\BibitemShut {NoStop}%
\bibitem [{\citenamefont {Hohenberg}\ and\ \citenamefont
  {Halperin}(1977)}]{halperin}%
  \BibitemOpen
  \bibfield  {author} {\bibinfo {author} {\bibfnamefont {P.~C.}\ \bibnamefont
  {Hohenberg}}\ and\ \bibinfo {author} {\bibfnamefont {B.~I.}\ \bibnamefont
  {Halperin}},\ }\bibfield  {title} {\bibinfo {title} {Theory of dynamic
  critical phenomena},\ }\href {https://doi.org/10.1103/RevModPhys.49.435}
  {\bibfield  {journal} {\bibinfo  {journal} {Rev. Mod. Phys.}\ }\textbf
  {\bibinfo {volume} {49}},\ \bibinfo {pages} {435} (\bibinfo {year}
  {1977})}\BibitemShut {NoStop}%
\bibitem [{\citenamefont {Basu}\ \emph {et~al.}(2022)\citenamefont {Basu},
  \citenamefont {Démery},\ and\ \citenamefont {Gambassi}}]{wellGauss}%
  \BibitemOpen
  \bibfield  {author} {\bibinfo {author} {\bibfnamefont {U.}~\bibnamefont
  {Basu}}, \bibinfo {author} {\bibfnamefont {V.}~\bibnamefont {Démery}},\ and\
  \bibinfo {author} {\bibfnamefont {A.}~\bibnamefont {Gambassi}},\ }\bibfield
  {title} {\bibinfo {title} {{Dynamics of a colloidal particle coupled to a
  Gaussian field: from a confinement-dependent to a non-linear memory}},\
  }\href {https://doi.org/10.21468/SciPostPhys.13.4.078} {\bibfield  {journal}
  {\bibinfo  {journal} {SciPost Phys.}\ }\textbf {\bibinfo {volume} {13}},\
  \bibinfo {pages} {078} (\bibinfo {year} {2022})}\BibitemShut {NoStop}%
\bibitem [{\citenamefont {Venturelli}\ \emph
  {et~al.}(2022{\natexlab{a}})\citenamefont {Venturelli}, \citenamefont
  {Ferraro},\ and\ \citenamefont {Gambassi}}]{ioeferraro}%
  \BibitemOpen
  \bibfield  {author} {\bibinfo {author} {\bibfnamefont {D.}~\bibnamefont
  {Venturelli}}, \bibinfo {author} {\bibfnamefont {F.}~\bibnamefont
  {Ferraro}},\ and\ \bibinfo {author} {\bibfnamefont {A.}~\bibnamefont
  {Gambassi}},\ }\bibfield  {title} {\bibinfo {title} {{Nonequilibrium
  relaxation of a trapped particle in a near-critical Gaussian field}},\ }\href
  {https://doi.org/10.1103/PhysRevE.105.054125} {\bibfield  {journal} {\bibinfo
   {journal} {Phys. Rev. E}\ }\textbf {\bibinfo {volume} {105}},\ \bibinfo
  {pages} {054125} (\bibinfo {year} {2022}{\natexlab{a}})}\BibitemShut
  {NoStop}%
\bibitem [{\citenamefont {Gross}(2021)}]{gross}%
  \BibitemOpen
  \bibfield  {author} {\bibinfo {author} {\bibfnamefont {M.}~\bibnamefont
  {Gross}},\ }\bibfield  {title} {\bibinfo {title} {Dynamics and steady states
  of a tracer particle in a confined critical fluid},\ }\href
  {https://doi.org/10.1088/1742-5468/abffce} {\bibfield  {journal} {\bibinfo
  {journal} {J. Stat. Mech. Theor. Exp.}\ }\textbf {\bibinfo {volume} {2021}},\
  \bibinfo {pages} {063209} (\bibinfo {year} {2021})}\BibitemShut {NoStop}%
\bibitem [{\citenamefont {Martínez}\ \emph {et~al.}(2017)\citenamefont
  {Martínez}, \citenamefont {Devailly}, \citenamefont {Petrosyan},\ and\
  \citenamefont {Ciliberto}}]{ciliberto}%
  \BibitemOpen
  \bibfield  {author} {\bibinfo {author} {\bibfnamefont {I.~A.}\ \bibnamefont
  {Martínez}}, \bibinfo {author} {\bibfnamefont {C.}~\bibnamefont {Devailly}},
  \bibinfo {author} {\bibfnamefont {A.}~\bibnamefont {Petrosyan}},\ and\
  \bibinfo {author} {\bibfnamefont {S.}~\bibnamefont {Ciliberto}},\ }\bibfield
  {title} {\bibinfo {title} {Energy transfer between colloids via critical
  interactions},\ }\href {https://www.mdpi.com/1099-4300/19/2/77} {\bibfield
  {journal} {\bibinfo  {journal} {Entropy}\ }\textbf {\bibinfo {volume}
  {19(2)}},\ \bibinfo {pages} {77} (\bibinfo {year} {2017})}\BibitemShut
  {NoStop}%
\bibitem [{\citenamefont {Schlesener}\ \emph {et~al.}(2003)\citenamefont
  {Schlesener}, \citenamefont {Hanke},\ and\ \citenamefont
  {Dietrich}}]{Schlesener2003}%
  \BibitemOpen
  \bibfield  {author} {\bibinfo {author} {\bibfnamefont {F.}~\bibnamefont
  {Schlesener}}, \bibinfo {author} {\bibfnamefont {A.}~\bibnamefont {Hanke}},\
  and\ \bibinfo {author} {\bibfnamefont {S.}~\bibnamefont {Dietrich}},\
  }\bibfield  {title} {\bibinfo {title} {{Critical Casimir forces in colloidal
  suspensions}},\ }\href {https://doi.org/10.1023/A:1022184508016} {\bibfield
  {journal} {\bibinfo  {journal} {J. Stat. Phys.}\ }\textbf {\bibinfo {volume}
  {110}},\ \bibinfo {pages} {981} (\bibinfo {year} {2003})}\BibitemShut
  {NoStop}%
\bibitem [{\citenamefont {Gambassi}\ \emph {et~al.}(2009)\citenamefont
  {Gambassi}, \citenamefont {Macio\l{}ek}, \citenamefont {Hertlein},
  \citenamefont {Nellen}, \citenamefont {Helden}, \citenamefont {Bechinger},\
  and\ \citenamefont {Dietrich}}]{Gambassi_2009_PRE}%
  \BibitemOpen
  \bibfield  {author} {\bibinfo {author} {\bibfnamefont {A.}~\bibnamefont
  {Gambassi}}, \bibinfo {author} {\bibfnamefont {A.}~\bibnamefont
  {Macio\l{}ek}}, \bibinfo {author} {\bibfnamefont {C.}~\bibnamefont
  {Hertlein}}, \bibinfo {author} {\bibfnamefont {U.}~\bibnamefont {Nellen}},
  \bibinfo {author} {\bibfnamefont {L.}~\bibnamefont {Helden}}, \bibinfo
  {author} {\bibfnamefont {C.}~\bibnamefont {Bechinger}},\ and\ \bibinfo
  {author} {\bibfnamefont {S.}~\bibnamefont {Dietrich}},\ }\bibfield  {title}
  {\bibinfo {title} {{Critical Casimir effect in classical binary liquid
  mixtures}},\ }\href {https://doi.org/10.1103/PhysRevE.80.061143} {\bibfield
  {journal} {\bibinfo  {journal} {Phys. Rev. E}\ }\textbf {\bibinfo {volume}
  {80}},\ \bibinfo {pages} {061143} (\bibinfo {year} {2009})}\BibitemShut
  {NoStop}%
\bibitem [{\citenamefont {Onuki}(2002)}]{onuki}%
  \BibitemOpen
  \bibfield  {author} {\bibinfo {author} {\bibfnamefont {A.}~\bibnamefont
  {Onuki}},\ }\href {https://doi.org/10.1017/CBO9780511534874} {\emph {\bibinfo
  {title} {Phase Transition Dynamics}}}\ (\bibinfo  {publisher} {Cambridge
  University Press},\ \bibinfo {year} {2002})\BibitemShut {NoStop}%
\bibitem [{Note1()}]{Note1}%
  \BibitemOpen
  \bibinfo {note} {We adopt here and in the following the Fourier convention
  $f(\vb {x}) = \DOTSI \intop \ilimits@ [\dd [d]{q}/(2\pi )^d] e^{i\vb {q}\cdot
  \vb {x}} f_{\vb {q}}$, and we normalize the delta distribution in Fourier
  space as $\DOTSI \intop \ilimits@ [\dd [d]{q}/(2\pi )^d] \delta
  ^d(q)=1$.}\BibitemShut {Stop}%
\bibitem [{\citenamefont {Risken}\ and\ \citenamefont {Haken}(1989)}]{risken}%
  \BibitemOpen
  \bibfield  {author} {\bibinfo {author} {\bibfnamefont {H.}~\bibnamefont
  {Risken}}\ and\ \bibinfo {author} {\bibfnamefont {H.}~\bibnamefont {Haken}},\
  }\href@noop {} {\emph {\bibinfo {title} {{The Fokker-Planck equation: methods
  of solution and applications}}}},\ \bibinfo {edition} {2nd}\ ed.\ (\bibinfo
  {publisher} {Springer},\ \bibinfo {year} {1989})\BibitemShut {NoStop}%
\bibitem [{\citenamefont {H{\"a}nggi}(1978)}]{Hanggi1978}%
  \BibitemOpen
  \bibfield  {author} {\bibinfo {author} {\bibfnamefont {P.}~\bibnamefont
  {H{\"a}nggi}},\ }\bibfield  {title} {\bibinfo {title} {{Correlation functions
  and masterequations of generalized (non-Markovian) Langevin equations}},\
  }\href {https://doi.org/10.1007/BF01351552} {\bibfield  {journal} {\bibinfo
  {journal} {Z. Phys. B Con. Mat.}\ }\textbf {\bibinfo {volume} {31}},\
  \bibinfo {pages} {407} (\bibinfo {year} {1978})}\BibitemShut {NoStop}%
\bibitem [{\citenamefont {Giuggioli}\ and\ \citenamefont
  {Neu}(2019)}]{Giuggioli2019}%
  \BibitemOpen
  \bibfield  {author} {\bibinfo {author} {\bibfnamefont {L.}~\bibnamefont
  {Giuggioli}}\ and\ \bibinfo {author} {\bibfnamefont {Z.}~\bibnamefont
  {Neu}},\ }\bibfield  {title} {\bibinfo {title} {{Fokker-Planck
  representations of non-Markov Langevin equations: application to delayed
  systems}},\ }\href {https://doi.org/10.1098/rsta.2018.0131} {\bibfield
  {journal} {\bibinfo  {journal} {Philos. T. R. Soc. A}\ }\textbf {\bibinfo
  {volume} {377}},\ \bibinfo {pages} {20180131} (\bibinfo {year}
  {2019})}\BibitemShut {NoStop}%
\bibitem [{\citenamefont {Bimonte}\ \emph {et~al.}(2022)\citenamefont
  {Bimonte}, \citenamefont {Emig}, \citenamefont {Graham},\ and\ \citenamefont
  {Kardar}}]{Bimonte_2022}%
  \BibitemOpen
  \bibfield  {author} {\bibinfo {author} {\bibfnamefont {G.}~\bibnamefont
  {Bimonte}}, \bibinfo {author} {\bibfnamefont {T.}~\bibnamefont {Emig}},
  \bibinfo {author} {\bibfnamefont {N.}~\bibnamefont {Graham}},\ and\ \bibinfo
  {author} {\bibfnamefont {M.}~\bibnamefont {Kardar}},\ }\href@noop {}
  {\bibinfo {title} {{Something can come of nothing: quantum fluctuations and
  the Casimir force}}} (\bibinfo {year} {2022}),\ \Eprint
  {https://arxiv.org/abs/2202.05386} {arXiv:2202.05386 [quant-ph]} \BibitemShut
  {NoStop}%
\bibitem [{\citenamefont {Hanke}\ \emph {et~al.}(1998)\citenamefont {Hanke},
  \citenamefont {Schlesener}, \citenamefont {Eisenriegler},\ and\ \citenamefont
  {Dietrich}}]{dietrich98}%
  \BibitemOpen
  \bibfield  {author} {\bibinfo {author} {\bibfnamefont {A.}~\bibnamefont
  {Hanke}}, \bibinfo {author} {\bibfnamefont {F.}~\bibnamefont {Schlesener}},
  \bibinfo {author} {\bibfnamefont {E.}~\bibnamefont {Eisenriegler}},\ and\
  \bibinfo {author} {\bibfnamefont {S.}~\bibnamefont {Dietrich}},\ }\bibfield
  {title} {\bibinfo {title} {{Critical Casimir forces between spherical
  particles in fluids}},\ }\href {https://doi.org/10.1103/PhysRevLett.81.1885}
  {\bibfield  {journal} {\bibinfo  {journal} {Phys. Rev. Lett.}\ }\textbf
  {\bibinfo {volume} {81}},\ \bibinfo {pages} {1885} (\bibinfo {year}
  {1998})}\BibitemShut {NoStop}%
\bibitem [{\citenamefont {Täuber}(2014)}]{Tauber}%
  \BibitemOpen
  \bibfield  {author} {\bibinfo {author} {\bibfnamefont {U.~C.}\ \bibnamefont
  {Täuber}},\ }\href {https://doi.org/10.1017/CBO9781139046213} {\emph
  {\bibinfo {title} {Critical Dynamics: A Field Theory Approach to Equilibrium
  and Non-Equilibrium Scaling Behavior}}}\ (\bibinfo  {publisher} {Cambridge
  University Press},\ \bibinfo {year} {2014})\BibitemShut {NoStop}%
\bibitem [{\citenamefont {Martin}\ \emph {et~al.}(1973)\citenamefont {Martin},
  \citenamefont {Siggia},\ and\ \citenamefont {Rose}}]{MSR}%
  \BibitemOpen
  \bibfield  {author} {\bibinfo {author} {\bibfnamefont {P.~C.}\ \bibnamefont
  {Martin}}, \bibinfo {author} {\bibfnamefont {E.~D.}\ \bibnamefont {Siggia}},\
  and\ \bibinfo {author} {\bibfnamefont {H.~A.}\ \bibnamefont {Rose}},\
  }\bibfield  {title} {\bibinfo {title} {Statistical dynamics of classical
  systems},\ }\href {https://doi.org/10.1103/PhysRevA.8.423} {\bibfield
  {journal} {\bibinfo  {journal} {Phys. Rev. A}\ }\textbf {\bibinfo {volume}
  {8}},\ \bibinfo {pages} {423} (\bibinfo {year} {1973})}\BibitemShut {NoStop}%
\bibitem [{\citenamefont {De~Dominicis}(1978)}]{DeDominicis}%
  \BibitemOpen
  \bibfield  {author} {\bibinfo {author} {\bibfnamefont {C.}~\bibnamefont
  {De~Dominicis}},\ }\bibfield  {title} {\bibinfo {title} {Dynamics as a
  substitute for replicas in systems with quenched random impurities},\ }\href
  {https://doi.org/10.1103/PhysRevB.18.4913} {\bibfield  {journal} {\bibinfo
  {journal} {Phys. Rev. B}\ }\textbf {\bibinfo {volume} {18}},\ \bibinfo
  {pages} {4913} (\bibinfo {year} {1978})}\BibitemShut {NoStop}%
\bibitem [{\citenamefont {Janssen}(1976)}]{Janssen1976}%
  \BibitemOpen
  \bibfield  {author} {\bibinfo {author} {\bibfnamefont {H.-K.}\ \bibnamefont
  {Janssen}},\ }\bibfield  {title} {\bibinfo {title} {{On a Lagrangean for
  classical field dynamics and renormalization group calculations of dynamical
  critical properties}},\ }\href {https://doi.org/10.1007/BF01316547}
  {\bibfield  {journal} {\bibinfo  {journal} {Z. Phys. B Con. Mat.}\ }\textbf
  {\bibinfo {volume} {23}},\ \bibinfo {pages} {377} (\bibinfo {year}
  {1976})}\BibitemShut {NoStop}%
\bibitem [{\citenamefont {Magazzù}\ \emph {et~al.}(2019)\citenamefont
  {Magazzù}, \citenamefont {Callegari}, \citenamefont {Staforelli},
  \citenamefont {Gambassi}, \citenamefont {Dietrich},\ and\ \citenamefont
  {Volpe}}]{casimirColloids}%
  \BibitemOpen
  \bibfield  {author} {\bibinfo {author} {\bibfnamefont {A.}~\bibnamefont
  {Magazzù}}, \bibinfo {author} {\bibfnamefont {A.}~\bibnamefont {Callegari}},
  \bibinfo {author} {\bibfnamefont {J.~P.}\ \bibnamefont {Staforelli}},
  \bibinfo {author} {\bibfnamefont {A.}~\bibnamefont {Gambassi}}, \bibinfo
  {author} {\bibfnamefont {S.}~\bibnamefont {Dietrich}},\ and\ \bibinfo
  {author} {\bibfnamefont {G.}~\bibnamefont {Volpe}},\ }\bibfield  {title}
  {\bibinfo {title} {{Controlling the dynamics of colloidal particles by
  critical Casimir forces}},\ }\href {https://doi.org/10.1039/C8SM01376D}
  {\bibfield  {journal} {\bibinfo  {journal} {Soft Matter}\ }\textbf {\bibinfo
  {volume} {15}},\ \bibinfo {pages} {2152} (\bibinfo {year}
  {2019})}\BibitemShut {NoStop}%
\bibitem [{\citenamefont {Zakine}\ \emph {et~al.}(2018)\citenamefont {Zakine},
  \citenamefont {Fournier},\ and\ \citenamefont {van Wijland}}]{fournier_2018}%
  \BibitemOpen
  \bibfield  {author} {\bibinfo {author} {\bibfnamefont {R.}~\bibnamefont
  {Zakine}}, \bibinfo {author} {\bibfnamefont {J.-B.}\ \bibnamefont
  {Fournier}},\ and\ \bibinfo {author} {\bibfnamefont {F.}~\bibnamefont {van
  Wijland}},\ }\bibfield  {title} {\bibinfo {title} {Field-embedded particles
  driven by active flips},\ }\href
  {https://doi.org/10.1103/PhysRevLett.121.028001} {\bibfield  {journal}
  {\bibinfo  {journal} {Phys. Rev. Lett.}\ }\textbf {\bibinfo {volume} {121}},\
  \bibinfo {pages} {028001} (\bibinfo {year} {2018})}\BibitemShut {NoStop}%
\bibitem [{\citenamefont {Venturelli}\ \emph
  {et~al.}(2022{\natexlab{b}})\citenamefont {Venturelli}, \citenamefont
  {Basu},\ and\ \citenamefont {Gambassi}}]{active}%
  \BibitemOpen
  \bibfield  {author} {\bibinfo {author} {\bibfnamefont {D.}~\bibnamefont
  {Venturelli}}, \bibinfo {author} {\bibfnamefont {U.}~\bibnamefont {Basu}},\
  and\ \bibinfo {author} {\bibfnamefont {A.}~\bibnamefont {Gambassi}},\
  }\bibfield  {title} {\bibinfo {title} {{Active particles in contact with a
  near-critical Gaussian field}},\ }\href@noop {} {\bibfield  {journal}
  {\bibinfo  {journal} {In preparation}\ } (\bibinfo {year}
  {2022}{\natexlab{b}})}\BibitemShut {NoStop}%
\bibitem [{\citenamefont {Bender}\ and\ \citenamefont {Orszag}(1978)}]{bender}%
  \BibitemOpen
  \bibfield  {author} {\bibinfo {author} {\bibfnamefont {C.~M.}\ \bibnamefont
  {Bender}}\ and\ \bibinfo {author} {\bibfnamefont {S.~A.}\ \bibnamefont
  {Orszag}},\ }\href@noop {} {\emph {\bibinfo {title} {{Advanced Mathematical
  Methods for Scientists and Engineers}}}}\ (\bibinfo  {publisher}
  {McGraw-Hill},\ \bibinfo {year} {1978})\BibitemShut {NoStop}%
\bibitem [{\citenamefont {Onsager}\ and\ \citenamefont
  {Machlup}(1953)}]{Onsager_1953}%
  \BibitemOpen
  \bibfield  {author} {\bibinfo {author} {\bibfnamefont {L.}~\bibnamefont
  {Onsager}}\ and\ \bibinfo {author} {\bibfnamefont {S.}~\bibnamefont
  {Machlup}},\ }\bibfield  {title} {\bibinfo {title} {Fluctuations and
  irreversible processes},\ }\href {https://doi.org/10.1103/PhysRev.91.1505}
  {\bibfield  {journal} {\bibinfo  {journal} {Phys. Rev.}\ }\textbf {\bibinfo
  {volume} {91}},\ \bibinfo {pages} {1505} (\bibinfo {year}
  {1953})}\BibitemShut {NoStop}%
\bibitem [{\citenamefont {Aron}\ \emph {et~al.}(2010)\citenamefont {Aron},
  \citenamefont {Biroli},\ and\ \citenamefont {Cugliandolo}}]{Aron_2010}%
  \BibitemOpen
  \bibfield  {author} {\bibinfo {author} {\bibfnamefont {C.}~\bibnamefont
  {Aron}}, \bibinfo {author} {\bibfnamefont {G.}~\bibnamefont {Biroli}},\ and\
  \bibinfo {author} {\bibfnamefont {L.~F.}\ \bibnamefont {Cugliandolo}},\
  }\bibfield  {title} {\bibinfo {title} {{Symmetries of generating functionals
  of Langevin processes with colored multiplicative noise}},\ }\href
  {https://doi.org/10.1088/1742-5468/2010/11/p11018} {\bibfield  {journal}
  {\bibinfo  {journal} {J. Stat. Mech. Theor. Exp.}\ }\textbf {\bibinfo
  {volume} {2010}},\ \bibinfo {pages} {P11018} (\bibinfo {year}
  {2010})}\BibitemShut {NoStop}%
\bibitem [{\citenamefont {Aron}\ \emph {et~al.}(2016)\citenamefont {Aron},
  \citenamefont {Barci}, \citenamefont {Cugliandolo}, \citenamefont {Arenas},\
  and\ \citenamefont {Lozano}}]{Aron_2016}%
  \BibitemOpen
  \bibfield  {author} {\bibinfo {author} {\bibfnamefont {C.}~\bibnamefont
  {Aron}}, \bibinfo {author} {\bibfnamefont {D.~G.}\ \bibnamefont {Barci}},
  \bibinfo {author} {\bibfnamefont {L.~F.}\ \bibnamefont {Cugliandolo}},
  \bibinfo {author} {\bibfnamefont {Z.~G.}\ \bibnamefont {Arenas}},\ and\
  \bibinfo {author} {\bibfnamefont {G.~S.}\ \bibnamefont {Lozano}},\ }\bibfield
   {title} {\bibinfo {title} {{Dynamical symmetries of Markov processes with
  multiplicative white noise}},\ }\href
  {https://doi.org/10.1088/1742-5468/2016/05/053207} {\bibfield  {journal}
  {\bibinfo  {journal} {J. Stat. Mech. Theor. Exp.}\ }\textbf {\bibinfo
  {volume} {2016}},\ \bibinfo {pages} {053207} (\bibinfo {year}
  {2016})}\BibitemShut {NoStop}%
\bibitem [{\citenamefont {Venturelli}\ and\ \citenamefont
  {Gross}(2022)}]{Venturelli_2022_confined}%
  \BibitemOpen
  \bibfield  {author} {\bibinfo {author} {\bibfnamefont {D.}~\bibnamefont
  {Venturelli}}\ and\ \bibinfo {author} {\bibfnamefont {M.}~\bibnamefont
  {Gross}},\ }\href@noop {} {\bibinfo {title} {Tracer particle in a confined
  correlated medium: an adiabatic elimination method}} (\bibinfo {year}
  {2022}),\ \Eprint {https://arxiv.org/abs/2209.10834} {arXiv:2209.10834
  [cond-mat.stat-mech]} \BibitemShut {NoStop}%
\bibitem [{\citenamefont {Gradshteyn}\ and\ \citenamefont
  {Ryzhik}(2007)}]{table}%
  \BibitemOpen
  \bibfield  {author} {\bibinfo {author} {\bibfnamefont {I.~S.}\ \bibnamefont
  {Gradshteyn}}\ and\ \bibinfo {author} {\bibfnamefont {I.~M.}\ \bibnamefont
  {Ryzhik}},\ }\href@noop {} {\emph {\bibinfo {title} {Table of integrals,
  series, and products}}},\ \bibinfo {edition} {7th}\ ed.\ (\bibinfo
  {publisher} {Elsevier/Academic Press, Amsterdam},\ \bibinfo {year}
  {2007})\BibitemShut {NoStop}%
\bibitem [{\citenamefont {Le~Bellac}(1991)}]{LeBellac}%
  \BibitemOpen
  \bibfield  {author} {\bibinfo {author} {\bibfnamefont {M.}~\bibnamefont
  {Le~Bellac}},\ }\href@noop {} {\emph {\bibinfo {title} {Quantum and
  statistical field theory}}}\ (\bibinfo  {publisher} {{Clarendon Press}},\
  \bibinfo {year} {1991})\BibitemShut {NoStop}%
\bibitem [{Note2()}]{Note2}%
  \BibitemOpen
  \bibinfo {note} {See Ref.~\cite {bender}, \protect \textit {Laplace's method
  for integrals with movable maxima}.}\BibitemShut {Stop}%
\bibitem [{\citenamefont {Frenkel}\ and\ \citenamefont {Smit}(2002)}]{frenkel}%
  \BibitemOpen
  \bibfield  {author} {\bibinfo {author} {\bibfnamefont {D.}~\bibnamefont
  {Frenkel}}\ and\ \bibinfo {author} {\bibfnamefont {B.}~\bibnamefont {Smit}},\
  }\href@noop {} {\emph {\bibinfo {title} {Understanding Molecular Simulation:
  From Algorithms to Applications}}},\ \bibinfo {edition} {2nd}\ ed.,\ \bibinfo
  {series} {Computational Science Series}, Vol.~\bibinfo {volume} {1}\
  (\bibinfo  {publisher} {Academic Press},\ \bibinfo {address} {San Diego},\
  \bibinfo {year} {2002})\BibitemShut {NoStop}%
\bibitem [{\citenamefont {Honeycutt}(1992)}]{stochasticRK}%
  \BibitemOpen
  \bibfield  {author} {\bibinfo {author} {\bibfnamefont {R.~L.}\ \bibnamefont
  {Honeycutt}},\ }\bibfield  {title} {\bibinfo {title} {{Stochastic Runge-Kutta
  algorithms. I. White noise}},\ }\href
  {https://doi.org/10.1103/PhysRevA.45.600} {\bibfield  {journal} {\bibinfo
  {journal} {Phys. Rev. A}\ }\textbf {\bibinfo {volume} {45}},\ \bibinfo
  {pages} {600} (\bibinfo {year} {1992})}\BibitemShut {NoStop}%
\end{thebibliography}%

\appendix

\section{Independent processes}
\label{par:correlators}
We revise here the well-known solutions of the independent processes which we obtain by setting the coupling constant $\lambda=0$. These are also the $\order{\lambda^0}$ expressions in our perturbative calculation. Averages over the independent processes are denoted as $\expval*{\dots}_0$ in the main text.
 
\subsection{Brownian motion in a harmonic potential}
\label{par:freepart}
The motion of a Brownian particle in a (possibly moving) harmonic potential is ruled by the Ornstein-Uhlenbeck process. Its Langevin equation reads
\begin{equation}
    \dot{\vb{X}}(t)= -\nu k \left[\vb{X}-\vb{X}_F(t) \right] + \bm{\xi}(t) \, ,
    \label{eq:freeparticle}
\end{equation}
where $\bm{\xi}(t)$ is a Gaussian variable with zero mean and
\begin{equation}
    \expval{\xi_{i}(t) \xi_{j}(t') } = 2\nu T \delta_{ij}\delta(t-t') \, .
\end{equation}
Each component $X_j$ of the particle position $\vb{X}$ is ruled by an independent Gaussian and Markovian process. The propagator $P_{1|1}(\vb{X},t|\vb{X}_0,t_0)$ is thus Gaussian, with
\begin{equation}
    P_{1|1}(\vb{X},t|\vb{X}_0,t_0) = \left[ \frac{1}{\sqrt{2\pi}\sigma(t)} \right]^d \exp[-\frac{|\vb{X}-\vb{m}(t)|^2}{2\sigma^2(t)}] \, ,
    \label{eq:gaussian_prop}
\end{equation}
where the symbol $(\dots |\dots)$ indicates a conditional average. This expression contains the expectation value $\vb{m}(t)$ of the particle position 
\begin{align}
    \vb{m}(t) &\equiv \expval{\vb{X}(t) | \vb{X}(t_0) = \vb{X}_0} \n\\
    &=\vb{X}_0e^{-\gamma(t-t_0)}+ \gamma \int_{t_0}^t \dd{s} e^{-\gamma(t-s)}\vb{X}_F(s)  \, ,
    \label{eq:m(t)}
\end{align}
and its variance which is, due to the isotropy of the problem, the same for each component $X_j$:
\begin{align}
    \sigma^2(t) &\equiv \expval{X_j^2(t) | X_j(t_0) = (\vb{X}_0)_j} - m_j^2(t) \n\\
    &=\frac{T}{k}\left[ 1-e^{-2\gamma(t-t_0)} \right] \, .
    \label{eq:sigma_freepart}
\end{align}
Above we called for brevity $\gamma\equiv \nu k$ and assumed the particle to start at time $t=t_0$ at position $\vb{X}(t=t_0)=\vb{X}_0$.

Note that, in general,
\begin{equation}
    P_{1|1}(\vb{X},t|\vb{X}_0,t_0) \neq P_{1|1}(\vb{X},t-t_0|\vb{X}_0,0) \, ,
\end{equation}
because of the explicit time dependence in $\vb{X}_F(t)$, which breaks the time-translational invariance of the problem.

Let us also compute here, by means of the Langevin equation \eqref{eq:freeparticle}, the connected two-time correlation function
\begin{align}
        &C(t_1,t_2) \equiv \expval{X_j\z (t_1) X_j\z (t_2)}_c \n \\
        &= \expval{\left[ X_j\z (t_1) - \expval{X_j\z (t_1)} \right]\left[ X_j\z (t_2) - \expval*{X_j\z (t_2)} \right] } \n\\
        & = \frac{T}{k} \left[ e^{-\gamma|t_2-t_1|} - e^{-\gamma(t_1+t_2-2t_0)} \right] \, .
    \label{eq:2point}
\end{align}

\subsubsection{Periodic forcing}
\label{par:app_periodic}
Consider now the motion of $\vb{X}(t) \mapsto \vb{Z}(t)$ when it is forced sinusoidally as in \cref{eq:forcing}. Setting $\vb{X}_0=0$, it is straightforward to obtain
\begin{align}
        \vb{m}(t) =&\expval{\vb{Z}(t)}_0 = \vb{\Delta} \left[1- e^{-\gamma_z (t-t_0)}\right] + \n\\
        &+  \vb{A}\left[ \sin (\Omega t - \theta_z) - \sin (\Omega t_0-\theta_z)e^{-\gamma_z (t-t_0)} \right] \n\\
        &\xrightarrow[t_0 \rightarrow - \infty]{} \vb{\Delta} + \vb{A}\sin (\Omega t - \theta_z) \, ,
\end{align}
where we defined the phase shift $\theta_z$ as in \cref{eq:phase_shift}. In the deterministic limit where $k_z \rightarrow \infty$, the particle simply follows the external forcing with no delay ($\theta_z \to 0$) and we recover $\expval{\vb{Z}(t)}_0= \vb{Z}_F(t)$.

\subsubsection{$n$-time correlation functions}
\label{par:expaverages}
The knowledge of the one- and two-time correlation functions is sufficient to write down the generating functional $\cor{Z}[j]$ for any Gaussian process: for each scalar component $X_i(t)\mapsto x(t)$, it reads
\begin{align}
        \cor{Z}[j] &= \expval*{e^{\int \dd{s} j(s)x(s)}} \n \\
        &= \int \cor{D}x(s) e^{-\cor{S}_\T{OM}[x(\tau)]+ \int \dd{s}j(s)x(s) } \n \\
        &= e^{\frac{1}{2} \int \dd{s_1}\dd{s_2} j(s_1)C(s_1,s_2)j(s_2) + \int \dd{s}j(s)m(s) } \, ,
    \label{eq:genfunctional}
\end{align}
where we averaged the source term $j(x)$ over the Onsager-Machlup dynamical functional \cite{Onsager_1953}
\begin{align}
    &\cor{S}_\T{OM}[x(\tau)] \\
    &\equiv \frac{1}{2} \int \dd{s_1}\dd{s_2} [x(s_1)-m(s_1)]C^{-1}(s_1,s_2)[x(s_2)-m(s_2)] \n
\end{align}
and where we normalized the integration measure $\cor{D}x(s)$ so that $\cor{Z}[j=0]=1$. We can use the generating functional to compute a generic $n$-time expectation value over the independent process, and in particular
\begin{equation}
    \expval*{e^{i \vb{q}\cdot \vb{X} (t) }}_0 = \prod_{j=1}^{d} \expval*{e^{i q_j X_j (t) }}_0 \, ,
\end{equation}
which enters the master equation \eqref{eq:master_general}. Indeed, each factor can be computed as
\begin{equation}
        \expval*{e^{i q_j X_j (t)}}_0 = \cor{Z} \left[ j(s) = iq_j \delta(s-t)   \right] 
        = e^{-\frac{q_j^2}{2} C(t,t)} e^{i q_j m_j(t)}
\end{equation}
and thus we find
\begin{equation}
    \expval*{e^{i \vb{q}\cdot \vb{X} (t) }}_0 = e^{-\frac{q^2}{2} C(t,t)} e^{i \vb{q} \cdot \vb{m}(t)} \, .
    \label{eq:exp1point}
\end{equation}
Similarly, the average
\begin{align}
    Q_q(s_1,s_2) &\equiv \expval*{ e^{i \vb{q} \cdot [\vb{X}^{(0)}(s_2)- \vb{X}^{(0)}(s_1)]} } \n\\
    &= \prod_{n=1}^d \expval*{ e^{i q_n [X_n(s_2)- X_n(s_1)]} }_0
\end{align}
which intervenes in the derivation of $\expval*{\vb{Y}(t)}$ in Appendix \ref{par:weakcoupling} can be dealt with as
\begin{equation}
   \expval*{ e^{i q_n [X_n(s_2)- X_n(s_1)]} }_0 = \cor{Z}[j=j^*] \, ,
\end{equation}
where we introduced
\begin{equation}
    j^*(s) \equiv iq_n \left[ \delta(s-s_2)-\delta(s-s_1)\right] \, ,
\end{equation}
and a straightforward calculation gives
\begin{align}
    Q_q(s_1,s_2)=& e^{i \vb{q} \cdot [\vb{m}(s_2)- \vb{m}(s_1)]} \times \n \\ &\times e^{-\frac{q^2}{2}\left[ C(s_1,s_1)+C(s_2,s_2)-2C(s_1,s_2) \right] } \, .
    \label{eq:Qq}
\end{align}
Here the expectation value of the position $\vb{m}(t)$ is given in \cref{eq:m(t)}, while we may write explicitly, in terms of the correlation function $C(s_1,s_2)$ defined in \cref{eq:2point},
\begin{align}
     &e^{-\frac{q^2}{2}\left[ C(s_1,s_1)+C(s_2,s_2)-2C(s_1,s_2) \right] } \n \\
     &\xrightarrow[t_0\rightarrow -\infty]{}
         \exp[-\frac{Tq^2}{k} \left( 1- e^{-\gamma|s_2-s_1|} \right) ] \, .
\label{eq:Qq2}
\end{align}

\subsection{Dynamics of the free-field}
\label{par:freefield}
The Langevin equation \eqref{eq:field_eom} for the field reads, at $\order{\lambda^0}$ and in Fourier space,
\begin{equation}
    \dot{\phi}_q = -\alpha_q \phi_q + \zeta_q \, ,
\end{equation}
where $\alpha_q$ is defined in \cref{eq:tau_phi} and noise correlations are given in \cref{eq:noise_corr_field_fourier}.
The problem is formally identical to that of the Ornstein-Uhlenbeck particle, so that by setting for simplicity $\phi_q(t_0)\equiv 0$ (a choice which is inconsequential in the long-time periodic state on which we will focus below) one can easily derive
\begin{equation}
    \expval*{\phi_q(s_1)\phi_p(s_2)}_0 = \delta^d(p+q)C_q(s_1,s_2) \, , \\
\end{equation}
where $C_q(s_1,s_2)$ is the free-field correlator defined in \cref{eq:freefieldcorrelator}. By construction, $C_q(s_1,s_2)= C_q(s_2,s_1)$; by formally taking the limit $t_0\rightarrow -\infty$ in \cref{eq:freefieldcorrelator}, one obtains the equilibrium correlator $C_q(\tau)$ given in \cref{eq:C_eq}, which is a function of the time difference $\tau=s_2-s_1$ only. 

It is also customary \cite{Tauber} to define the response function $G_q(t)$ and linear susceptibility $\chi_q(t)$ of the free-field as in \cref{eq:field-prop,eq:field-susc}, respectively; their time-translational invariance derives from that of the equation of motion. They are linked to the equilibrium correlator in \cref{eq:C_eq} by the fluctuation-dissipation theorem
\begin{equation}
    T \chi_q(\tau) = -\Theta(\tau) \pdv{\tau} C_q(\tau) \, ,
\end{equation}
where we indicated by $\Theta(s)$ the Heaviside theta function. It is finally straightforward to derive the relation
\begin{equation}
    C_q(s_1,s_2) = \Omega_\phi(\vb{q}) \int_{t_0}^{\text{min}(s_1,s_2)}\dd{u} G_q(s_1-u)G_{-q}(s_2-u) \, ,
    \label{eq:magicrelation_realspace}
\end{equation}
where we named the noise amplitude in \cref{eq:noise_corr_field_fourier}
\begin{equation}
    \Omega_\phi(\vb{q}) \equiv 2DTq^\alpha \, ,
    \label{eq:Omega_phi}
\end{equation}
and which becomes, in equilibrium and in Fourier space,
\begin{equation}
    C_q(\omega) = \Omega_\phi(\vb{q}) G_q(\omega)G_{-q}(-\omega) \, .
    \label{eq:magicrelation}
\end{equation}

\section{Weak-coupling expansion}
\label{par:weakcoupling}
In this Appendix we illustrate how the coupled dynamics of the field and the particles can be studied by expanding their coordinates in powers of the coupling constant $\lambda$, as in \cref{eq:series_expansion}, and similarly to what was done in Refs. \cite{ioeferraro,wellGauss}. Plugging these expansions into the equations of motion of the particles, \cref{eq:particle_eom,eq:particleZ}, one finds
\begin{align}
    \dot{\vb{X}}_a\z(t) &= -\nu k \vb{X}_a\z(t) + \bm{\xi}(t) \, , \label{eq:partzero} \\
    \dot{\vb{X}}_a^{(n)}(t) &= -\nu k \vb{X}_a^{(n)}(t) + \nu \vb{f}_a^{(n-1)}(t) \, ,
    \label{eq:perturb}
\end{align}
with $a=y,z$, and where we defined
\begin{equation}
    \vb{f}_a^{(n)}(t) \equiv \frac{1}{n!}\eval{\dv[n]{\lambda}}_{\lambda=0} \vb{f}_a(t) \, .
\end{equation}
Equation \eqref{eq:partzero} is solved by the Ornstein-Uhlenbeck process, as discussed in Appendix \ref{par:correlators}, while the higher-order corrections can be formally expressed as
\begin{equation}
    \vb{X}_a^{(n)}(t) = \nu \int_{t_0}^t \dd{s} e^{-\gamma (t-s)} \vb{f}_a^{(n-1)}(s) \, .
\end{equation}
Similarly, the Langevin equation \eqref{eq:field_eom} for the field in Fourier space becomes, order by order in $\lambda$,
\begin{align}
    \partial_t\phi^{(0)}_q(t) 
        =& - \alpha_q \phi^{(0)}_q(t) + \zeta_q(t) \, ,
    \label{eq:phi0} \\
    \partial_t\phi^{(n)}_q(t) =& - \alpha_q \phi^{(n)}_q(t) \label{eq:phi1}\\ 
    &+\frac{D q^\alpha}{(n-1)!} \sum_{a=y,z} V_q^{(a)} \eval{\dv[n-1]{\lambda}}_{\lambda=0} e^{-i\vb{q}\cdot \vb{X}_a} \, . \n
\end{align}
The dynamics of the decoupled field $\phi^{(0)}_q(t) $ has been discussed in Appendix \ref{par:correlators}, while the equation of motion for $\phi_q \o (t)$ can be formally solved as
\begin{equation}
    \phi_q \o (s) = Dq^\alpha \sum_{a=y,z} V_q^{(a)}  \int_{t_0}^s \dd{\tau} e^{-\alpha_q (s-\tau)} e^{-i\vb{q}\cdot \vb{X}_a\z (\tau)} \, .
\end{equation}

Let us derive, as an example, the average particle position $\expval*{\vb{Y}(t)}$, which is given by
\begin{equation}
    \expval*{\vb{Y}(t)} = \expval*{\vb{Y}\z (t)} + \lambda \expval*{\vb{Y}\o (t)} + \lambda^2 \expval*{\vb{Y}\t (t)} + \order{\lambda^3} \, .
\end{equation}
In the setting described in \cref{par:Model} we have $\expval*{\vb{Y}\z (t)}=0$, and one can argue by symmetry under
\begin{equation}
    \{ \lambda\leftrightarrow -\lambda \,, \; \phi\leftrightarrow -\phi \} 
    \label{eq:symmetry}
\end{equation}
that $\expval*{\vb{Y}\o (t)} = 0$ \cite{wellGauss}. The leading order contribution then reads
\begin{equation}
     \expval*{\vb{Y}\t(t)} = \cor{S}_1 + \cor{S}_2 + \cor{S}_3 \, ,
\end{equation}
with
\begin{align}
    \cor{S}_1  =& i \nu_y  \int \frac{\dd[d]{q}}{(2\pi)^d} \vb{q} V_{q}\y V_{-q}\y \int_{t_0}^t \dd{s_2} e^{-\gamma_y (t-s_2)} \n\\
    &\times \int_{t_0}^{s_2} \dd{s_1} \chi_q(s_2-s_1) Q_q\y(s_1,s_2) \, ,  \label{eq:S1} \\
    \cor{S}_2  =& i \nu^2_y \int \frac{\dd[d]{q}}{(2\pi)^d} \vb{q} q^2 V_{q}\y V_{-q}\y \int_{t_0}^t \dd{s_2} \n\\  &\times \int_{t_0}^{s_2} \dd{s_1} e^{-\gamma_y (t-s_1)} C_q(s_1,s_2) Q_q\y(s_1,s_2) \, ,  \label{eq:S2} \\
    \cor{S}_3 =& i \nu_y \int \frac{\dd[d] {q}}{(2\pi)^d} \vb{q} V^{\T{(z)}}_{q} V_{-q}\y  \int_{t_0}^t \dd{s_2} e^{-\gamma_y (t-s_2)} \n\\
    &\times \int_{t_0}^{s_2} \dd{s_1}   \chi_q(s_2-s_1) \varphi_{-q}^{\T{(z)}}(s_1) \varphi_q\y(s_2) \, .
    \label{eq:S3}
\end{align}
The only subtle point in this calculation is the observation that, as we take the expectation values over the stochastic noises,
\begin{align}
    &\expval*{\phi_q \z (s_2)\phi_q \z (s_1)  e^{i \vb{q} \cdot [\vb{Y}\z(s_2)- \vb{Y}\z(s_1)]}} \n \\
    &= \expval*{\phi_q (s_2)\phi_q (s_1)}_0 \expval*{  e^{i \vb{q} \cdot [\vb{Y}(s_2)- \vb{Y}(s_1)]}}_0 \, ,
\end{align}
because at $\order{\lambda^0}$ the various processes are independent. Hence we defined
\begin{align}
    Q_q\y (s_1,s_2) &\equiv \expval*{ e^{i \vb{q} \cdot [\vb{Y}(s_2)- \vb{Y}(s_1)]} }_0 \, , \\
    \varphi_q\y(t) &= \expval*{e^{i \vb{q}\cdot \vb{Y} (t) }}_0 \, , 
\end{align}
which have been computed in Appendix \ref{par:expaverages}, while the functions $\chi_q(s_1,s_2)$ and $C_q(s_1,s_2)$ are the dynamical susceptibility of the free-field and its correlator introduced in \cref{eq:field-susc,eq:freefieldcorrelator}, respectively.

We now observe that in the absence of a second colloid $\vb{Z}$ one expects $\expval*{\vb{Y}(t\rightarrow \infty)}=0$ at equilibrium. Indeed, it was proved in Ref.~\cite{ioeferraro} (and we will discuss this further below in Appendix \ref{par:machefortuna}) that the equilibrium distribution of a single particle is not affected by the presence of the field and it remains the canonical one, given by $P_\T{eq}(\vb{Y}) = \exp( -\beta k_y Y^2/2)$. But in fact the quantities $\cor{S}_1$ and $\cor{S}_2$ in \cref{eq:S1,eq:S2} do not depend on $\vb{Z}(t)$, so they are the same whether we add a second particle to the problem or not. Accordingly, we conclude that
\begin{equation}
    \cor{S}_1 \, , \, \cor{S}_2 \xrightarrow[t\rightarrow +\infty]{} 0 \, .
\end{equation}
The average position in the periodic state attained at long times is then described by the third contribution only, \ie, $\expval*{\vb{Y}\t (t)}=\cor{S}_3$, which coincides with the expression we derived in Section \ref{par:PS}, see \cref{eq:dyn_avg_prior}. 

\section{Master equation}
In this Appendix we provide details on the derivation and solution of the master equation for $P_1(\vb{y},t)$ discussed in \cref{par:masterequation}.

\subsection{Derivation of the master equation}
\label{par:masterequation_conti}
The master equation \eqref{eq:master_general} can be derived from \cref{eq:master_initial} by evaluating each of the terms which appear on its RHS. The first one reads simply
\begin{equation}
    \expval{\delta(\vb{y}-\vb{Y}(t)) \vb{Y}(t) } = \vb{y} P_1(\vb{y},t) \, ,
    \label{eq:term1}
\end{equation}
where the average is intended over all possible realizations of the stochastic noises $\zeta_q(t)$ and $\bm{\xi}_{y,z}(t)$, and similarly
\begin{equation}
    \expval{\delta(\vb{y}-\vb{Y}(t)) e^{i\vb{q}\cdot \vb{Y}(t)} } = e^{i\vb{q}\cdot \vb{y}} P_1(\vb{y},t) \, .
    \label{eq:term2}
\end{equation}
In order to obtain the first nontrivial correction of $\order{\lambda^2}$, it is sufficient to compute up to $\order{\lambda^0}$ the term
\begin{equation}
    \expval{\delta(\vb{y}-\vb{Y}(t)) e^{i\vb{q}\cdot \vb{Z}(s)} }_0 = \expval{e^{i\vb{q}\cdot \vb{Z}(s)}}_0 P_1(\vb{y},t) \, ,
    \label{eq:avg_over_z}
\end{equation}
where we used the fact that the processes for $\vb{Y}$ and $\vb{Z}$ with $\lambda=0$ are independent, and the remaining average on the r.h.s. of \cref{eq:avg_over_z} is meant over the noise $\bm{\xi}\zz(t)$ only. Expectation values involving the noises $\bm{\xi}\y(t)$ and $\zeta(t)$ can be handled by taking path-integrals over the stochastic actions \cite{Tauber}
\begin{align}
    &\cor{S}_\xi[\bm{\xi}] = \frac{1}{2\Omega_y} \sum_{i=1}^d \int \dd{\tau} \xi_i^2(\tau) \, , \\
    &\cor{S}_\zeta[\zeta] = \frac{1}{2} \int \dslash{q} \int \dd{\tau} \frac{\zeta_q(\tau)\zeta_{-q}(\tau)}{\Omega_\phi(\vb{q})} \, ,
\end{align}
where $\Omega_y \equiv 2\nu_y T$ and $\Omega_\phi(\vb{q})$ is given in \cref{eq:Omega_phi}.
For instance, calling for brevity $\bm{\xi}\y \equiv \bm{\xi}$,
\begin{align}
    &\expval{\delta(\vb{y}-\vb{Y}(t)) \xi_i(t) } = \int \cor{D}\bm{\xi} \, \delta(\vb{y}-\vb{Y}(t)) \xi_i(t) e^{-\cor{S}_\xi[\bm{\xi}]} \n \\
    &= -\Omega_y \int \cor{D}\bm{\xi} \, \delta(\vb{y}-\vb{Y}(t)) \fdv{\xi_i(t)} e^{-\cor{S}_\xi[\bm{\xi}]} \n \\
    &= \Omega_y \expval{ \fdv{\xi_i(t)}  \delta(\vb{y}-\vb{Y}(t))} \n \\
    &= -\Omega_y \grad_{\vb{y}} \cdot
    \expval{\delta(\vb{y}-\vb{Y}(t)) \fdv{\vb{Y}(t)}{\xi_i(t)} } \, .
\end{align}
Using the equation of motion \eqref{eq:particle_eom} for $\vb{Y}(t)$, it is then easy to derive
\begin{align}
    \fdv{Y_j(t)}{\xi_i(t)} = \int_{t_0}^t \dd{s} e^{-\gamma_y (t-s)} \delta_{ij} \delta(t-s) = \frac{1}{2} \delta_{ij} \, ,
    \label{eq:strat}
\end{align}
whence
\begin{equation}
    \expval{\delta(\vb{y}-\vb{Y}(t)) \bm{\xi}(t) } = -\frac{\Omega_y}{2} \grad_{\vb{y}} P_1(\vb{y},t) \, .
    \label{eq:term4}
\end{equation}
We have adopted the Stratonovich convention in \cref{eq:strat}, but this does not affect the resulting Fokker-Planck equation because the noise $\bm{\xi}$ enters additively in the Langevin equation \eqref{eq:particle_eom} for the particle $\vb{Y}$ \cite{risken}. Similarly,
\begin{align}
    &\expval{\delta(\vb{y}-\vb{Y}(t)) \zeta_q(s) }  \n\\
    &= -\Omega_\phi(\vb{q}) \int \cor{D}\bm{\zeta} \, \delta(\vb{y}-\vb{Y}(t)) \fdv{\zeta_{-q}(s)} e^{-\cor{S}_\zeta[\bm{\zeta}]} \n \\
    &=- \Omega_\phi(\vb{q})  \grad_{\vb{y}} \cdot
    \expval{\delta(\vb{y}-\vb{Y}(t)) \fdv{\vb{Y}(t)}{\zeta_{-q}(s)} } \, ,
\end{align}
and using the effective Langevin equation for $\vb{Y}(t)$ one obtains
\begin{align}
    \fdv{\vb{Y}(t)}{\zeta_{-q}(s)} =& i \nu_y \lambda  \int_{t_0}^t \dd{s_2} e^{-\gamma_y (t-s_2)} \int \dslash{p} \vb{p}  V_{-p}\y  \n\\
    & \times \fdv{\zeta_{-q}(s)} \left[ \phi_p(s_2) e^{i\vb{p}\cdot \vb{Y}(s_2)} \right] \, .
\end{align}
From \cref{eq:particle_eom,eq:phi_risolto} it follows that
\begin{align}
    &\fdv{\zeta_{-q}(s)} \left[ \phi_p(s_2) e^{i\vb{p}\cdot \vb{Y}(s_2)} \right] =  e^{i\vb{p}\cdot \vb{Y}(s_2)} \fdv{\phi_p(s_2)}{\zeta_{-q}(s)} + \order{\lambda} \n \\
    &=  e^{i\vb{p}\cdot \vb{Y}(s_2)} \int_{t_0}^{s_2} \dd{s_1} G_p(s_2-s_1) \delta^d(p+q)\delta(s_1-s) + \order{\lambda} \n \\
    &=  e^{i\vb{p}\cdot \vb{Y}(s_2)} G_p(s_2-s) \delta^d(p+q) + \order{\lambda} \, ,
\end{align}
so we can express
\begin{align}
    &\expval{\delta(\vb{y}-\vb{Y}(t)) \zeta_q(s) } = i \nu_y \lambda \Omega_\phi(\vb{q}) V_{q}\y \vb{q}
    \int_{t_0}^t \dd{s_2} e^{-\gamma_y (t-s_2)} \n\\ 
    &\times G_q(s_2-s)   \grad_{\vb{y}} \expval{\delta(\vb{y}-\vb{Y}(t)) e^{-i\vb{q}\cdot \vb{Y}(s_2)}  } + \order{\lambda^2} \, .
    \label{eq:term5}
\end{align}
Note that in the calculation above there has been no need to specify the Itô or Stratonovich interpretation, because the noise $\zeta_{q}(s)$ gets integrated over the past times in the effective Langevin equation for $\vb{Y}$ derived as explained in \cref{par:masterequation_2}: the non-Markovianity renders such a specification unnecessary \cite{Aron_2010,Aron_2016}. Finally, we interpret
\begin{align}
    &\expval{\delta(\vb{y}-\vb{Y}(t)) e^{i\vb{q}\cdot \vb{Y}(s)}} = \int_{\vb{Y}(t)=\vb{y}} \cor{D}\vb{Y}(\tau) \, e^{i\vb{q}\cdot \vb{Y}(s)} \n\\
    &= \int \dd{\vb{x}} e^{i\vb{q}\cdot \vb{x}} P_2(\vb{y},t ; \vb{x}, s) \, ,
    \label{eq:term6}
\end{align}
where the path integral is intended over all possible realizations of the process $\vb{Y}(\tau)$, conditioned to the constraint $\vb{Y}(t)=\vb{y}$. Putting together the various terms (\eg, \cref{eq:term1,eq:term2,eq:avg_over_z,eq:term4,eq:term5,eq:term6}) and using \cref{eq:magicrelation_realspace}, we finally arrive at the master equation in \cref{eq:master_general}. Note that a term of $\order{\lambda^3}$ in the marginal distribution $P(\vb{y},t)$ is forbidden by the symmetry in \cref{eq:symmetry}, so that the next perturbative correction is at least of $\order{\lambda^4}$.

A hierarchy of master equations linking the $n$-time correlation function $P_n$ with $P_{n+1}$ can be obtained starting from the definition \cite{Giuggioli2019}
\begin{equation}
    P_n(\vb{x}_n,t_n; \dots ; \vb{x}_1,t_1)
    = \expval{ \delta(\vb{x}_n-\vb{Y}(t_n)) \dots \delta(\vb{x}_1-\vb{Y}(t_1)) }
\end{equation}
and acting as
\begin{align}
    &\partial_{t_j} P_n(\vb{x}_n,t_n; \dots ; \vb{y},t_j ; \dots ; \vb{x}_1,t_1) \\
    &=  - \grad_{\vb{y}} \cdot \expval{ \delta(\vb{x}_n-\vb{Y}(t_n)) \dots  \delta(\vb{x}_1-\vb{Y}(t_1)) \dot{\vb{Y}}(t_j)} \, .\n
\end{align}
The result of this procedure is completely analogous to \cref{eq:master_general} upon replacing $P_1 \rightarrow P_n$ and $P_2 \rightarrow P_{n+1}$.

In order to check the accuracy of the master equation \eqref{eq:master_general}, we can use it to predict the expectation value of the position $\expval{\vb{Y}(t)}$ when $\vb{Y}(t=t_0)\neq 0$ and the second particle $\vb{Z}$ is decoupled from the system (\ie, with $V_q\zz=0$). This quantity was recently derived in \ccite{ioeferraro} via a weak-coupling expansion such as the one in \cref{par:weakcoupling}. One starts by replacing the two-point function $P_2(\vb{y},t ; \vb{x}, s)$ in \cref{eq:master_general} by its $\order{\lambda^0}$ approximation,
\begin{equation}
    P_2(\vb{y},t ; \vb{x}, s) = P_{1|1}(\vb{y},t | \vb{x}, s)P_1( \vb{x}, s) +\order{\lambda^2} \, ,
    \label{eq:2point_approx}
\end{equation}
where we used the fact that the independent ($\lambda=0$) process is Markovian, $P_{1|1}$ is the Ornstein-Uhlenbeck propagator given in \cref{eq:gaussian_prop,eq:m(t),eq:sigma_freepart}, and $P_1( \vb{x}, s)$ is chosen to be the thermal equilibrium distribution of the particle $\vb{Y}$ in its harmonic trap (see, c.f., \cref{eq:eq_dist_y}). By using \cref{eq:2point_approx}, the master equation \eqref{eq:master_general} becomes a Fokker-Planck equation which can be used to compute $\expval{\vb{Y}(t)}$: a straightforward calculation renders the same result as in Eq.~(24) of \ccite{ioeferraro}, as expected.

\subsection{Irrelevance of the memory kernel in the periodic state up to $\order{\lambda^2}$}
\label{par:machefortuna}
Here we prove that the non-Markovian term in the master equation \eqref{eq:master_general} containing the memory kernel $\cor{L}(t-s)$ can be discarded in the periodic state. In order to do this, we need to use a result derived in \ccite{ioeferraro,wellGauss}, which we briefly report here.

Consider a \textit{single} particle coupled to a fluctuating scalar field via a translationally invariant coupling (for instance, consider the setup studied in this work, but with the second particle $\vb{Z}$ decoupled from the field). Suppose that the joint Hamiltonian has the form
\begin{equation}
    \cor{H}[\phi,\vb{X}] = \cor{H}_\phi[\phi] + \cor{U}(\vb{X}) - \lambda \cor{H}_\T{int}[\phi,\vb{X}] \, ,
\end{equation}
where $\cor{H}_\phi[\phi]$ describes the field in the bulk (not necessarily Gaussian), and $\cor{U}(\vb{X})$ is a confining particle potential, for example $\cor{U}(\vb{X}) = k X^2/2$ in the case considered here. Finally, $\cor{H}_\T{int}$ describes the interaction between the field and the particle via a possibly nonlinear coupling
\begin{equation}
    \cor{H}_\T{int}[\phi,\vb{X}] = \int \dd[d]{x} F[\phi(\vb{x})]V(\vb{x}-\vb{X}) \, ,
    \label{eq:generic_interaction}
\end{equation}
where $F[\phi(\vb{x})]$ is a quasi-local functional of $\phi$. Importantly, $\cor{H}_\T{int}$ is translationally invariant, in the sense that
\begin{equation}
    \cor{H}_\T{int}[\phi(\vb{x}),\vb{X}] = \cor{H}_\T{int}[\phi(\vb{x}-\vb{a}),\vb{X}+\vb{a}] \, .
    \label{eq:Hint_symmetry}
\end{equation}
Note that the interacting Hamiltonian in \cref{eq:Hint} satisfies these requirements.

Under these hypotheses, one can show \cite{ioeferraro,wellGauss} that the marginal equilibrium distribution of the particle alone reads simply
\begin{align}
    P_\T{eq}(\vb{X}) \propto \exp[-\beta\cor{U}(\vb{X})] \, ,
    \label{eq:boltzmann_single}
\end{align}
\ie, that the interaction with the field does not affect the equilibrium distribution of the particle. This argument does not apply if the system is not translationally invariant, as it happens, for instance, in the presence of boundaries or confinement \cite{gross,Venturelli_2022_confined}. On the other hand, this result does \textit{not} rely on the linearity of the field-particle coupling, nor on the fact that the field is Gaussian, and not even on the particular choice of a quadratic particle potential $\cor{U}(\vb{X})$.

Turning back to the master equation for $\vb{Y}(t)$ in \cref{eq:master_general}, let us now set initially $V_q\zz\equiv 0$, so that the second particle is decoupled from the problem. The master equation then reads at long times
\begin{align}
    &\partial_t P_1(\vb{y},t) = \cor{L}_0 P_1(\vb{y},t) \\
    &+ \lambda^2 \int_{0}^\infty \dd{u} \int \dd{\vb{x}} \cor{L}(\vb{y}-\vb{x},u)   P_2(\vb{y},t ; \vb{x}, t-u) + \order{\lambda^4} \, ,\n
\end{align}
where we called $u \equiv t-s$ and sent $t_0 \rightarrow -\infty$, and where the operators $\cor{L}_0$ and $\cor{L}$ were given in \cref{eq:L_0,eq:memory}, respectively. In the absence of any external forcing, the system will reach a state of thermal equilibrium with a stationary probability distribution $ P_\T{1,eq}$ satisfying $ \partial_t P_\T{1,eq}(\vb{y}) \equiv 0 $: this has to be the case order by order in the coupling constant $\lambda$. In particular, we read at $\order{\lambda^2}$
\begin{align}
    0\equiv \; &\cor{L}_0 P_\T{1,eq}^\T{(2)}(\vb{y}) \\
    &+ \int_{0}^\infty \dd{u} \int \dd{\vb{x}} \cor{L}(\vb{y}-\vb{x},u)  P_\T{2,eq}^\T{(0)}(\vb{y},t ; \vb{x}, t-u) \, ,\n
\end{align}
where the superscript indicates the order in the expansion in powers of $\lambda$. On the other hand, we know \textit{a priori} (see discussion above) that the stationary distribution of a single particle in thermal equilibrium with a fluctuating field reads simply
\begin{equation}
    P_\T{1,eq}\left(\vb{Y}\right) \propto \int \cor{D}\phi \, e^{-\beta \cor{H} \left[\phi,\vb{Y} \right] } \propto e^{-\beta k_y Y^2/2} \, .
\end{equation}
Here we deduce in particular that $P_\T{1,eq}^\T{(2)}(\vb{y}) =0$, and thus we can conclude that
\begin{align}
    \int_{0}^\infty \dd{u} \int \dd{\vb{x}} \cor{L}(\vb{y}-\vb{x},u)  P_\T{2,eq}^\T{(0)}(\vb{y},t ; \vb{x}, t-u) \equiv 0 \, .
    \label{eq:PS_kernel}
\end{align}
Switching on the coupling $V_q\zz$, so as to include the second particle into the problem, has actually no effect on $P_\T{2,eq}(\vb{y},t ; \vb{x}, t-u)$ at $\order{\lambda^0}$ -- this can be deduced by looking at its master equation, see Appendix \ref{par:masterequation_conti}. Accordingly, we conclude that \cref{eq:PS_kernel} must still hold true in the periodic state, up to $\order{\lambda^2}$.

\subsection{Solution of the master equation in the periodic state}
\label{par:solutionME}
In this Appendix we look for a perturbative solution of the master equation \eqref{eq:master_PS} in powers of the coupling constant $\lambda$. To lighten the notation, we will drop the subscript $y$ from the constants $\nu$ and $k$, and simply add the subscript $z$ when we are referring to the second colloid $\vb{Z}$.
Notice first that \cref{eq:master_PS} is solved at the lowest order by the stationary distribution of the Ornstein-Uhlenbeck process (see Appendix \ref{par:freepart}):
\begin{equation}
    P_1\z(\vb{y}) = \left( 2\pi T/k \right)^{-d/2}  \exp(-ky^2/2T) \, .
    \label{eq:eq_dist_y}
\end{equation}
The effect of the external perturbation only appears at the next perturbative order as
\begin{align}
    \partial_t P_1\t(\vb{y},t) = \cor{L}_0 P_1\t(\vb{y},t) + \cor{L}_z(t) P_1\z(\vb{y},t) \, ,
    \label{eq:master_PS_2}
\end{align}
with $\cor{L}_0$ and $\cor{L}_z$ given in \cref{eq:L_0,eq:L_z}, respectively. The Green function of the operator $\cor{L}_\T{OU} \equiv \partial_t- \cor{L}_0$ is simply the Ornstein-Uhlenbeck propagator in \cref{eq:gaussian_prop}, henceforth denoted as $P_{1|1}\z$, so that the solution of \cref{eq:master_PS_2} after an initial transient will read
\begin{equation}
    P_1\t(\vb{y},t) = \int \dd{\vb{x}} \int^t_{-\infty} \dd{t'} P_{1|1}\z(\vb{y},t|\vb{x},t') f_s(\vb{x},t') \, ,
    \label{eq:convolution}
\end{equation}
where we introduced the source term
\begin{equation}
    f_s(\vb{y},t) \equiv \cor{L}_z(t) P_1\z(\vb{y},t) \, .
\end{equation}
Using the definition of $\cor{L}_z(t)$ given in \cref{eq:L_z} and integrating by parts, it is straightforward to check that
\begin{equation}
    \int \dd{\vb{y}} P_1\t(\vb{y},t) = 0 \, ,
\end{equation}
which shows that the normalization condition $\int \dd{\vb{y}} P_1(\vb{y},t)=1$ is still satisfied.

Since the operator $\cor{L}_\T{OU} $ is time-translational invariant, then so will be its propagator $P_{1|1}\z(\vb{y},t|\vb{x},t')= P_{1|1}\z(\vb{y},\tau|\vb{x},0)$, with $\tau\equiv t-t'$. As a result, \cref{eq:convolution} takes the form of a convolution over the time domain. The integration over the spatial degrees of freedom can be readily performed by noting that
\begin{align}
    &\int \dd{\vb{x}} P_{1|1}\z(\vb{y},\tau|\vb{x},0) \grad_{\vb{x}} \left[ e^{-i\vb{q}\cdot \vb{x}} P_1\z(\vb{x}) \right] \\
    &= -\int \dd{\vb{x}}e^{-i\vb{q}\cdot \vb{x}} P_1\z(\vb{x}) \grad_{\vb{x}} P_{1|1}\z(\vb{y},\tau|\vb{x},0) \n \\
    &= e^{-\gamma \tau}  \grad_{\vb{y}} \int \dd{\vb{x}}e^{-i\vb{q}\cdot \vb{x}} P_1\z(\vb{x})  P_{1|1}\z(\vb{y},\tau|\vb{x},0) \n \\
    & = e^{-\gamma \tau}  \grad_{\vb{y}} P_1\z(\vb{y})\exp[-i\vb{q}\cdot \vb{y} e^{-\gamma \tau}-q^2\sigma^2(\tau)/2] \, , \n
\end{align}
where $\sigma(\tau)$ is given in \cref{eq:sigma_freepart} and where we used Gaussian integration in the last line. We thus find
\begin{align}
    P_1\t(\vb{y},t) =&  \grad_{\vb{y}} \cdot \Bigg[ \nu P_1\z(\vb{y}) \int \dslash{q} i \vb{q} v(\vb{q})  \label{eq:p2}\\
    &\times \int^t_{-\infty} \dd{t'} F_q\zz(t')  e^{-\gamma \tau}  e^{-i\vb{q}\cdot \vb{y} e^{-\gamma \tau}-q^2 \sigma^2(\tau)/2} \Bigg] \, ,\n
\end{align}
with $F_q\zz(t)$ and $v(\vb{q})$ defined in \cref{eq:F_q(t),eq:v(q)}, respectively.

As usual, knowing $P_1(\vb{y},t)$ allows one to compute expectation values of one-time observables $\cor{O}(\vb{Y})$ as
\begin{equation}
    \expval{\cor{O}(\vb{Y})(t)} = \int \dd{\vb{y}} \cor{O}(\vb{y}) \left[ P_1\z(\vb{y}) + \lambda^2 P_1\t(\vb{y},t) \right] +\order{\lambda^3} \, .
    \label{eq:expectation_value}
\end{equation}
Instead of applying $\grad_{\vb{y}}$ to the r.h.s. of \cref{eq:p2}, we notice that we can simply integrate by parts in $\dd{\vb{y}}$ and trade it for $\grad_{\vb{y}}\cor{O}(\vb{y})$ in \cref{eq:expectation_value}, which is generally simpler. For instance, the expectation value of the position $\cor{O}(\vb{y})=\vb{y}$ will be given by
\begin{align}
    \expval{\vb{Y}(t)} =& -\nu \lambda^2 \int \dslash{q} i \vb{q} v(\vb{q}) e^{-\frac{Tq^2}{2k}}  \n\\ 
    &\times \int^t_{-\infty} \dd{t'} F_q\zz(t')  e^{-\gamma (t-t')} \, ,
\end{align}
and its variance $\cor{O}(\vb{y})=y_j^2$ by
\begin{align}
    \expval{Y_j^2(t)} =& \frac{T}{k} \Big[  1 -2\nu \lambda^2 \int \dslash{q} q_j^2 v(\vb{q}) e^{-\frac{Tq^2}{2k}} \n\\
    &\times \int^t_{-\infty} \dd{t'} F_q\zz(t')  e^{-2\gamma (t-t')} \Big] \, .
\end{align}
This shows how the preliminary result in \cref{eq:p2} can be used in practical calculations for a generic choice of $F_q\zz(t)$. Below we will focus instead on the particular case of periodic driving.

\subsubsection{Periodic driving}
We have already observed that the time integral in \cref{eq:p2} is a convolution between $F_q\zz(t)$ and a non-periodic function which we will denote as
\begin{equation}
    H(\tau) \equiv \Theta(\tau) h(\tau) \equiv \Theta(\tau) e^{-\gamma \tau}  e^{-i\vb{q}\cdot \vb{y} e^{-\gamma \tau}-q^2\sigma^2(\tau)/2} \, ,
\end{equation}
the Fourier transform of which reads
\begin{equation}
    \tilde{H}(\omega) \equiv \int_{-\infty}^\infty \dd{\tau} e^{-i\omega \tau} h(\tau) \Theta(\tau) =  \int_0^\infty \dd{\tau} e^{-i\omega \tau} h(\tau) \, .
\end{equation}
Now let us choose $F_q\zz(t)$ to be periodic with period $T=2\pi / \Omega$, so that we can expand it in Fourier series as
\begin{equation}
    F_q\zz(t) = \sum_{n\smallin \mathbb{Z}} a_n(\vb{q}) e^{i n \Omega t} \, ,
    \label{eq:forcing_expanded}
\end{equation}
where the values of the coefficients $a_n(\vb{q})$ depend on the specific form of the external forcing applied to the particle $\vb{Z}(t)$ in \cref{eq:F_q(t)} (further below we will focus on the specific case of monochromatic forcing). The Fourier transform of $F_q\zz(t)$ will then read
\begin{equation}
    F(\omega) = \sum_{n \smallin \mathbb{Z}} a_n(\vb{q}) \delta(\omega -n\Omega) \, ,
\end{equation}
and we can transform the convolution in \cref{eq:p2} into a product in Fourier space. This gives
\begin{widetext}
\begin{align}
    P_1\t(\vb{y},t) &= \grad_{\vb{y}} \cdot \Bigg[ \nu P_1\z(\vb{y}) \int \dslash{q} i \vb{q} v(\vb{q})  \int^\infty_{-\infty} \dd{t'} F_q\zz(t')  H(t-t') \Bigg] \n \\
    &=  \grad_{\vb{y}} \cdot \Bigg[ \nu P_1\z(\vb{y}) \int \dslash{q} i \vb{q} v(\vb{q}) \int \frac{\dd{\omega}}{2\pi} \sum_{n} a_n(\vb{q})\delta(\omega -n\Omega) \tilde{H}(\omega) e^{i\omega t} \Bigg] \n\\
    &= \sum_{n\smallin \mathbb{Z}} \left[ \grad_{\vb{y}} \cdot \nu P_1\z(\vb{y}) \int \dslash{q} i \vb{q} v(\vb{q})  a_n(\vb{q})  \tilde{H}(n\Omega) \right] e^{i n\Omega t} \, ,
\end{align}
\end{widetext}
where in $\tilde{H}(\omega)$ we understand a further dependence on $\vb{q}$ and $\vb{y}$. One can also obtain an expression for the moment generating function by using Gaussian integration,
\begin{equation}
    \expval{e^{-i\vb{p}\cdot \vb{Y}(t)}} = e^{-\frac{Tp^2}{2k}} \left[ 1-\nu \lambda^2 \sum_{n} C_n(\vb{p}) e^{i n\Omega t} \right] +\order{\lambda^4}\, ,
\end{equation}
where we introduced
\begin{equation}
    C_n(\vb{p}) \equiv \int \dslash{q}  e^{-\frac{Tq^2}{2k}} v(\vb{q}) a_n(\vb{q}) A_n(\vb{p}\cdot \vb{q}) \, , 
\end{equation}
and where the function $A_n(\vb{p}\cdot \vb{q})$ was given in \cref{eq:A_n}. We can use the moment generating function to compute the mean displacement of the colloid,
\begin{align}
    &\expval{\vb{Y}(t)} = i \eval{\grad_{\vb{p}} \expval{e^{-i\vb{p}\cdot \vb{Y}(t)}}}_{\vb{p}=0} \\
    &\simeq -\nu \lambda^2 \sum_{n\smallin\mathbb{Z}} \left[\int \dslash{q} i \vb{q} e^{-\frac{Tq^2}{2k}} v(\vb{q}) \frac{a_n(\vb{q})}{\gamma + i n \Omega} \right]e^{i n\Omega t} \, .\n
\end{align}
Connected correlations can be obtained from the cumulant generating function, which reads, up to the first nontrivial order in $\lambda$,
\begin{align}
    \ln \expval{e^{-i\vb{p}\cdot \vb{Y}(t)}} \simeq -\frac{Tp^2}{2k}
    -\nu \lambda^2 \sum_{n\smallin \mathbb{Z}} C_n(\vb{p}) e^{i n\Omega t} \, .
\end{align}
For instance, the variance can then be retrieved as
\begin{align}
    &\expval{Y_j^2(t)}_c = -\eval{\pdv[2]{p_j} \ln \expval{e^{-i\vb{p}\cdot \vb{Y}(t)}}}_{\vb{p}=0}  \\
    &\simeq\frac{T}{k}\lgraf 1 -\nu \lambda^2 \sum_{n\smallin\mathbb{Z}} \left[\int \dslash{q} q_j^2  e^{-\frac{Tq^2}{2k}}  \frac{v(\vb{q}) a_n(\vb{q})}{2\gamma + i n \Omega} \right]e^{i n\Omega t} \rgraf \, . \n
\end{align}

\subsubsection{Monochromatic forcing}
Motivated by the setting described in \cref{par:Model}, we consider here a sinusoidal forcing term $\vb{Z}_F(t)$ as in \cref{eq:forcing}. In order to calculate explicitly the various quantities discussed above, we need to determine the coefficients of the Fourier series of the function $F_q\zz(t)$, \ie,
\begin{equation}
    a_n(\vb{q}) \equiv \ps{e^{i n \Omega t}}{F_q\zz(t)} \, ,
    \label{eq:a_n(q)}
\end{equation}
where we introduced the scalar product
\begin{equation}
    \ps{f(t)}{g(t)} = \frac{\Omega}{2\pi} \int_0^{2\pi/\Omega} \dd{t} 
    f^*(t) g(t) \, .
\end{equation}
Recall the definition of $F_q\zz(t)$ in \cref{eq:F_q(t)}, where the expectation value of $\expval*{ \exp[i\vb{q}\cdot \vb{Z}(t)]}_0$ was computed in Appendix \ref{par:freepart} and is given in \cref{eq:expZ} for the case of a sinusoidal forcing. Using the properties of the Bessel functions of the first kind $J_n(z)$ \cite{table}, one can prove the relation
\begin{equation}
    \ps{e^{i n \Omega t}}{e^{i z \sin(\Omega (t-u))}} = e^{-i n \Omega u}J_n(z) \, ,
    \label{eq:fourier1}
\end{equation}
so that the Fourier coefficients in \cref{eq:a_n(q)} take the form
\begin{equation}
    a_n(\vb{q}) = Dq^\alpha  \frac{ J_n(\vb{q}\cdot \vb{A}) }{\alpha_q+i n \Omega}  e^{i\vb{q}\cdot \Delta}\exp(-\frac{Tq^2}{2k_z}-i n \theta_z) \, .
\end{equation}
For $n=0$ the coefficient does not depend on the dynamics of the field ($\alpha=0$ or $\alpha=2$), and one recovers the adiabatic mean value in \cref{eq:meanvalue}.
Notice also that, in the deterministic limit $k_z\rightarrow \infty$, one has $\exp[-Tq^2/(2k_z)-i n \theta_z] \rightarrow 1$.

\section{Effective field picture}
\label{par:effectivefield}
In this Appendix we analyze the dynamics of the colloid $\vb{Y}$ as if it were immersed into the effective field
\begin{equation}
    \phi_q^\T{eff} (t) = \int_{-\infty}^{t} \dd{s} G_q(t-s) \left[ \lambda D q^\alpha V_q\zz e^{-i\vb{q}\cdot \vb{Z}(s)} + \zeta_q(s) \right] \, .
    \label{eq:phi_eff}
\end{equation}
In this expression the second colloid $\vb{Z}(t)$ is treated as a source, on the same footing as the noise $\zeta_q(t)$.
We take for simplicity the deterministic limit $k_z \rightarrow \infty$ for the motion of the second colloid, so that it appears clearly that the field in \cref{eq:phi_eff} is Gaussian with mean value
\begin{align}
    \expval*{\phi_q^\T{eff}(t)} &= \lambda  V_q\zz \int_{0}^{\infty} \dd{u} \chi_q(u) e^{-i\vb{q}\cdot \vb{Z}_F(t-u)} \n\\
    &= \lambda  V_q\zz F_{-q}\zz(t) \, ,
\end{align}
and (connected) correlations which are analogous to those of the free-field (see Appendix \ref{par:freefield}). The function $F_{-q}\zz(t)$ was defined in \cref{eq:F_q(t)}. Plugging this expression for the average field $\expval*{\phi_q^\T{eff}(t)}$ into the Langevin equation \eqref{eq:particle_eom} for the colloid $\vb{Y}$, we get
\begin{align}
    \dot{\vb{Y}}(t)=& -\gamma_y \vb{Y}(t)  + \bm{\xi}^{(y)}(t) \n\\
    &+ \nu_y \lambda \int \dslash{q} i \vb{q}  V_{-q}\y \expval*{\phi^\T{eff}_q(t)} e^{i\vb{q}\cdot \vb{Y}(t)}\, .
    \label{eq:particle_effective}
\end{align}
Notice that we are treating the field $\phi^\T{eff}$ as if it were independent of the variable $\vb{Y}(t)$, and that by using $\expval*{\phi^\T{eff}_q(t)}$ in place of $\phi^\T{eff}_q(t)$ we are practically ignoring its thermal fluctuations. It is however rather straightforward (see, \eg, Ref.~\cite{risken}) to show that \cref{eq:particle_effective} is equivalent to the Fokker-Planck equation \eqref{eq:master_PS} satisfied by the colloid up to $\order{\lambda^2}$ in the periodic state. That the thermal fluctuations of the field do not enter at all the Fokker-Planck equation (up to and including $\order{\lambda^2}$) may look surprising at first sight. However, this is actually consistent with the fact that such fluctuations do not modify the equilibrium distribution of the colloid in the absence of any external forcing (see discussion in \cref{par:machefortuna}). Indeed, the field does not know that the particle $\vb{Y}$ is not in equilibrium, being its displacement already of $\order{\lambda^2}$: any feedback effect would only appear at higher perturbative orders in the coupling constant.

\section{Upper bound on the value of $\lambda$}
\label{par:upperbound}
In Section \ref{par:PS} we derived an expression for the variance of the particle position, \cref{eq:dyn_var}, which takes the form
\begin{equation}
    \expval{Y_j^2(t)}_c =\frac{T}{k_y}\left( 1 -\lambda^2 \cor{A} \right)
    \label{eq:variance_bound}
\end{equation}
upon calling
\begin{align}
    \cor{A} \equiv & \sum_{n\smallin\mathbb{Z}} \frac{\nu D }{2\gamma_y + i n \Omega}e^{i n(\Omega t-\theta_z)}\\
    &\times \left[\int \dslash{q}  \frac{q_j^2 q^\alpha  v(\vb{q})J_n(\vb{q}\cdot \vb{A})}{\alpha_q+ i n \Omega}e^{-\frac{Tq^2}{2k_p}+i\vb{q}\cdot \bm{\Delta}} \right] \, . \n
\end{align}
Up to this order in $\lambda$, a necessary condition for the variance to be positive is $\lambda^2 \cor{A} \leq 1$. Calling $g(\vb{q}) \equiv q_j^2  v(\vb{q}) \exp[-Tq^2/(2k_p)]$, it is simple to derive an upper bound for
\begin{align}
    |\cor{A} | & \leq \sum_{n\smallin\mathbb{Z}} \int \dslash{q}  \frac{D q^\alpha g(\vb{q}) J_n(\vb{q}\cdot \vb{A})}{\sqrt{\alpha_q^2+ ( n \Omega)^2} \sqrt{(2\gamma_y)^2+ ( n \Omega)^2}}  \n \\
    & \leq \int \dslash{q}  \frac{D q^\alpha g(\vb{q})}{2\gamma_y \alpha_q } \sum_{n\smallin\mathbb{Z}} J_n(\vb{q}\cdot \vb{A}) \n\\
    &= \frac{1}{2\gamma_y} \int \dslash{q} \frac{g(\vb{q})}{q^2+r} \, , \label{eq:bound1}
\end{align}
where in the second line we set $\Omega=0$, and in the third we used the identity \cite{table}
\begin{equation}
    \sum_{n=-\infty}^\infty J_n(x) = 1 \, .
\end{equation}
The last integral in \cref{eq:bound1} is a decreasing function of the parameter $r$ and it can be computed in closed form for some elementary functional forms of the interaction potentials $V_q^{(a)}$ contained in $v(\vb{q})$ (see \cref{eq:v(q)}). Choosing, for instance, Gaussian interacting potentials as in \cref{eq:gaussianpotential} and setting $r=0$ (for which the integral is maximum), we get the upper bound reported in \cref{eq:upperbound}. For $\lambda$ smaller than this upper bound it is guaranteed that the variance in \cref{eq:variance_bound} is positive, a necessary condition for the perturbative expansion to provide meaningful results.

\section{Equilibrium effective potential}
\label{par:effectivepotential}
In this Appendix we study the effective induced interaction between the two particles due to the presence of the field.

\subsection{Derivation of the potential}
Let us start by considering the joint probability distribution of the two colloids, which at equilibrium is the canonical one given in \cref{eq:canonical}. Under the functional integral we recognize, up to a normalization factor, the stationary distribution of the field at fixed colloids positions
\begin{align}
        &\cor{P}_\T{st}\left[\phi| \vb{Y},\vb{Z}\right] = \frac{1}{\cor{Z}_\T{st}(\vb{Y},\vb{Z})} e^{-\beta \left( \cor{H}_\phi - \lambda \cor{H}_\T{int} \right) } \\
        &= \frac{1}{\cor{Z}_\T{st}} e^{-\beta \int \dslash{q}  \lgraf \frac{1}{2}(q^2+r) \phi_q\phi_{-q}-\lambda \phi_q \left[ V_{-q}\y e^{i\vb{q}\cdot\vb{Y}} + V_{-q}\zz e^{i\vb{q}\cdot\vb{Z}} \right] \rgraf} \, .\n
\end{align}
The coupling to the field is linear, so the Gaussian functional integral over $\cor{P}_\T{st}\left[\phi| \vb{Y},\vb{Z}\right]$ in \cref{eq:canonical} can be calculated exactly. To this end, we first bring it in the form
\begin{align}
        &\int \cor{D}\phi \, e^{-\beta \left( \cor{H}_\phi - \lambda \cor{H}_\T{int} \right)  } = \int \cor{D}\phi \, e^{-\frac{\beta}{2} \left(\phi,\hat{A}\phi \right) +\beta \lambda \left(h\y +h\zz,\, \phi\right) } \n \\
        &\propto  e^{-\frac{\beta \lambda^2}{2}\left( h\y +h\zz ,\, \hat{A}^{-1} (h\y +h\zz ) \right) } \, ,
\end{align}
where we introduced the vectors $h^{(a)} (\vb{x}) \equiv V\a(\vb{x}-\vb{X}_a)$ and the scalar product
\begin{equation}
    \ps{f}{g} = \int \dd[d]{x} f(\vb{x})g(\vb{x}) \, .
\end{equation}
The operator $\hat{A}$ is defined by its kernel
\begin{align}
    A(\vb{x},\vb{y}) &= \left( -\laplacian +r \right) \delta (\vb{x}-\vb{y}) \, , \\
    \hat{A}\phi(\vb{x}) &= \int  \dd[d]{y} A(\vb{x},\vb{y}) \phi(\vb{y}) \, .
\end{align}
In Fourier space, these become  $\tilde{h}\a(\vb{q}) = V\a_q \exp[-i \vb{q}\cdot \vb{X}_a(t)]$ and
\begin{equation}
    \tilde{A}(\vb{q},\vb{p}) = \left( q^2 +r \right) \delta (\vb{q}+\vb{p}) \; \rightarrow \; \tilde{A}^{-1}(\vb{q},\vb{p}) = \frac{\delta (\vb{q}+\vb{p})}{ q^2 +r } \, .
\end{equation}
Integrating over the dummy variables (momenta) as
\begin{equation}
    \ps{f}{\hat{A}g} = \int \frac{\dd[d]{q}\dd[d]{p}}{(2\pi)^{2d}} \tilde{f}(-\vb{q}) \tilde{A}(\vb{q},-\vb{p}) \tilde{g}(\vb{p}) \, ,
\end{equation}
we finally get the effective Hamiltonian $\cor{H}_\T{eff}(\vb{Y},\vb{Z})$ given in \cref{eq:effectiveH}, featuring the field-induced interaction potential $V_c(\vb{Y}-\vb{Z})$ of \cref{eq:inducedpotential} (up to a constant that we fix by requiring $V_c(\vb{x}\to \infty)= 0$). We notice that $V_c(\vb{x})$ is translational invariant, as expected, so that the induced force is given by $ \vb{F}_c(\vb{x}) = - \lambda^2 \grad_{\vb{x}} V_c(\vb{x})$. The latter is in general a non-monotonic function of $\vb{x}$, and the location of its extremal points $\vb{x}=\vb{x}_*$ along the various spatial directions is found by inspecting the Hessian matrix
\begin{equation}
    0 \equiv \eval{\pdv{ V_c(\vb{x})}{x_i}{x_j} }_{\vb{x}=\vb{x}_{*}} = \int \dslash{q} \frac{v(\vb{q})}{q^2+r} q_i q_j e^{i \vb{q}\cdot \vb{x}_{*}}  \, .
    \label{eq:minimum_condition_d}
\end{equation}

\subsection{Analysis of the induced potential for the isotropic case}
If the interaction potentials of the two colloids are equal, \ie, $V\y_q = V\zz_q \equiv V_q$, then $v(\vb{q}) \equiv |V_q|^2$ (see \cref{eq:v(q)}). Moreover, if $V(\vb{x})$ is isotropic (which is a sensible requirement if the colloids are assumed to be spherically symmetric particles), then $v(\vb{q})=v(q)$ and we can rewrite \cref{eq:inducedpotential} in polar coordinates as
\begin{equation}
    V_c(x) = -\int_0^\infty \dd{q} \frac{q^{d-1}}{q^2+r} v(q) \int \frac{\dd{\Omega_d}}{(2\pi)^d} e^{i \vb{q}\cdot \vb{x}} \, .
\end{equation}
Using the property of the Bessel functions \cite{table}, one can prove that
\begin{equation}
    \int \frac{\dd{\Omega_d}}{(2\pi)^d} e^{i \vb{q}\cdot \vb{x}} = \frac{J_{d/2-1}(qx) }{(2\pi)^{d/2} (qx)^{d/2-1} } \, ,
\end{equation}
and introducing the dimensionless variable $z\equiv qx$ we find
\begin{equation}
    V_c(x) = - \frac{x^{2-d}}{(2\pi)^{d/2}} \int_0^\infty \dd{z} \frac{z^{d/2}}{z^2+rx^2} v\left(z/x \right)J_{d/2-1}(z) \, .
\end{equation}
If we assume for $V_q$ a Gaussian form as in \cref{eq:gaussianpotential}, then this expression becomes
\begin{equation}
    V_c(x) = R^{2-d} f(x/\xi,x/R) \, , \label{eq:potential_scaling}
\end{equation}
where the scaling function
\begin{equation}
    f(\Theta,\Lambda) \equiv - \frac{\Lambda^{2-d}}{(2\pi)^{d/2}} \int_0^\infty \dd{z} \frac{z^{d/2}e^{-\left( z/\Lambda\right)^2}}{z^2+\Theta^2}  J_{d/2-1}(z)
\end{equation}
depends on the dimensionless parameters $\Theta = x\sqrt{r}= x/\xi$ and $\Lambda=x/R$ (in accordance with the scaling form in Eq.~(1) of \ccite{dietrich98}). Note that $\Theta$ and $\Lambda$ actually play the role of an IR and a UV cutoff respectively.

Similarly, the resulting induced force is
\begin{equation}
    \vb{F}_c(x) = -\lambda^2 \grad_{\vb{x}} V_c(x) = -\vu{x} \lambda^2 R^{1-d} f'(\frac{x}{\xi},\frac{x}{R})  \, , \label{eq:force_scaling}
\end{equation}
with
\begin{equation}
    f'(\Theta,\Lambda) \equiv   \frac{\Lambda^{1-d}}{(2\pi)^{d/2}} \int_0^\infty \dd{z} \frac{z^{d/2+1} e^{-\left(z/\Lambda\right)^2}}{z^2+\Theta^2}  J_{d/2}(z) \, .
\end{equation}
We now look for the asymptotic behavior for large $x$ of the induced potential in \cref{eq:inducedpotential} with the Gaussian interaction potential $v(q) = \exp(-q^2R^2)$. This can be obtained by first using the identity \cite{LeBellac}
\begin{equation}
    \frac{1}{q^2+r}= \int_0^\infty \dd{\mu} e^{-\mu(q^2+r)} \, ,
    \label{eq:feynman}
\end{equation}
and then performing the Gaussian integration in $\dd[d]{q}$. Changing variables to $s\equiv (\mu+R^2)/x$ gives \footnote{See Ref.~\cite{bender}, \textit{Laplace's method for integrals with movable maxima}.}
\begin{equation}
    V_c(x) = -\frac{x^{1-d/2}}{(4\pi )^{d/2}} e^{R^2 r} \int_{R^2/x}^\infty \frac{\dd{s}}{s^{d/2}} e^{-x \left[ r s + 1/(4s) \right]} \, .
\end{equation}
Finally, the integral over $s$ can be estimated for large $x$ by using the Laplace method, leading to
\begin{equation}
    V_c(x) \sim  -\frac{(2\pi x)^{(1-d)/2}}{2 r^{(3-d)/4}} e^{R^2 r-x\sqrt{r}}  \, ,
    \label{eq:potential_asymptotic}
\end{equation}
which presents the familiar exponential tails $\sim \exp(-x/\xi)$, being $\xi = r^{-1/2}$ the field correlation length. One can check that a similar asymptotic behavior is shared by the induced force, since for large $x$ one finds $\vb{F}_c(x) \sim \vu{x}\lambda^2 \sqrt{r} V_c(x)$.

In $d=1$, the expressions above become
\begin{align}
    V_c(x) &= -\frac{x}{\pi} \int_0^\infty \dd{z} \frac{\cos(z)e^{-z^2\left( R/x\right)^2}}{z^2+rx^2} \, ,
    \label{eq:potential1d} \\
    F_c(x) &= -  \frac{\lambda^2}{\pi} \int_0^\infty \dd{z} \frac{z \sin(z) e^{-z^2\left( R/x\right)^2}}{z^2+rx^2} \, ,
    \label{eq:casimirforce1d}
\end{align}
which are plotted in \cref{fig:potential} (rescaled by the $R$-dependent part of their asymptotic amplitude found in \cref{eq:potential_asymptotic}). It appears that the induced force is small for both small and large $x$, while it presents a maximum defined by the condition
\begin{equation}
     \eval{\partial_x F_c(x)}_{x_\T{max}} \propto \int_{-\infty}^\infty \dd{q} \frac{q^2 e^{-q^2 R^2}}{q^2+r}  e^{i q x_\T{max} } \equiv 0 \, .
    \label{eq:max_condition}
\end{equation}
Notice that the induced potential in \cref{eq:potential1d} diverges for $r=0$, but the force in \cref{eq:casimirforce1d} does not. Equivalently, the induced potential in \cref{eq:potential1d} is regularized by subtracting its value in $x=0$,
\begin{equation}
    V_c(x=0) = -\frac{1}{\pi} \int_0^\infty \dd{q} \frac{v(q)}{q^2+r} \, ,
\end{equation}
which is just a constant shift in energy. However, the induced force $F_c(x)$ in $d=1$ and for $r=0$ is still somewhat pathological, in that it saturates to a constant value at large distance $x$ instead of decaying to zero. To understand why, we note that at large distances $x$ the cutoff $R$ in the induced potential $V_c(\vb{x})$ is expected to play no role (apart from taming possible UV divergences which can arise for sufficiently large $d$). If we set $R\simeq 0$ in \cref{eq:inducedpotential}, we obtain 
\begin{equation}
    V_c(\vb{x}) \simeq -\int \dslash{q} \frac{1}{q^2+r}  e^{i \vb{q}\cdot \vb{x}} = - \expval*{\phi(\vb{x})\phi(0)} \, ,
\end{equation}
where we recognized the two-point correlation function of a scalar Gaussian field in $d$ spatial dimensions \cite{LeBellac}. At the critical point $r=0$, this behaves generically as
\begin{equation}
    \expval*{\phi(\vb{x})\phi(0)} \sim |\vb{x}|^{2-d} \, ,
\end{equation}
and in particular in $d=1$ it grows linearly with $x$. This explains why the force $F_c(x)\propto \partial_x V_c(x)$ saturates to a constant value for large $x$. However, we will simply interpret this phenomenon as a pathology of the model for $d=1$ and $r=0$, which does not affect our results since we always assume the field to have a (possibly small but) finite correlation length $\xi=r^{-1/2}$.

Different choices of the interaction potential $V(x)$ lead to qualitatively similar results. For instance, a more realistic representation of a spherical colloid requires
\begin{equation}
    V(\vb{x}) = \frac{1}{V_d} \Theta(R-|\vb{x}|) \, ,
\end{equation}
where $V_d$ is the volume of a $d$-dimensional sphere and $\Theta(z)$ is the Heaviside distribution. Its Fourier transform reads
\begin{equation}
    V_q = \left( \frac{2}{qR} \right)^{d/2} \Gamma \left(\frac{d}{2}+1\right) J_{d/2}(qR) \, .
\end{equation}
This leads to the same scaling forms as in \cref{eq:potential_scaling,eq:force_scaling} for the induced potential and force, with different scaling functions
\begin{align}
    f_2(\Theta,\Lambda) &\equiv -\Lambda^2 c_d  \int_0^\infty \dd{z} \frac{z^{-d/2} J_{d/2-1}(z) }{z^2+\Theta^2} \left[ J_{d/2}\left(\frac{z}{\Lambda}\right) \right]^2 \, , \n \\
    f'_2(\Theta,\Lambda) &\equiv \Lambda c_d  \int_0^\infty \dd{z} \frac{z^{1-d/2} J_{d/2}(z) }{z^2+\Theta^2} \left[ J_{d/2}\left(\frac{z}{\Lambda}\right) \right]^2 \, , \n  \\
    c_d &\equiv \left(2/\pi\right)^{d/2}  \left[\Gamma \left(d/2+1 \right)\right]^2 \, ,
\end{align}
which are qualitatively similar to the Gaussian case shown in \cref{fig:potential}. In particular, the induced force still presents a maximum as a function of the distance $x$, which can give rise to the phenomenon of frequency doubling in the adiabatic response (see \cref{par:adiabatic_amplitude,par:freq_doubling}).

\section{Particle dynamics within the adiabatic approximation}
\label{par:adiabatic_dynamics}
In this Appendix we derive the colloid dynamics at lowest order within the adiabatic approximation. This is achieved by averaging the equations of motion \eqref{eq:particle_eom} and \eqref{eq:particleZ} of $\vb{Y}(t)$ and $\vb{Z}(t)$, respectively, over the stationary distribution $\cor{P}_\T{st}\left[\phi| \vb{Y},\vb{Z}\right]$ of the field $\phi$ at fixed colloids positions given in \cref{eq:stationarydist}. This is analogous to the Born-Oppenheimer approximation in condensed matter physics, where the wavefunction of the electrons orbiting around a nucleus is obtained by exploiting the separation of their dynamical timescales.

\subsection{Derivation of the Langevin equation within the adiabatic approximation}
Let us focus on the motion of $\vb{Y}(t)$, the colloid in the fixed trap, which is ruled by \cref{eq:particle_eom}. We average each of the terms which appear in \cref{eq:particle_eom} over the stationary distribution in \cref{eq:stationarydist}. The terms proportional to $\vb{Y}(t)$ and $\dot{\vb{Y}}(t)$ yield trivially
\begin{equation}
    \expval{\vb{Y}(t)}_\T{st} = \int \cor{D}\phi\, \vb{Y}(t) \cor{P}_\T{st}\left[\phi| \vb{Y},\vb{Z}\right] = \vb{Y}(t) \, ,
\end{equation}
and similarly for $\dot{\vb{Y}}(t)$, while
\begin{align}
    \expval{\lambda \vb{f}_y}_\T{st} &=  \lambda \expval{ \nabla_y \cor{H}_\T{int}}_\T{st} \n\\
    &= \frac{\lambda}{\cor{Z}_\T{st}} \int \cor{D}\phi\, \nabla_y \cor{H}_\T{int} e^{-\beta \left( \cor{H}_\phi - \lambda \cor{H}_\T{int} \right) } \n\\
    &= \frac{1}{\beta} \nabla_y \log \cor{Z}_\T{st} = \lambda^2 \nabla_y V_c(\vb{Y},\vb{Z}) \, ,
\end{align}
where in the last passage we used \cref{eq:Vc_partitionfunction}. This leads to the effective Langevin equation \eqref{eq:adiabaticlangevin}. 

Now we look for a perturbative solution of \cref{eq:adiabaticlangevin} which is valid up to $\order{\lambda^2}$, and which we will denote as $\vb{Y}_\T{ad}(t)$. To this end, we average each of its terms over the noises $\bm{\xi}\y(t)$ and $\bm{\xi}\zz(t)$ by bearing in mind that
\begin{align}
    &\expval{e^{i \vb{q}\cdot \left( \vb{Z}(t) -\vb{Y}(t)\right)}} = \expval{e^{i \vb{q}\cdot \left( \vb{Z}(t) -\vb{Y}(t)\right)}}_0 +\order{\lambda} \n \\
    &=\expval{e^{i\vb{q}\cdot \vb{Z}(t)}}_0 \expval{e^{-i\vb{q}\cdot \vb{Y}(t)}}_0 +\order{\lambda} \, ,
\end{align}
where we used the fact that the two independent processes for $\vb{Y}(t)$ and $\vb{Z}(t)$ factorize. Specializing \cref{eq:exp1point} to the present case gives
\begin{equation}
    \expval{e^{-i\vb{q}\cdot \vb{Y}(t)}}_0 = \exp[-Tq^2/(2k_y)] \, ,
\end{equation} 
which leads to 
\begin{align}
    \partial_t \expval{\vb{Y}_\T{ad}}=& -\nu_y k_y \expval{\vb{Y}_\T{ad}} \n \\
    &- \nu_y \lambda^2 \int \dslash{q} \frac{i \vb{q} v(\vb{q} )}{q^2+r}  e^{-\frac{Tq^2}{2k_y}} \expval{e^{i \vb{q}\cdot \vb{Z}}}_0 \, .
\end{align}
Solving this differential equation with the initial condition $\expval{\vb{Y}_\T{ad}(t=t_0)}=0$, we finally obtain \cref{eq:y_adiabatic}.

\subsection{Adiabatic limit from the master equation}
\label{par:adiabaticlimit}
Consider the free-field susceptibility $\chi_q(t-s)$ in \cref{eq:field-susc} and assume $\alpha_q \neq 0$ (see \cref{eq:tau_phi}). One can take the formal limit $D\to \infty$, finding
\begin{equation}
    \chi_q(t-s) \xrightarrow[D \rightarrow \infty]{} \frac{\delta(t-s)}{q^2+r} \, ,
    \label{eq:adiabaticlimit}
\end{equation}
and inserting this expression into \cref{eq:dyn_avg_prior} for the average displacement of the colloid, we immediately recover its adiabatic approximation in \cref{eq:y_adiabatic}.

Conversely (and more generally), it is straightforward to check \cite{risken} that the Fokker-Planck equation corresponding to the adiabatic Langevin equation \eqref{eq:adiabaticlangevin} is exactly the master equation \eqref{eq:master_PS}. To see this, one can use the adiabatic limit in \cref{eq:adiabaticlimit} in the expression \eqref{eq:F_q(t)} for the function $F_q(t)$, which appears in the operator $\cor{L}_z(t)$ of the master equation. The key observation is then that
\begin{equation}
    F_q\zz(t) \xrightarrow[D \rightarrow \infty]{} \frac{\expval*{e^{i\vb{q}\cdot \vb{Z}(t)}}_0}{q^2+r} \, .
\end{equation}

\subsection{Frequency doubling in the adiabatic response}
\label{par:freq_doubling}
Looking at the Fourier coefficients of the adiabatic response in \cref{eq:adiabatic_reduced_coeff}, it appears that $|\vb{b}_1|=0$ in correspondence of a certain value $r_1$ of the parameter $r$. This value can be approximately found, in $d=1$, by writing
\begin{equation}
    \frac{q}{q^2+r} = \frac{1}{q} \left( 1-\frac{r}{q^2+r} \right)
\end{equation}
in the condition $|b_1|= 0$, which gives
\begin{align}
    &\int \dd{q}  e^{-q^2 \widetilde{R}^2 }   \frac{J_1(qA)}{q} \cos(q \Delta) \n\\
    &=
    r_1\int \dd{q} e^{-q^2 \widetilde{R}^2 } \frac{J_1(qA)}{q(q^2+r_1)}    \cos(q \Delta) \n \\
    &= \int \dd{y} \exp(-y^2 \widetilde{R}^2 r_1 )  \frac{J_1(yA \sqrt{r_1})}{y(y^2+1)}    \cos(y \Delta \sqrt{r_1}) \, ,
\end{align}
where we changed variable as $q=y\sqrt{r_1}$ in the last line. Assuming $r_1$ to be small, which can be verified \textit{a posteriori}, one can expand for small $r_1$, finding
\begin{equation}
    r_1= \left[ \frac{4}{\pi A} \int_0^\infty \dd{q}  e^{-q^2 \widetilde{R}^2 }  \frac{J_1(qA)}{q} \cos(q \Delta)  \right]^2 + \order{r_1^{3/2}} \, .
    \label{eq:frequencydoubling}
\end{equation}
One can check numerically that $r_1$ determined above is an increasing function of the forcing amplitude $A$. A further expansion for small $A$ gives
\begin{equation}
    r_1\simeq \frac{e^{-\Delta^2/(2\widetilde{R}^2)}}{\pi\widetilde{R}^2}+ \order{r_1^{3/2}} \, ,
\end{equation}
which is finite even for $A=0$, in agreement with the physical interpretation we proposed in Section \ref{par:adiabatic_analysis}.

\section{Linear response}
\label{par:linear_response}
Here we compute the response of the particle in the fixed trap to a perturbation of amplitude $A$ of the position of the other particle, within the linear response regime. The resulting expressions can be easily analyzed even in spatial dimensionality $d>1$.

\subsection{Adiabatic response}
The linear response approximation for the average position in the adiabatic case is formally recovered from \cref{eq:y_adiabatic} as
\begin{align}
    \expval*{\vb{Y}_\T{ad}(t)}_\T{LR} &\equiv  \eval*{ \expval*{\vb{Y}_\T{ad}(t)}}_{\vb{A}=0} + \eval{ \left(\vb{A} \cdot \grad_{\vb{A}}\right) \expval*{\vb{Y}_\T{ad}(t)}  }_{\vb{A}=0} \n\\
    &\equiv \vb{Y}_\T{stat} + \vb{A}\cdot \hat{\bm{\chi}}^\T{eq}(t) \, ,
    \label{eq:linearresponse}
\end{align}
and it is made of a static plus an oscillating part. In particular,
\begin{equation}
     \vb{Y}_\T{stat} \equiv \frac{\lambda^2}{k_y} \int \frac{\dd[d]{q}}{(2\pi)^d} \vb{q} \frac{e^{- \widetilde{R}^2 q^2}}{q^2+r} \sin(\vb{q} \cdot \bm{\Delta}) \, ,
     \label{eq:staticpart}
\end{equation}
while
\begin{align}
    \chi_{ij}^\T{eq}(t) &=\frac{\lambda^2}{k_y} (\widetilde{\cor{I}}_0)_{ij}  \frac{\sin( \Omega t -\theta_z -\theta_y)}{\sqrt{1+\left(\Omega /\gamma_y \right)^2}} \, , \label{eq:linear_adiabatic} \\
    (\widetilde{\cor{I}}_0)_{ij} &\equiv \int \frac{\dd[d]{q}}{(2\pi)^d} \frac{q_iq_j e^{- \widetilde{R}^2 q^2}}{q^2+r} \cos(\vb{q} \cdot \bm{\Delta}) \, ,
\end{align}
 and all the higher harmonics are suppressed, since they contain higher powers of $A$. Note that $\vb{Y}_\T{stat}$ coincides with the equilibrium position we obtained in \cref{eq:equilibriumposition} in the static limit. If the particle $\vb{Z}$ oscillates in a direction parallel to the separation $\bm{\Delta}$ between the two colloids, then one can focus on the component $\expval*{Y_j(t)}$ of the particle displacement parallel to $\vb{A}$ and $\bm{\Delta}$. This is controlled in linear response by
\begin{align}
    (\widetilde{\cor{I}}_0)_{jj} &= \int_0^\infty \dd{q} \frac{q^{d+1} e^{- \widetilde{R}^2 q^2}}{q^2+r} F(q \Delta)\, , \label{eq:integral_lr} \n \\
    F(z) &= \frac{z^{-d/2}}{(2\pi)^{d/2}}\left[ z J_{d/2-1}(z) + (1-d)  J_{d/2}(z) \right] \, ,
\end{align}
and this integral can be evaluated numerically. In $d=1$ and for sufficiently large $r$, the integration over momenta in \cref{eq:integral_lr} returns a negative number, while it changes sign for very small $r$. This property carries over to the nonlinear case, thus producing the frequency doubling described in \cref{par:adiabatic_analysis} and \cref{par:freq_doubling}. In $d=2$ and $d=3$ one observes a qualitatively similar behavior, with $(\cor{I}_0)_{jj}$ changing sign for some small value of $r$ depending on the choices of $\Delta$ and $\widetilde{R}=\widetilde{R}(R,T)$.

\subsection{Dynamical response}
\label{par:linear_dyn}
Starting from the expression of the average position in \cref{eq:dyn_avg} describing the actual dynamics and using \cref{eq:linearresponse}, we can derive a linear response expression
\begin{align}
    \expval*{\vb{Y}(t)}_\T{LR} = \vb{Y}_\T{stat} + \vb{A}\cdot \hat{\bm{\chi}}(t) \, ,
\end{align}
which again consists of a static part $\vb{Y}_\T{stat}$ -- which turns out to be the same as in the static case, see \cref{eq:staticpart} -- and of an oscillating part. It is easier in this case to start from the Fourier expansion in \cref{eq:coeff_dynamic} and write
\begin{align}
    \chi_{ij}(t) &=\sum_{n=\pm1}\frac{-i\lambda^2\nu_y D}{\gamma_y+in\Omega} (\tilde{I}_n)_{ij}  e^{in( \Omega t -\theta_z)} \, , \\
    (\tilde{I}_n)_{ij} &\equiv \int \frac{\dd[d]{q}}{(2\pi)^d} \frac{q_iq_jq^\alpha e^{- \widetilde{R}^2 q^2}}{\alpha_q+in\Omega} \cos(\vb{q} \cdot \bm{\Delta}) \, .
\end{align}
For $\Omega=0$, we easily get
\begin{align}
    |\chi_{ij}^{(\Omega=0)}| =\frac{\lambda^2}{k_y} \int \frac{\dd[d]{q}}{(2\pi)^d} \frac{q_iq_j e^{- \widetilde{R}^2 q^2}}{q^2+r} \cos(\vb{q} \cdot \bm{\Delta}) \, ,
\end{align}
which coincides with the one obtained from the linear response of the adiabatic case, see \cref{eq:linear_adiabatic}, in the limit of $\Omega\rightarrow 0$, and which makes contact with  the amplitude of the quasi-static response in \cref{eq:staticamp}.

\begin{figure}
    \centering
    \resizebox{\columnwidth}{!}{\includegraphics[]{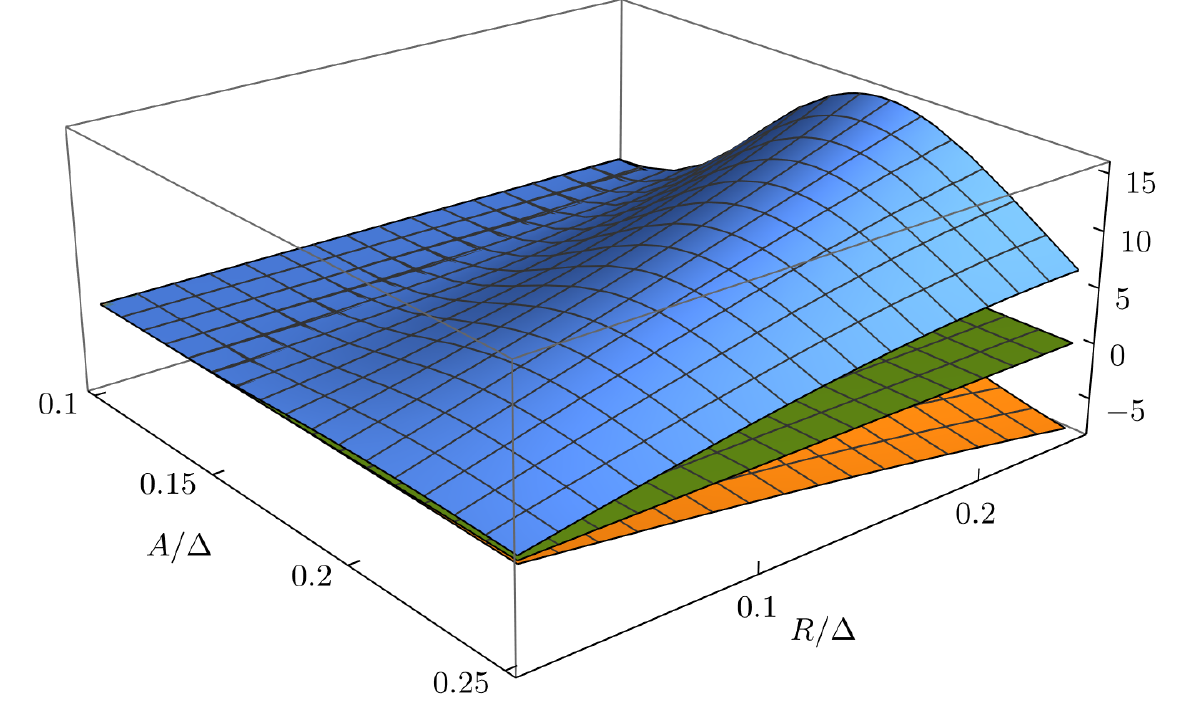}}
    \caption{Plot of the integral in \cref{eq:sign} in the case of model A (orange, below, rescaled for graphical convenience) and model B (blue, above), compared with the plane $z= 0$ (green, center) to check the zero crossings.}
    \label{fig:sign}
\end{figure}

\section{Phase of the dynamical response}
\label{app:dynamical_phase}
Here we derive some of the results concerning the phase of the dynamical response anticipated in \cref{par:dyn_phase_main}.

\subsection{Large-$\Omega$ behavior}
Let us focus first on the limit in which the frequency $\Omega$ of the external driving is large: then the Fourier coefficients in \cref{eq:coeff_dynamic} become
\begin{equation}
    \vb{c}_n e^{i n\theta_z} \sim i\frac{\lambda^2\nu_y D} {(n \Omega)^2} \int \frac{\dd[d] {q}}{(2\pi)^d} \vb{q} q^\alpha  J_n(\vb{q}\cdot \vb{A})  e^{-q^2 \widetilde{R}^2 +i\vb{q}\cdot \Delta} \, ,
    \label{eq:large_omega}
\end{equation}
where we factored out the phase $\theta_z$ of the driven colloid. As explained in the main text, this approximation works if the condition in \cref{eq:phase_asymptotic_condition} is met. Notice that the quantity on the r.h.s. of \cref{eq:large_omega} is purely imaginary, which means that for large $\Omega$ the dynamical response of $\vb{Y}$ is either in phase or in counterphase with the motion of the colloid $\vb{Z}(t)$ (see \cref{eq:forced_phase}). To determine its sign, one has to evaluate the integral in \cref{eq:large_omega}. In $d=1$ and focusing on the first harmonic $n=1$, we can rescale $z\equiv q\Delta$ and write
\begin{equation}
    c_1 e^{i \theta_z} \propto \frac{i}{\Omega^2} \int_0^\infty \dd{z} z^{\alpha+1} \cos(z) J_1(z\beta_1)  e^{-(z\beta_2)^2} \, ,
    \label{eq:sign}
\end{equation}
where we called $\beta_1\equiv \widetilde{R}/\Delta$ and $\beta_2 \equiv A/\Delta$ the small parameters of our problem (see setup in \cref{fig:setup}). Figure~\ref{fig:sign} shows that, for $\beta_1$ and $\beta_2\ll 1$, this integral is positive for model B ($\alpha=2$) and negative for model A ($\alpha=0$). This corresponds to the behavior of the phase observed in \cref{fig:phase}. Although the sign may change in $d>1$, one would in any case observe a $\pi/2$ phase shift with respect to the adiabatic prediction at large $\Omega$ (see \cref{fig:adiabatic_omega}).

\subsection{Dependence of the phase $\varphi_1$ on $\Delta$}
\label{app:slope}
\begin{figure}
    \centering
    \resizebox{\columnwidth}{!}{\includegraphics[]{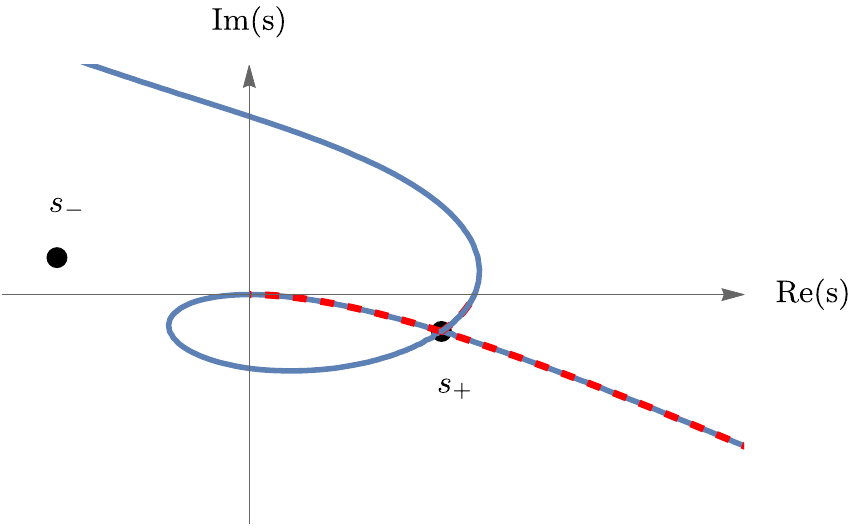}}
    \caption{Integration contour for the function in \cref{eq:saddle_integral} analytically continued to the complex plane $s$. The black dots indicate the stationary points $s_\pm$ of $f(s)$, and we plotted in solid blue the contour lines of $v(s)=\Im[f(s)]$ passing through $s_+$. We deform the original integration contour, \ie, the positive real axis, to a portion of the curve above indicated by the red-dashed line. Indeed, the integrand in \cref{eq:saddle_integral} vanishes for large $|s|$ in the region $\{\Re(s)>0 \, , \, \Im(s)<0\}$, so the integration contour can be closed at infinity and the Cauchy theorem applies. In this plot we set all the parameters in $f(s)$ to unity, for the sake of illustration.}
    \label{fig:contour}
\end{figure}

Let us now study the dependence of the phase $\varphi_1$ of the dynamical response defined in \cref{eq:varphi1} on the average separation $\Delta$ between the two traps. To this end, we will consider the case of model A ($\alpha=0$) and examine the behavior for large $\Delta$ of the integral $I_1$ which appears in \cref{eq:varphi1_integral}. Focusing on the component $j$ parallel to $\vb{A}$ and $\bm{\Delta}$ we notice that \cref{eq:varphi1_integral} can be rewritten, up to first order in the driving amplitude $A$, as
\begin{equation}
    I_1 = -\frac{A}{2 D R^{d+4}}\pdv[2]{x} \int \frac{\dd[d] {y}}{(2\pi)^d}  \frac{e^{-y^2 +i y_j x}}{y^2 + R^2r+i \Omega/\Omega_0}   \, ,
\end{equation}
where we used $J_1(x)\simeq x/2$ and we rescaled momenta as $q=y/R$. One can check \textit{a posteriori} that including higher orders in $A$ will not change our conclusions as long as $A\ll \Delta$. We also defined the quantities $x\equiv \Delta/R$ and $\Omega_0\equiv D/R^2$, which we recognize from \cref{eq:tau_phi} as the inverse timescale of relaxation of the field $\phi$ over a length scale $\sim R$. The integral in $I_1$ finally contains a further dependence on $R^2r=(R/\xi)^2$. Using again \cref{eq:feynman} and computing the Gaussian integral we find
\begin{equation}
    I_1 = -\frac{A}{2 D R^{d+4}}\pdv[2]{x} \int_0^\infty \dd{\mu} \frac{ e^{-\mu c - x^2/4(1+\mu)}  }{[4\pi(1+\mu)]^{d/2}}   \, ,
\end{equation}
where we defined $c\equiv R^2r+i \Omega/\Omega_0$. In order to avoid the trivial saddle-point $\mu=\infty$, we change variables as $\mu = s x$, leading to 
\begin{align}
    I_1 &= -\frac{A}{2 D R^{d+4}}\pdv[2]{x} \left[ \frac{x^{1-d/2}}{(4\pi)^{d/2}} Q(x) \right] \, , \label{eq:Q(x)} \\
    Q(x) &= \int_0^\infty \dd{s} g(s) e^{-x f(s)} \, , \label{eq:saddle_integral}
\end{align}
with
\begin{align}
    f(s) &\equiv s c - \frac{1}{4 s} \, ,\\
    g(s) &\equiv \left(s+1/x \right)^{-d/2} \exp[ \frac{x}{4s(sx+1)} ] \, . \label{eq:g(s)}
\end{align}
Since the function $g(s)$ is regular and $x$-independent for large $x$, the integral in \cref{eq:saddle_integral} can be estimated using the method of steepest descent \cite{bender}. To this end, one considers the analytic continuation in the complex plane $s=a+ib$ of the function $f(s)=u(a,b)+i v(a,b)$, and then deforms the original integration path (\ie, the positive real axis) to a level curve of $v(a,b)$ passing through a stationary point of $u(a,b)$. By the Cauchy-Riemann conditions, these stationary points coincide with the extrema of the function $f(s)$, given by $s_\pm = \pm 1/(2\sqrt{c})$. The relevant integration contour is shown in \cref{fig:contour}, and it passes through the saddle-point $s_+$. By standard methods, one then finds
\begin{equation}
    Q(x) \simeq \sqrt{\frac{2 \pi}{x|f''(s_+)|}} g(s_+) e^{-x f(s_+) -i3 \theta/2} \, ,
    \label{eq:saddled}
\end{equation}
where $f(s_+)=\sqrt{c}\equiv \rho \exp(i \theta)$ and we introduced
\begin{align}
    \rho &= R \left[ r^2 +(\Omega/D)^2 \right]^{1/4} \, , \\
    \theta &= \frac12 \arctan(\frac{\Omega}{Dr}) \, . 
\end{align}
We also see that $|f''(s_+)|=1/(2\rho^6)$, while
\begin{equation}
    g(s_+) = \left(s+1/x \right)^{-d/2} \exp[ \frac{x}{4s(sx+1)} ] \, .
    \label{eq:g(s+)}
\end{equation}
Notice that $g(s_+)$ still contains $x$, and this may have affected the position of the saddle-point. We must then make sure \textit{a posteriori} that $x$ is sufficiently large so that $g(s)$ has saturated to its asymptotic value, $g(s)\to s^{-d/2}\exp[1/(4s^2)]$, at the saddle-point $s=s_+$. From \cref{eq:g(s+)}, this amounts at requiring $x\gg 2|\sqrt{c}|$, or equivalently $\Delta \gg 2 R \rho(\Omega)$. When $r$ is negligible, this condition becomes $\Omega \ll \Omega_0 (\Delta/R)^2$: this sets a limit to the values of $\Omega$ for which the saddle-point estimate is valid.

We finally plug \cref{eq:saddled} back into \cref{eq:Q(x)} to get
\begin{equation}
    I_1 \propto e^{-i x \Im[f(s_+)] } = e^{-i x \rho \sin\theta } \, ,
\end{equation}
where we omitted a complex prefactor and extracted the $\Delta$-dependent part of the phase. This justifies the result reported in \cref{eq:effective_wavenumber}.

\section{Numerical simulation}
\label{par:appnumerical}
Numerical simulations are performed by direct integration of the coupled Langevin equations of motion \eqref{eq:field}, \eqref{eq:particle_eom} and \eqref{eq:particleZ} in real space. Field variables are discretized as $\{ \phi_i(t) \}_{i=1}^N$ with $\phi_i(t)\equiv \phi(\vb{x}_i,t) \smallin\mathbb{R}$, and they sit on the $N=L^d$ sites of a $d$-dimensional hypercubic lattice with side $L$. Space is measured in units of the lattice spacing $a$, which we retain for clarity in the following formulas, but which will be eventually set to unity. On the contrary, we will take the particle coordinates $\vb{Y}(t)\,,\, \vb{Z}(t) \smallin\mathbb{R}^d$ to be real-valued, \ie, not constrained to move on the lattice sites only. Upon integration by parts, we may rewrite the equation of motion for $\vb{Y}(t)$ as
\begin{align}
        \dot{\vb{Y}}(t) & = -\nu k\vb{Y} + \nu\lambda \int \dd[d]{x} V(\vb{x}-\vb{Y}) \grad \phi(\vb{x}) + \bm{\xi}(t) \n \\
        &\simeq -\nu k\vb{Y} + \nu\lambda \sum_{i=1}^N V(\vb{x}_i-\vb{Y}) \widetilde{\grad} \phi_i + \bm{\xi}(t) \, ,
    \label{eq:num_particle}
\end{align}
where we introduced the discrete gradient
\begin{equation}
    \widetilde{\grad}_j \phi_i = \frac{ \phi(\vb{x}_i+\hat{\bm{\mu}}_j)-\phi(\vb{x}_i-\hat{\bm{\mu}}_j)}{2a} \, ,
\end{equation}
with $\hat{\bm{\mu}}_j$ locating the position of the $2$ neighbouring sites of each $\vb{x}_i$ along direction $j$. The second particle, $\vb{Z}(t)$, is moved deterministically as in the infinite trap strength limit $k_z \rightarrow \infty$. The discretized equation of motion for the field in model A reads
\begin{align}
    \partial_t\phi_i(t)=&
    -D \big[ (r-\widetilde{\Delta})\phi_i(t)-\lambda V(\vb{x}_i-\vb{Y}(t)) \n\\ 
    &-\lambda V(\vb{x}_i-\vb{Z}(t)) \big] + \zeta_i(t) \, ,
    \label{eq:num_A}
\end{align}
where $\zeta_i(t)$ is a Gaussian random variable with variance $\expval*{\zeta_i(t)\zeta_j(t')}= 2DT a^{-1} \delta_{ij}\delta(t-t')$. We also defined the discrete Laplacian
\begin{equation}
    \widetilde{\Delta} \phi_i = \frac{1}{a^2}  \sum_{\expval{k,i}} \left( \phi_k - \phi_i \right) \, ,
\end{equation}
where the sum runs over the $2d$ neighbouring sites of $\vb{x}_i$. Similarly, the discretized equation of motion for the field in model B reads
\begin{align}
        \partial_t\phi_i(t)=& D\widetilde{\Delta} \big[(r-\widetilde{\Delta})\phi_i(t)-\lambda V(\vb{x}_i-\vb{Y}(t)) \n\\ &-\lambda V(\vb{x}_i-\vb{Z}(t)) \big] + \widetilde{\grad} \cdot \bm{\eta}_i(t) \, ,
    \label{eq:num_B}
\end{align}
where $\bm{\eta}_i(t)$ is a vectorial noise with zero mean and variance $\expval*{\eta_i^{(\alpha)}(t)\eta_j^{(\beta)}(t')}= 2DT a^{-1} \delta_{ij}\delta_{\alpha \beta}\delta(t-t')$, and we take its discrete divergence $\widetilde{\grad}_\alpha \eta_i^{(\alpha)}(t)$. We chose in both cases a Gaussian interaction potential $V_\T{G}(\vb{x})$ as in \cref{eq:gaussianpotential}, which yields a smooth expression for its Laplacian
\begin{equation}
    \laplacian V_\T{G}(\vb{x}) = \frac{\abs{\vb{x}}^2-R^2 d}{R^4} V_\T{G}(\vb{x}) \, .
\end{equation}
Equations \eqref{eq:num_particle} and \eqref{eq:num_A} or \eqref{eq:num_B} represent a set of $(N+d)$ stochastic differential equations which can now be integrated by standard methods in real space. We choose a simple Euler-Maruyama scheme (order $\Delta t^{1/2}$ \cite{frenkel}) for the evolution of the field variables and a more refined method, Stochastic Runge-Kutta (order $\Delta t^{3/2}$, see \ccite{stochasticRK}), for the particle coordinate. We expect this to improve the stability of the particle dynamics in spite of the lower-order algorithm adopted for the field, because the latter only contributes at $\order{\lambda\ll 1}$ to the evolution of the particle.

Once we start the simulation, we have to wait until the system has reached its long-time periodic state, which can be recognized by the fact that the mean value of the oscillations of $\vb{Y}(t)$ stops growing, and from its independence of the field dynamics (model A or B). This process takes longer as we approach criticality, $r=0$, but it is never infinite because the system size $L$ is finite. We may estimate the relaxation time by using \cref{eq:tau_phi} and inserting $q\simeq 2\pi/L$. Once the non-equilibrium periodic state is reached, we record the trajectory of $\vb{Y}(t)$ and use its periodicity in order to average the relevant observables over each period $T=2\pi/\Omega$. This allows to improve the statistics without the need to repeat the initial relaxation for each run.

\section{Dynamical functional for the many-particle problem}
\label{app:many_body}
In this Appendix we derive the Martin-Siggia-Rose dynamical functional \cite{MSR,DeDominicis,Janssen1976} which describes the many-particle problem introduced in \cref{par:many-body}.
The dynamical functionals corresponding to \cref{eq:langevin_many_parts,eq:field} can be obtained by standard methods \cite{Tauber}, leading in the \textit{free} case $\lambda=0$ to
\begin{align}
    &\cor{S}_a[ \vb{X}_a, \widetilde{\vb{X}}_a ] = \label{eq:free_action_parts} \\
    &\int \dd{t} \lgraf \widetilde{\vb{X}}_a(t) \left[ \dot{\vb{X}}_a(t) -  \vb{F}_a(\vb{X}_a(t),t) \right]-\frac{\Omega_a}{2}|\widetilde{\vb{X}}_a(t)|^2 \rgraf \, , \n \\
    &\cor{S}_\phi[ \phi, \tilde{\phi}] = \label{eq:free_action_field} \\
    &\int \dslash{q} \int\dd{t} \Big[ \tilde{\phi}_{-q}(t) \left(\partial_t+\alpha_q \right) \phi_q(t) -\frac{\Omega_\phi(\vb{q})}{2}\tilde{\phi}_q^2(t)\Big] \, . \n
\end{align}
Above we have introduced $\Omega_a \equiv 2\nu_a T$, while $\Omega_\phi(\vb{q})$ was given in \cref{eq:Omega_phi}, and $\widetilde{\vb{X}}_a$, $\tilde{\phi}$ are the auxiliary variables (response fields \cite{Tauber}) conjugate to $\vb{X}_a$ and $\phi$, respectively. Choosing $\lambda\neq 0$ leads to the total action
\begin{align}
    \cor{S}[\phi,\tilde{\phi},\{ \vb{X}_a, \widetilde{\vb{X}}_a\} ] =& \sum_{a=1}^N \cor{S}_a[ \vb{X}_a, \widetilde{\vb{X}}_a ] + \cor{S}_\phi[ \phi, \tilde{\phi}]\n\\
    &-\lambda \cor{S}_\T{int}[\phi,\tilde{\phi},\{ \vb{X}_a, \widetilde{\vb{X}}_a\} ] \, ,
    \label{eq:total_action}
\end{align}
where the interaction terms proportional to the coupling $\lambda$ gave rise to
\begin{align}
    &\cor{S}_\T{int}[\phi,\tilde{\phi},\{ \vb{X}_a, \widetilde{\vb{X}}_a\} ] = \sum_{a=1}^N \int \dd{t} \int \dslash{q} \Big[ Dq^\alpha \tilde{\phi}_q(t) \n\\
    &+ i \nu_a \phi_q(t)\, \vb{q} \cdot \widetilde{\vb{X}}_a(t)  \Big] V_{-q}^{(a)} e^{i \vb{q} \cdot \vb{X}_a(t)} \, .
\end{align}
It is useful at this point to introduce the vector notation
\begin{equation}
    \vb{\Psi} = \vb{\Psi}_q(t) \equiv \mqty( \phi_q(t) \\ \tilde{\phi}_q(t) ) \;, \;\;\; \vb{\Psi}^T \equiv \mqty( \phi_{-q}(t) &  \tilde{\phi}_{-q}(t) ) \, ,
\end{equation}
so as to rewrite in a compact form
\begin{align}
    &\cor{S}_\phi[ \phi, \tilde{\phi}] = \frac{1}{2} \vb{\Psi}^T \hat{A} \vb{\Psi} \, ,\\
    &\cor{S}_\T{int}[\phi,\tilde{\phi},\{ \vb{X}_a, \widetilde{\vb{X}}_a\} ] = \frac{1}{\lambda} \vb{b}^T \vb{\Psi} \, ,
\end{align}
where we introduced the matrix
\begin{equation}
        \hat{A}_{q,p}(t,t') = \mqty( 0 & -\partial_t + \alpha_q \\ \partial_t + \alpha_q & -\Omega(\vb{q})) \delta^d(\vb{q}+\vb{p}) \delta(t-t') \, ,
\end{equation}
and the vector
\begin{equation}
    \vb{b}_q(t) \equiv \sum_{a=1}^N \mqty( -i \nu_a  \vb{q} \cdot \widetilde{\vb{X}}_a(t)  \\ Dq^\alpha ) V_q^{(a)} e^{-i \vb{q} \cdot \vb{X}_a(t)} \, .
\end{equation}
The effective action $\cor{S}_\T{eff}[\{ \vb{X}_a, \widetilde{\vb{X}}_a\} ] $ which describes the particles alone will have the form of \cref{eq:effective_dynamical_action}, where the free part $\cor{S}_0[\{ \vb{X}_a, \widetilde{\vb{X}}_a\} ]$ is simply the sum of the single-particle actions $\cor{S}_a[ \vb{X}_a, \widetilde{\vb{X}}_a ]$ given in \cref{eq:free_action_parts}.
In order to obtain the interacting part $\cor{S}_\lambda[\{ \vb{X}_a, \widetilde{\vb{X}}_a\} ]$, we marginalize over the field $\phi$ and its conjugate variable $\tilde{\phi}$ as
\begin{align}
    &e^{-\cor{S}_\lambda[\{ \vb{X}_a, \widetilde{\vb{X}}_a\} ]} \equiv \int \cor{D}\phi \, \cor{D}\tilde{\phi }\, e^{-\cor{S}_\phi+\lambda \cor{S}_\T{int}} \n\\
    &= \int \cor{D}\phi \, \cor{D}\tilde{\phi } \, e^{-\frac{1}{2} \vb{\Psi}^T \hat{A} \vb{\Psi}+\vb{b}^T \vb{\Psi} } \propto e^{ \frac12 \vb{b}^T \hat{A}^{-1} \vb{b}} \, .
    \label{eq:gaussian_integration}
\end{align}
The result of the Gaussian integration involves the inverse matrix \cite{Tauber}
\begin{equation}
        \hat{A}^{-1}_{q,p}(t,t') = \mqty( C_q(t,t') & G_q(t-t') \\ G_q(t'-t) & 0) \delta^d(\vb{q}+\vb{p})  \, ,
\end{equation}
where $C_q(t,t')$ and $G_q(t)$ are the correlator and the linear response function of the field given in \cref{eq:field-susc,eq:field-prop}, respectively. Equation \eqref{eq:gaussian_integration} is only formal, but it can be made explicit by integrating over the dummy variables (times and momenta) as
\begin{equation}
    \vb{b}^T \hat{A}^{-1} \vb{b} \equiv \int \frac{\dd[d]{q}\dd[d]{p}}{(2\pi)^{2d}} \int \dd{t} \dd{t'}\vb{b}_{-q}(t) \hat{A}^{-1}_{q,-p}(t,t')\vb{b}_{p}(t') \, .
\end{equation}
The resulting $\cor{S}_\lambda[\{ \vb{X}_a, \widetilde{\vb{X}}_a\} ]$ can then be expressed as in \cref{eq:int_action}, with
\begin{widetext}
\begin{equation}
    \cor{S}_{ab}[ \vb{X}_a, \widetilde{\vb{X}}_a ,\vb{X}_b, \widetilde{\vb{X}}_b ] 
    = \nu_a \int \dslash{q} V_q^{(b)} V_{-q}^{(a)} \int \dd{t} \int \dd{t'} [\vb{q} \cdot \widetilde{\vb{X}}_a(t)] \, e^{i \vb{q} \cdot [\vb{X}_a(t)- \vb{X}_b(t')]} \left\lbrace i \chi_q(t-t')  +\frac{\nu_b}{2} C_q(t,t')  [\vb{q} \cdot \widetilde{\vb{X}}_b(t')] \right\rbrace ,
    \label{eq:int_action_parts}
\end{equation}
\end{widetext}
and where $\chi_q(t)$ is the linear susceptibility of the field given in \cref{eq:field-susc}. We note that the terms with $a=b$ in \cref{eq:int_action} describe the self-interaction of the particle $\vb{X}_a$ mediated by the field $\phi$, while the terms with $a\neq b$ describe the field-induced interaction between pairs of particles. Finally, we recognize in \cref{eq:int_action_parts} a \textit{drift} term (\ie, the one containing $\chi_q(t-t')$) which is non-local in time, and a colored \textit{noise} term (\ie, the one containing $C_q(t,t')$): they both result from having integrated out the field degrees of freedom from the dynamics.
\end{document}